\documentclass[a4paper,11pt]{article}
\pdfoutput=1 

\usepackage{jcappub} 
\usepackage{xspace}
\usepackage{xcolor}
\usepackage{booktabs}
\usepackage[noabbrev]{cleveref}
\usepackage{lineno}
\usepackage{hyperref}
\usepackage{orcidlink}

\defcitealias{dMdB2020}{\texttt{dMdB20}}
\newcommand{\dMdB}{\citetalias{dMdB2020}\xspace}

\newcommand{\lya}{Ly$\alpha$\xspace}
\newcommand{\lyaf}{Ly$\alpha$ forest\xspace}
\newcommand{\ion}[2]{#1\thinspace{}#2\xspace}
\newcommand{\ciii}{C\thinspace{}III\xspace}
\newcommand{\angs}{\textup{\AA}}
\newcommand{\lyaxlya}{Ly$\alpha\times$Ly$\alpha$\xspace}
\newcommand{\lyaxlyaA}{Ly$\alpha$(A)$\times$Ly$\alpha$(A)\xspace}
\newcommand{\lyaxlyaB}{Ly$\alpha$(A)$\times$Ly$\alpha$(B)\xspace}
\newcommand{\lyaBxlyaB}{Ly$\alpha$(B)$\times$Ly$\alpha$(B)\xspace}
\newcommand{\lyaxqso}{Ly$\alpha\times$QSO\xspace}
\newcommand{\lyaxqsoA}{Ly$\alpha$(A)$\times$QSO\xspace}
\newcommand{\lyaxqsoB}{Ly$\alpha$(B)$\times$QSO\xspace}

\newcommand{\Mpc}{\mathrm{Mpc}}
\newcommand{\hMpc}{h^{-1}\,\mathrm{Mpc}}
\newcommand{\ap}{\alpha_\parallel}
\newcommand{\at}{\alpha_\perp}
\newcommand{\zeff}{z_{\rm eff}}

\title{DESI 2024 IV: Baryon Acoustic Oscillations from the Lyman Alpha Forest}

\author{{DESI Collaboration}:}
\emailAdd{spokespersons@desi.lbl.gov}
\affiliation{Affiliations are in Appendix \ref{sec:affiliations}}

\author[1]{{A.~G.~Adame},}
\author[2]{{J.~Aguilar},}
\author[3]{{S.~Ahlen}\orcidlink{0000-0001-6098-7247},}
\author[4]{{S.~Alam}\orcidlink{0000-0002-3757-6359},}
\author[5,6]{{D.~M.~Alexander}\orcidlink{0000-0002-5896-6313},}
\author[2]{{M.~Alvarez},}
\author[7]{{O.~Alves},}
\author[2]{{A.~Anand}\orcidlink{0000-0003-2923-1585},}
\author[8,7]{{U.~Andrade}\orcidlink{0000-0002-4118-8236},}
\author[9]{{E.~Armengaud}\orcidlink{0000-0001-7600-5148},}
% ...

\abstract{
We present the measurement of Baryon Acoustic Oscillations (BAO) from the Lyman-$\alpha$ (\lya) forest of high-redshift quasars with the first-year dataset of the Dark Energy Spectroscopic Instrument (DESI). 
Our analysis uses over $420\,000$ \lya\ forest spectra and their correlation with the spatial distribution of more than $700\,000$ quasars.
An essential facet of this work is the development of a new analysis methodology on a blinded dataset. 
We conducted rigorous tests using synthetic data to ensure the reliability of our methodology and findings before unblinding. 
Additionally, we conducted multiple data splits to assess the consistency of the results and scrutinized various analysis approaches to confirm their robustness.
For a given value of the sound horizon ($r_d$), we measure the expansion at $\zeff=2.33$ with 2\% precision, $H(\zeff) = \left( 239.2 \pm 4.8 \right) \left(147.09~\Mpc /r_d \right)$ km/s/Mpc.
Similarly, we present a 2.4\% measurement of the transverse comoving distance to the same redshift, $D_M(\zeff) = \left( 5.84 \pm 0.14 \right) \left(r_d/147.09~\Mpc \right)$ Gpc.
Together with other DESI BAO measurements at lower redshifts, these results are used in a companion paper to constrain cosmological parameters.}

\begin{document}
\maketitle

\flushbottom

\section{Introduction} 
\label{sec:introduction}

Baryon Acoustic Oscillations (BAO) in the distribution of matter are a unique probe of the cosmic expansion history and the geometry of the Universe \cite{Weinberg2013}. By themselves, BAO measurements at different redshifts enable precise measurements of the energy density parameters. In combination with studies of the Cosmic Microwave Background (CMB), they allow us to better constrain extensions to the $\Lambda$CDM model~\cite{Planck2018,eBOSS2021}.

The Dark Energy Spectroscopic Instrument (DESI, \cite{DESI2016a.Science}) project aims to measure BAO with unprecedented precision over a wide range of redshifts. DESI is in the midst of a five-year campaign to obtain accurate redshifts for 40 million galaxies and quasars over $14\,000$ square degrees. 
The survey started in May 2021 and the upcoming DESI Data Release 1 (DESI DR1, \cite{DESI2024.I.DR1}), covering  data collected until June 14, 2022, contains about 13 million galaxies, 1.5 million quasars, and 4 million stars over an area of more than $9\,500$ square degrees\footnote{$9\,500$ square degrees is the area covered by the dark time survey where quasars are observed, the total area surveyed by DESI with other programs is larger.}. DESI has obtained seven BAO measurements at different redshifts with this DR1 dataset.

The BAO measurements from the clustering of DESI galaxies and quasars at $z<2$ are presented in \cite{DESI2024.III.KP4}. 
In this publication we present a BAO measurement at $z=2.33$ using the auto-correlation of the Lyman-$\alpha$ (\lya) forest dataset from DESI DR1 and its cross-correlation with quasar positions.
Neutral hydrogen along the line-of-sight towards high-redshift quasars cause absorption features in their spectra. While gas pressure dominates the distribution of gas on scales of tens of kiloparsecs \cite{McQuinn2016}, on larger scales the \lya forest is a powerful tracer of the density fluctuations \cite{Slosar2011}.
The cosmological interpretation of all DESI DR1 BAO measurements is presented in \cite{DESI2024.VI.KP7A}.

While the first BAO measurements \cite{Eisenstein2005,Cole2005} used the distribution of galaxies to trace the density fluctuations, the \lyaf can also be used to extend the BAO measurements to higher redshifts than those available with current galaxy surveys.
The Baryon Oscillation Spectroscopic Survey (BOSS, \cite{Dawson2013}) presented the first BAO measurement using the auto-correlation of the \lyaf \cite{Busca2013,Slosar2013,Kirkby2013} with measurements of 50 000 quasar spectra from the 9th data release of the Sloan Digital Sky Survey (SDSS DR9, \cite{Eisenstein2011, Ahn2012}).
With the 11th data release (DR11, \cite{Alam2015}), BOSS presented as well the first BAO measurement from the cross-correlation of quasars and the \lyaf \cite{FontRibera2014}, which doubled the amount of information available from the same dataset.

Updated \lya BAO measurements were published following subsequent SDSS data releases \cite{Delubac2015,Bautista2017,dMdB2017,dSA2019,Blomqvist2019}, including data from the Extended Baryon Oscillation Spectroscopic Survey (eBOSS, \cite{Dawson2016}).
The final \lya BAO results used $210\,000$ \lya forests from BOSS and eBOSS observations in the 16th data release (SDSS DR16, \cite{Ahumada2020}). 
These results were published in du Mas des Bourboux et al. (2020) \cite{dMdB2020} and are referred to as \dMdB in the rest of this article.
\dMdB has been the state-of-the art in \lya BAO measurements until this publication, and we compare the measurements and discuss the methodological differences in the next sections (notably the use of a cross-covariance between the auto-correlation of the \lya forest and its cross-correlation with quasars).
In this work we use over $420\,000$ \lya forests, doubling the number of lines of sight used in \dMdB.

The structure of this paper is designed to guide readers through the main aspects of the BAO measurements with the \lyaf using DESI's first year of data. 
We start in \cref{sec:data} with a description of the DESI survey and the datasets used in our analysis, i.e. the \lyaf absorption and quasar catalogues \citep{QSO.TS.Chaussidon.2023,VIQSO.Alexander.2023}, along with the masking of Broad Absorption Lines and Damped Lyman-$\alpha$ systems found in the spectra.
The methodology to extract the \lyaf fluctuations from the spectra of DESI quasars is presented in \cite{2023MNRAS.tmp.3626R}, and it is summarised in \cref{sec:cont_fit}.
We present the measurement of correlations in \cref{sec:correlations} and how we model them in \cref{sec:model}.
The methodology employed in both of these sections is described in more detail in \cite{2023JCAP...11..045G}.
In \cref{sec:model} we also summarize the (minor) contamination from sky residuals and other pipeline-induced systematics, which is described in detail in a companion paper \cite{KP6s5-Guy}.
In \cref{sec:results} we present the main result from this work: the most precise BAO measurement at $z>2$ to date.

In order to minimize confirmation (and other) biases, the DESI Collaboration used blinding strategies in the BAO measurements.
The analysis methodology was entirely developed on blinded measurements, and we were only allowed to \textit{unblind} the measurement once we had fulfilled a long list of validation tests. 
The blinding strategy used in the \lya BAO measurement is described in \cref{app:blinding}, and the analysis validation is presented in \cref{sec:validation}. 
In \cite{KP6s6-Cuceu} we present the validation tests in more detail. These tests are based on 150 synthetic realisations of our dataset (or mocks).
These mocks were generated following the methods described in \cite{2024arXiv240100303H}.
In \cref{sec:discussion} we contextualize our findings, compare them to the \lya BAO measurement from \dMdB, and discuss a few options to improve future \lya BAO analyses.
We conclude in \cref{sec:conclusions}.
\section{Data Sample} 
\label{sec:data}

We use quasar spectra collected during the first year of the DESI main survey.
The observations were conducted with the Mayall 4-m telescope at Kitt Peak National Observatory, in Arizona, with a new prime focus, multi-fiber spectrograph.
It consists of a new corrector equipped with an atmospheric dispersion compensator \cite{Corrector.Miller.2023} providing a 3.2 degree diameter field of view, and a focal plane composed of 5000 robotically actuated fibers \cite{FocalPlane.Silber.2023} that distribute the light to 10 spectrographs situated below the telescope, in a temperature-controlled room. Each spectrograph is composed of three arms, blue (3600-5930\AA), red (5600-7720\AA) and near infrared (7470-9800\AA). 
In this analysis we use only data from the blue arms. 
In each blue camera, the light from 500 fibers is dispersed and refocused, forming 500 spectral traces on a 4096x4096 pixel STA4150 CCD. 
Despite this dense fiber packing, the cross-contamination of the spectra from adjacent fibers was measured to be only of 0.2\% or smaller after spectral extraction (for wavelengths shorter than 8900\AA, see \cite{Spectro.Pipeline.Guy.2023} section 4.1). This effect is corrected in post-processing and we expect the residual cross-talk to be negligible.
The CCD pixel size is 15~$\mu$m, corresponding to 0.6~\AA\ in wavelength. 
The spectrograph line spread function is 1.8~\AA\ FWHM in the blue channel, which corresponds to a resolution from 2000 to 3400 depending on the wavelength. 
Many more details on the instrument can be found in \cite{DESI2022.KP1.Instr}.

The main survey started on May 14, 2021 after a survey validation period (see \cite{DESI2023a.KP1.SV} and references therein) used to tune the target selection, the fiber assignment, the exposure times, and exercise the data processing pipeline that is run every day to validate the observations. DESI is running two programs for bright and dark time, switching dynamically between them with the observation conditions (moonlight, but also seeing, sky brightness and sky transparency). The choice of pointing (called a {\it tile} when combined with a specific selection of targets to observe) and the allocation of fibers to targets is performed automatically during the night \citep{TS.Pipeline.Myers.2023,SurveyOps.Schlafly.2023}.
The exposure times are adjusted in real time in order to obtain the same signal to noise for a fiducial target in each pointing. The quasar sample is observed during dark time. The quasar targets that are spectroscopically confirmed and have a redshift $>2.1$ are scheduled for re-observation to build up the signal to noise necessary for the \lya\ forest measurements. The objective is to acquire four exposures for each quasar, each with an effective exposure time\footnote{The effective exposure time corresponds to the exposure time in ideal observing conditions: at zenith, with a nominal seeing, in photometric conditions \citep{SurveyOps.Schlafly.2023}.} of 1000~sec. 
We use data collected until June 14, 2022 for this analysis which will be made public in the DESI Data Release 1 (DESI DR1) \citep{DESI2024.I.DR1}.

Data collected on the mountain are transferred to NERSC (National Energy Research Scientific Computing Center) within minutes and processed by the offline pipeline \citep{Spectro.Pipeline.Guy.2023}. CCD images are preprocessed, the spectra extracted and calibrated for each exposure. It is worth noting that the extraction algorithm returns spectra with uncorrelated noise on the same wavelength grid of bin 0.8~\AA\ for all fibers and exposures, which simplifies the co-addition (or averaging) of spectra from the same target obtained in different exposures. The data processing pipeline provides 2 sets of co-added spectra, one per spectrograph and tile, combining data of several exposures and nights for the same pointing and fiber allocation, and one per \texttt{HEALPix} pixel \citep{Gorski2005} on the sky where the full co-added spectra (across exposures and tiles) of all the targets in a \texttt{HEALPix} pixel are saved. We use this later data set for this analysis.
These spectra include fluxes, a wavelength array (in vacuum, and in the solar system barycenter frame), an estimate of the flux variance, a mask to flag bad pixels, and a resolution matrix that encodes the line spread function of the instrument \citep{Spectro.Pipeline.Guy.2023}. 

In the subsequent sections, we delineate the processing applied to our quasar sample,  both at the catalog and spectral level, including the identification of Damped Lyman-$\alpha$ (DLAs) systems and  Broad Absorption Line (BAL) troughs, to yield the ultimate datasets for our analysis. While a comprehensive overview is presented in \citep{2023MNRAS.tmp.3626R}, here we emphasize on the small but yet significant adjustments made to enhance our methodology.

\subsection{The quasar catalog}
\label{subsec:quasar_catalog}

Studies of the \lya forest require a pure sample of quasars with accurately determined redshifts. To meet these stringent criteria, DESI employs a multi-step classification and redshift fitting procedure for the identification of quasar spectra. 
This process is extensively outlined in \citep{QSO.TS.Chaussidon.2023} (see their figure 9 summarizing the workflow) and complementary information is also provided in \citep{QSOPrelim.Yeche.2020,VIQSO.Alexander.2023,DESI2024.I.DR1}. 
The cornerstone of DESI's classification and redshift fitting is \texttt{Redrock}, a template fitting code \citep{Redrock.Bailey.2024}. It utilizes Principal Component Analysis (PCA) templates representing three broad object classes (stars, galaxies, and quasars) while scanning a range of redshifts. 
The determination of the most probable spectral class and redshift for a given spectrum relies on the lowest $\chi^{2}$ fit.
Additionally, DESI employs two quasar-specific classifiers, referred to as ``afterburners", to enhance the completeness of the quasar sample by $\sim 10\%$ \citep{VIQSO.Alexander.2023,QSO.TS.Chaussidon.2023}. One afterburner scrutinizes spectra not classified as quasars by \texttt{Redrock} for broad MgII emission; upon detection, it reclassifies the spectrum as a quasar without altering the assigned redshift. The second afterburner, known as \texttt{QuasarNET} \citep{Busca18,Farr20_QN}, employs a deep convolutional neural network to identify potential quasar emission features and estimate the redshift. If \texttt{QuasarNET} identifies a target as a quasar, there is another \texttt{Redrock} fitting iteration, although this time with only quasar templates and a redshift prior based on the \texttt{QuasarNET} result. 

The resulting DESI DR1 quasar catalog was produced by the three classifiers, and is slated for publication as part of first DESI Data Release (DR1) \citep{DESI2024.I.DR1}. This catalog boasts an estimated completeness and redshift purity that surpasses 95\% and 98\%, respectively \citep{RedrockQSO.Brodzeller.2023}. 

From this catalog we keep only quasars with \texttt{ZWARN=0} or \texttt{ZWARN=4}, rejecting quasars with any issue reported by the spectroscopic pipeline except for the low $\Delta \chi^2$ flag of \texttt{Redrock} that is irrelevant for QSOs.
We also use updated redshifts for this \lya analysis. These are based on new quasar templates tailored for high redshifts, specifically $1.4 < z < 7.0$. 
Additionally, a slightly modified version of the \texttt{Redrock} code was used that integrates a more recent optical depth model from \citep{Kamble2020}. 
These subtle adjustments were prompted by the identification of a bias for redshifts $z>2$, as identified by their impact on the \lya quasar cross-correlation (see \citep{KP6s4-Bault} for more details). 
The statistical precision of the updated redshifts is estimated to be better than 150~km~s$^{-1}$, with a catastrophic redshift failure rate of $\sim$2.5\% ($\sim$4\%) for redshifts in error by more than 3000 km~s$^{-1}$ (1000 km~s$^{-1}$).

\begin{table}
\centering
\begin{tabular}{ccccc}
Catalogue                & Num quasars              & Too short               
 & Negative continuum       & Valid forests         \\
\hline
\hline
Total                    & 1529530                  & -                        & -                        & -                     \\
Tracers $(z>1.77)$        & 709565                   & -                        & -                        & -                     \\

\hline
\lya, region A & 531000                & 83666 (15.8\%)                  & 18931 (3.6\%)                  & 428403 (80.7\%)               \\
\lya, region B & 199449                   & 53162 (26.7\%)                   & 8855 (4.4\%)                    & 137432 (68.9\%)               \\
\ion{C}{III} region  & 1183522                  & 66506 (5.6\%)                   & 5753 (0.5\%)                    & 1111263 (93.9\%)              \\
\end{tabular}
    \caption{Total number of quasars in DESI DR1 that pass our selection criteria. 
    Those with $z>1.77$, referred to as ``tracers'', contribute to the measurement of the \lyaxqso cross-correlation.
    The bottom part of the table shows the statistics of number of lines of sight available in each of the three rest-frame wavelength regions (or forests) discussed in \cref{subsec:forests}. 
    Some of the forests are discarded because they do not have 150 valid pixels (too short), or because they do not have a valid continuum fit (negative continuum).
    These cuts are described in \cref{sec:cont_fit}.
    }
    \label{tab:num_qso}
\end{table}

\begin{figure}
    \begin{minipage}{0.7\linewidth}
    \centering
    \includegraphics[width=\textwidth]{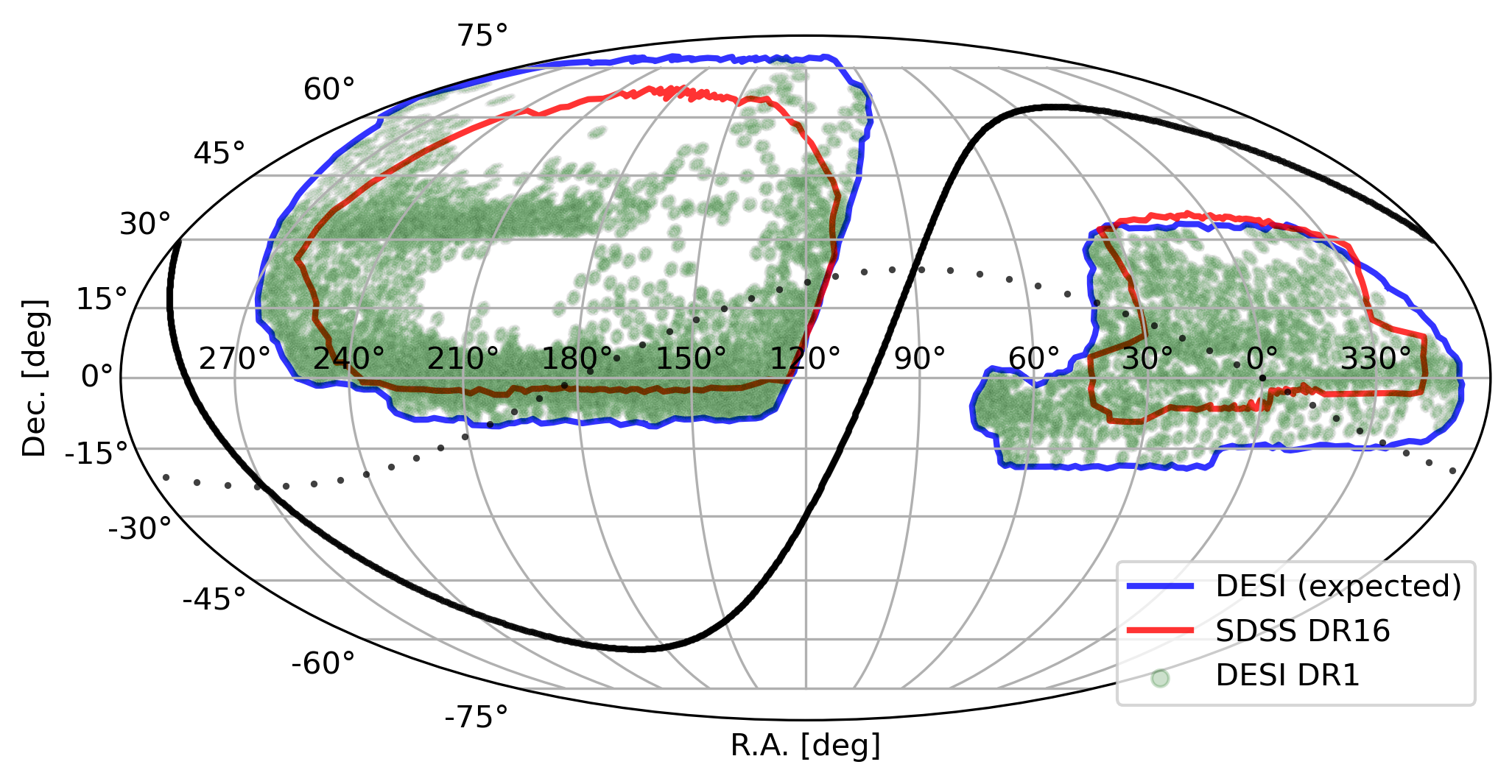}
\end{minipage}
\begin{minipage}{0.29\linewidth}
    \centering
\includegraphics[width=\textwidth]{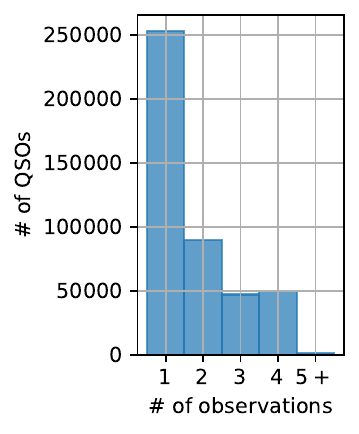}
\end{minipage}
\caption{Left: Expected final DESI (blue) and SDSS-DR16 footprint (red) together with the spatial distribution of DESI DR1 observed quasars (green). 
For reference we also show the Galactic plane (solid black) and the Ecliptic plane (dotted black). Right: Number of observations for \lya quasars in the DESI DR1 sample.}
   
\label{fig:eboss_desi_footprint}
\end{figure}

In the left panel of \cref{fig:eboss_desi_footprint} we show the spatial distribution of quasars in the DESI DR1 sample (green points), compared to the SDSS DR16 footprint \dMdB (red curve), and the expected final footprint DESI (blue curve), while the right panel shows the distribution of number of observations for \lya quasars in DESI DR1.
The number of quasars in the DESI DR1 sample is given in \cref{tab:num_qso}, together with the number of quasar spectra covering each of the rest-frame wavelength ``regions" described in \cref{subsec:forests}.

\begin{figure}
    \centering
    \includegraphics[width=0.95\textwidth]{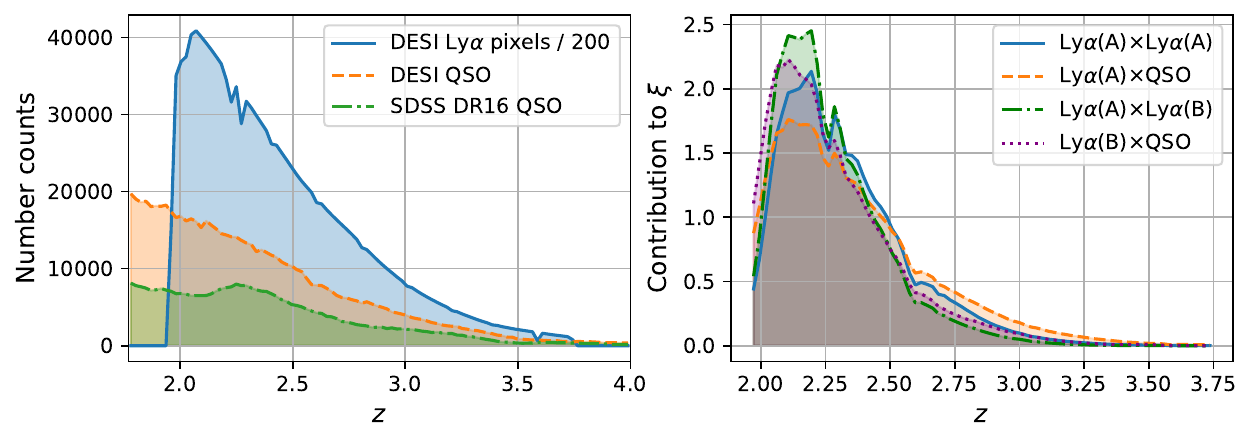}
    \caption {Left: Redshift distribution of quasars in DESI DR1 (orange), compared to the distribution in SDSS DR16 (green) and to the distribution of \lya pixels (blue, divided by 200). 
    Right: Contribution of different redshifts to the four measured correlation functions. 
    In particular, we show the sum of weights used in \cref{eqn:auto_corr,eqn:cross_corr} as a function of redshift, for large transverse separations (to reduce the biasing effect from the clustering of background quasars) and $r_\parallel=0$ (the contribution varies as a function of $r_\parallel$, especially for quasar cross-correlations).
    As described in \cref{sec:correlations}, we measure \lya correlations in two different rest-frame wavelength regions of the quasar spectra (A and B).}
    \label{fig:eboss_desi_data}
\end{figure}

The redshift distribution of DESI DR1 and SDSS DR16 quasars are compared in the left panel of \cref{fig:eboss_desi_data}, together with the redshift distribution of \lya pixels. 
In the right panel of the same figure we show the contribution of different redshifts to the measurement of the four correlation functions discussed in \cref{sec:correlations}.
Integrating the curve, one finds that 95\% of the measurement of the \lyaxlya auto-correlation comes from $1.96 < z < 2.8$, while 95\% of the \lyaxqso cross-correlation comes from $1.96 < z < 2.95$. 
As discussed in \cref{sec:correlations}, however, quasars at redshifts as low as $z=1.77$ and as high as $z=4.16$ can also contribute to the measurement of the cross-correlation.

\subsection{Damped \lya\ systems}\label{sec:dla}

The \lya forest BAO measurement exploits the fact that typical fluctuations in the forest trace matter fluctuations of moderate over density, or under density, along the line of sight.  DLA systems are produced by neutral hydrogen concentrations with column density, $N_{\rm HI} > 2 \times 10^{20} \, \rm{cm}^{-2}$, typically arising in the interstellar medium of high redshift galaxies. Because of the damping wings, each DLA can affect a noticeable fraction of a \lya forest spectrum (corresponding to thousands of km/s), so even though these systems are rare, it is important to mask the contaminated regions of the spectra, in order to make a robust  BAO measurement with uncertainties that can be easily modeled. 
In DESI we have used two DLA finders, one based on a convolutional neural network (CNN), an algorithm first proposed in \citep{2018MNRAS.476.1151P}, and another one based on a Gaussian Process (GP) model \citep{2020MNRAS.496.5436H}. 
Their adaptation to DESI and performance in DESI's simulated spectra is presented in \citep{Wang2022}.
The performance of both algorithms in the DESI DR1 dataset will be presented in an upcoming paper. 

For the purpose of this work, we construct a DLA catalog based on the combined result of the two DLA finders. We select DLAs found by the two algorithms with a threshold probability of 50\%, as done in \cite{2023MNRAS.tmp.3626R,2023JCAP...11..045G}. However, in order to ensure the purity of the DLA catalog we also limit it to DLAs detected in spectra with signal-to-noise (SNR) larger than 3 (mean SNR within the \lya forest region). 
The DLAs included in the final catalog are masked in the spectra before computing the forest fluctuations (see \cref{subsec:forests}) and the contamination from undetected ones is taken into account in the modelling of the correlation functions (see \cref{subsec:hcds}).

\subsection{Broad Absorption Line quasars}\label{sec:bal}

Numerous studies of large quasar samples have identified broad absorption line (BAL) troughs in 10 -- 30\% of quasars \citep[e.g.,][]{Foltz1990,Trump2006}. While these features are most commonly seen as blue shifted absorption relative to the \ion{C}{IV} emission feature, the absorption is also associated with many other features, including several that contribute absorption in the Ly$\alpha$ forest region \citep{Masribas2019}. 
Significant BAL features associated with \ion{C}{IV} and other strong emission lines are also expected to introduce some systematic redshift biases and increase redshift errors for a subset of BAL quasars \citep{Garcia2023}. We consequently need to identify and characterize BALs in the DESI quasar sample to mitigate their impact on the cosmological analysis. 

We catalog BALs in the DESI quasar sample following the same approach employed by \cite{Filbert2023} for the DESI Early Data Release (EDR). 
This approach fits a series of continuum components to each quasar, and iteratively identifies and masks regions that meet the Absorptivity Index \citep[AI,][]{Hall2002} and Balnicity Index \citep[BI,][]{Weymann1991} criteria. 
For the DR1 BAL catalog, we use the same components developed by \cite{Brodzeller2022}.
The catalog includes measurements of the velocity range of each trough, AI and BI values, and other quantities measured relative to the \ion{C}{IV} and \ion{Si}{IV} features, similar to the DESI EDR catalog.
This catalog is used in our analysis to mask the corresponding locations of BAL features before computing the \lya forest fluctuations (see \cref{subsec:forests}).

\subsection{Catalog of the \lya forest}
\label{subsec:forests}

\begin{figure}
    \centering
    \includegraphics[width=0.98\textwidth]{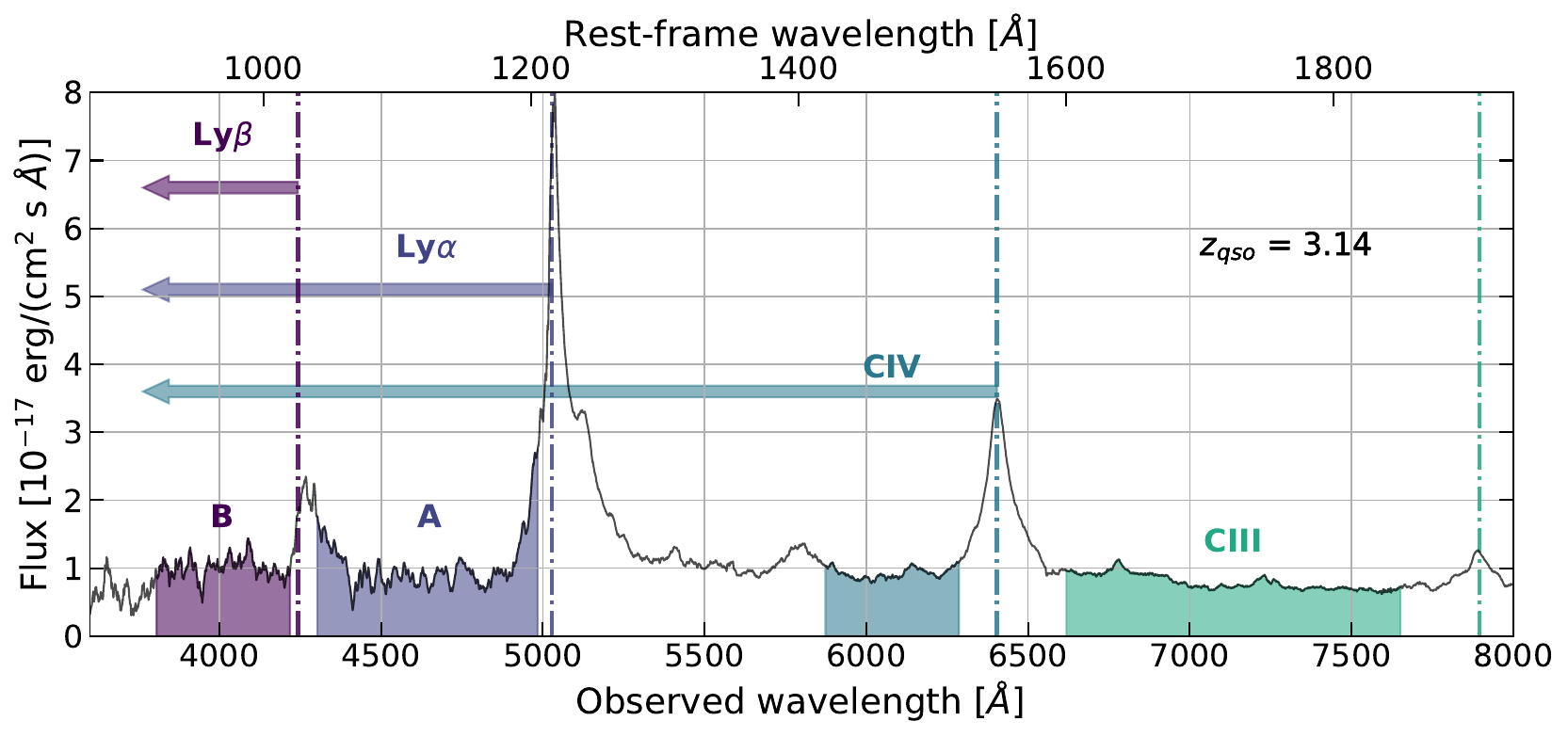}
    \caption{QSO spectrum from the first year of DESI data at redshift $z=3.14$ (TargetID = 39627581225438176).
    The region B is highlighted in purple. 
    The region A is highlighted in indigo. 
    The \ion{C}{IV} and \ion{C}{III} regions are highlighted in various shades of green. 
    While there is almost no \ion{C}{III} absorption, the \ion{C}{IV} absorption spans leftward of the \ion{C}{IV} doublet, contaminating the \lya\ regions A and B. 
    The \lya absorption extends into region B.
    For better visualization, we have chosen a relatively high signal-to-noise spectrum and we have smoothed it by averaging the pixels in groups of 25.
    }
    \label{fig:desi_spectrum}
\end{figure}

We discuss here the subset of quasar spectra that we use to study the fluctuations in the \lya forest.
We consider two different catalogues that correspond to \lya absorption in two different regions of the quasar spectra. Region ``A" is defined as the rest-frame wavelength range from $1040$ to $1205\angs$ \footnote{Following \cite{2023MNRAS.tmp.3626R}, we extend the A region to $1205\angs$, going a bit further than the $1200\angs$ limit used in \dMdB.}, while region ``B" covers the range from $920$ to $1020\angs$.
Pixels in the B region are also affected by absorption from other Lyman lines, since it extends beyond the Ly$\beta$, Ly$\gamma$ and Ly$\delta$ emission lines, and therefore it require special treatment
\footnote{Note that these regions were referred to as \lya and Ly$\beta$ regions in \dMdB.}.
See \cref{fig:desi_spectrum} for an illustration of these regions in a quasar observed with DESI.

In addition to the rest-frame wavelength cut, we also restrict our observed wavelength range from $3600\angs$ to $5772\angs$. 
The lower bound is the minimum wavelength provided by the pipeline; the upper bound corresponds to the middle of the overlap region between the blue and the red arms of the spectrographs\footnote{We could extend the analysis to pixels in the red spectrograph, but the gain would be marginal given the limited number of high redshift quasars.}.
These cuts in observer-frame and rest-frame wavelength limit the redshift range of the background quasars from $z=2.1$ to $z=4.4$ for the region A and $z=2.6$ to $z=5.1$ for the region B. The number of quasars available for these catalogues is given in \cref{tab:num_qso}.


Once we have the initial set of \lya forests, we apply four different masks to remove bad pixels and astrophysical contaminants. The first mask comes from the  data reduction pipeline, which flags bad pixels in the spectra \citep{Spectro.Pipeline.Guy.2023}. These are usually caused by cosmic rays or defects in the CCD cameras. As shown in \cite{2023MNRAS.tmp.3626R}, the fraction of pixels masked as a function of observed wavelength is generally constant and below 1\%. The second mask consists of a set of narrow windows in observed wavelength: $3933.0$ to $3935.8\angs$, $3967.3$ to $3971.0\angs$, and $5570.5$ to $5586.5\angs$ (see \cite{2023MNRAS.tmp.3626R}). This mask aims to remove galactic absorption not observed in the calibration stars (first two lines) and residual sky lines from the sky subtraction models \citep{Spectro.Pipeline.Guy.2023}.

The other two masks aim at removing the effects of astrophysical contaminants. In particular, we mask DLAs (see \cref{sec:dla}) and BALs (see \cref{sec:bal}). For DLAs, we mask the regions of the \lya{} forest where the presence of a DLA decreases the transmitted flux by 20\% or more. We then correct the remaining DLA wings using a Voigt profile.
For BALs, we follow the procedure described in \cite{Ennesser2022}, which we summarize here. 
We mask the expected locations of all potential BAL features associated with, irrespective of whether or not the absorption is apparent. 
These include \lya, \ion{N}{IV}, \ion{C}{III}, \ion{Si}{IV}, and \ion{P}{V} in region A, and \ion{O}{VI}, \ion{O}{I}, Ly$\beta$, Ly$\gamma$, \ion{N}{III} and Ly$\delta$ in region B.
Because of this, the mask may remove some path length that is unaffected by BALs. However, we note that there is a net increase in path length compared to completely discarding these spectra as done in previous analyses \citep[e.g.][]{dMdB2020}. The paper by \cite{KP6s9-Martini} investigates the impact of other masking strategies on the BAO analysis. 

At this stage, we discard forests that are too short (defined as having less than 150 valid pixels corresponding to a length of $120\angs$). 
This is necessary as we later fit the unabsorbed quasar continua to extract the transmitted flux field (see \cref{sec:cont_fit}), and having forests that are too short interferes with the continuum fitting procedure. 
The number of forests lost because of this in each of the samples is given in \cref{tab:num_qso}.

Finally, to check for unaccounted-for calibration residuals, we compute the mean transmitted flux fraction in a quasar rest-frame region where there is no \lya absorption. 
As in \cite{2023MNRAS.tmp.3626R}, we use the \ciii{} region ($\lambda_{\rm rf} \in \left[1600, 1850\right]\angs$). 
We note that previous analysis from eBOSS used longer rest-frame wavelengths,
but \cite{2023MNRAS.tmp.3626R} showed that the measured calibration residuals are the same irrespective of the metal region used to measure them, and there are more spectra available in the \ciii{} region.
We use this small correction (less than 5\% in the relevant wavelengths range) to re-calibrate our fluxes and instrumental noise estimates.

\subsection{Continuum fitting}\label{sec:cont_fit}

The initial step in our data analysis is to compute the transmitted flux field $\delta_{q}\left(\lambda\right)$. In general terms, it is defined as the ratio of the observed flux to the expected flux:
\begin{equation}
    \delta_{q}\left(\lambda\right) = \frac{f_{q}\left(\lambda\right)}{\overline{F}\left(\lambda\right)C_{q}\left(\lambda\right)} - 1~,
    \label{eqn:delta_definition}
\end{equation}
where $\overline{F}\left(\lambda\right)$ is the mean transmission and $C_{q}$ is the unabsorbed quasar continuum. 
The sub-index $q$ indicates that these are line-of-sight (quasar) dependent. We label the process of estimating their product as {\it continuum fitting}.

The continuum fitting procedure we use is described in detail in \cite{2023MNRAS.tmp.3626R} and we summarize it here. 
The product $\overline{F} C_{q}\left(\lambda\right)$ is taken to be a universal function of the quasar $\overline{C}\left(\lambda_{\rm rf}\right)$, corrected by a first degree polynomial in $\Lambda\equiv\log\lambda$ to account for quasar diversity and the redshift evolution of $\bar{F}(z)$:
\begin{equation}
    \overline{F}\left(\lambda\right) C_{q}\left(\lambda\right) = \overline{C}\left(\lambda_{\rm rf}\right) \left(a_{q} + b_{q}\frac{\Lambda - \Lambda_{\rm min}}{\Lambda_{\rm max} - \Lambda_{\rm min}}\right) ~.
    \label{eqn:aqbq_definition}
\end{equation}

Here, the universal function $\overline{C}\left(\lambda_{\rm rf}\right)$ is computed to be the weighted mean (inverse variance weighting) of all the lines of sight entering the analysis. The parameters $a_{q}$ and $b_{q}$ are estimated by maximizing the likelihood function
\begin{equation}
    2\ln{\mathcal L} = -\sum_{i}\frac{\left(f_{i} - \overline{F}C_{q}\left(\lambda_{i}, a_{q}, b_{q}\right)\right)^{2}}{\sigma^{2}_{q}\left(\lambda_{i}, a_q, b_q \right)} - \sum_{i}\ln{\sigma^{2}_{q}\left(\lambda_{i}, a_q, b_q \right)} ~,
    \label{eqn:aqbq_likelihood}
\end{equation}
where $\sigma_{q}^2$ is the total variance of the data, including the contribution from the noise estimated from the pipeline ($\sigma_{{\rm pip},q}^2$), modulated by a function $\eta_{\rm pip}(\lambda)$ to account for small inaccuracies in the noise estimation, and the intrinsic variance of the \lya forest, $\sigma^2_{\rm LSS}$, multiplied by $(\overline{F}C_{q})^2$:
\begin{equation}
\sigma^2_q\left(\lambda\right) = \eta_{\rm pip} \left(\lambda\right)\sigma_{{\rm pip}, q}^2\left(\lambda\right) +  \sigma^2_{\rm LSS}\left(\lambda\right) \left(\overline{F}C_{q}\right)^2\left(\lambda\right) \label{eqn:delta_variance} ~.
\end{equation}

We keep the original wavelength bin size of $0.8\angs$ when computing the $\delta_q(\lambda)$ following the study of~\cite{2023MNRAS.tmp.3626R}. 
We note that the continuum fitting procedure distorts the $\delta_{q}$ field, since line-of-sight fluctuations on scales comparable to the length of a given forest will be suppressed when fitting the ($a_q$, $b_q$) parameters.
This is taken into account with the definition of a projected flux transmission field in \cref{subsec:auto-corr-meas} and in the correlation function model in \cref{subsec:dmat}. 
We perform the continuum fit in each of the analysed regions independently. This means that we have up to 3 independent sets of ($a_{q}$, $b_{q}$) per quasar as each quasar can be used in the \lya regions A and B, and in the calibration region. 

Occasionally, the best-fit values of $a_{q}$ and $b_{q}$, combined with the shape of $\overline{C}\left(\lambda_{\rm rf}\right)$ result in a negative estimated product $\overline{F} C_{q}\left(\lambda\right)$. 
Whenever this happens for at least one pixel, we deem the continuum fit as problematic and discard the entire region. 
The number of forests lost because of this in each of the samples is given in \cref{tab:num_qso}. 
Note that, because the $a_{q}$ and $b_{q}$ are fitted independently in each of the regions of interest, a particular spectrum might be removed from one of the regions but kept in another.

\section{Measurement of correlations}
\label{sec:correlations}

In \cref{sec:data} we have described the three cosmological datasets that we use in our BAO measurement: the catalog of quasar positions (angles and redshifts) and the catalogs of \lya fluctuations in the rest-frame regions A and B. 
There are six (3x2) possible correlations that we could use, but following \dMdB we focus on the four correlations that have higher signal to noise: we ignore the auto-correlation of quasars, and the auto-correlation of \lya fluctuations in the B region.
For convenience, in some sections we group the correlations in two subsets:
we use the term \textit{auto-correlation} to refer to the combination of the auto-correlation of the \lya fluctuations in region A, \lyaxlyaA, and the correlation of \lya fluctuations in region A with the \lya fluctuations in region B, \lyaxlyaB
\footnote{While this is technically the cross-correlation of two disjoint datasets, we include it in the \textit{auto-correlation} since both datasets contain pixels with \lya fluctuations, and we use the same model as in the \lyaxlyaA correlation}; 
we use the term \textit{cross-correlation} to refer to the correlation of quasar positions with the \lya fluctuations in region A, \lyaxqsoA, and in region B, \lyaxqsoB.

In this section we summarise the methodology used to measure the different correlations and their covariance matrix, and we refer the reader to \cite{2023JCAP...11..045G} for more details.
We measure the correlations with the software \texttt{picca}\footnote{\url{https://github.com/igmhub/picca}. We used version \href{https://github.com/igmhub/picca/tree/v9.0.0}{v9.0.0}. We also acknowledge the use of the following packages: \texttt{numpy} \cite{Harris:2020}, \texttt{scipy} \cite{scipy:2020}, \texttt{astropy} \cite{astropy:2013,astropy:2018,astropy:2022}, \texttt{mpi4py} \cite{mpi4py}, \texttt{healpy} \cite{healpy}, \texttt{matplotlib} \cite{matplotlib}, \texttt{GetDist} \cite{Lewis:2019}, \texttt{numba} \cite{numba:2015}, and \texttt{fitsio}, \url{https://github.com/esheldon/fitsio}.}~\citep{picca} that was developed originally for the \lya analysis of BOSS and eBOSS data, and that was also used to estimate the \lya fluctuations.
The only relevant difference with respect to \cite{2023JCAP...11..045G} is that we now include the cross-covariance of the different correlations, which were considered independent in previous analyses.
We discuss this in \cref{subsec:covariance}.

We measure the correlations in bins of comoving separation along ($r_\parallel$) and across ($r_\perp$) the line of sight, computed from the angular and redshift separations using a fiducial cosmology based on the best-fit flat $\Lambda$CDM model from Planck 2018 \cite{Planck2018} (see \cref{tab:fid_cosmo}).
We present a single measurement of the correlations averaged over a wide range of redshifts (see \cref{fig:eboss_desi_data}).
This is motivated by the fact that the BAO scale in comoving coordinates does not vary much with redshift if the true cosmic expansion is not very far from the fiducial one for those redshifts
\footnote{If the fiducial cosmology was significantly different than the truth, the BAO peak would be smeared when averaging over a wide redshift range. However, as shown in the appendix B of \cite{KP6s6-Cuceu}, varying the fiducial cosmology by $1 \sigma$ (in terms of $\at$, $\ap$) causes a bias smaller than $0.1 \sigma$.}
, and that considering several redshift bins would make the BAO peak less significant and possibly more difficult to fit \cite{Cuceu2020}.
As discussed in \cref{sec:model}, we model the correlations at an effective redshift ($\zeff=2.33$) and report the BAO measurement at that redshift.
In our fiducial model and at our effective redshift, the BAO scale corresponds to an angular separation of $1.46$ degrees and an observed wavelength separation of $141\angs$.

\begin{table}
\centering
\begin{tabular}{cc}
\hline
\hline
Parameter                              & Planck (2018) cosmology             \\
                                       & (TT,TE,EE+lowE+lensing)                \\
\hline
$\Omega_{\rm m}h^2$ =                  & 0.14297                             \\
$+\Omega_{\rm c}h^2$                   & 0.12                                \\
$+\Omega_{\rm b}h^2$                   & 0.02237                             \\
$+\Omega_{\rm \nu}h^2$                 & 0.0006                              \\
$h$                                    & 0.6736                              \\
$n_{\rm s}$                            & 0.9649                              \\
$10^9 A_{\rm s}$                       & 2.100                               \\

\hline
$\Omega_{\rm m}$                       & 0.31509                              \\
$\Omega_{\rm r}$                       & 7.9638e-05                           \\
$\sigma_8 (z=0)$                     & 0.8119                               \\
$r_{\rm d} \; [\rm Mpc]$               & 147.09                               \\
$r_{\rm d} \; [h^{-1} \rm Mpc]$        & 99.08                                \\
$D_{\rm H}(\zeff=2.33)/r_{\rm d}$   & 8.6172                               \\
$D_{\rm M}(\zeff=2.33)/r_{\rm d}$   & 39.1879                              \\
$f(\zeff=2.33)$                     & 0.9703                               \\
\end{tabular}
    \caption{
        Parameters of the flat-$\Lambda$CDM cosmological model used in the analysis, both to compute comoving separations and in the modelling (described in \cref{sec:model}).
        The first part of the table gives the cosmological parameters,
        the second part gives derived quantities used in this paper.
        }
    \label{tab:fid_cosmo}
\end{table}

\subsection{Measurement of the auto-correlation}
\label{subsec:auto-corr-meas}

The correlation function measurement relies on a fiducial cosmology to convert the angular separations and redshifts into comoving separations. We use the same approach as in \cite{2023JCAP...11..045G} and earlier studies.
We first define the longitudinal and transverse separations $(r_\parallel,r_\perp)$ for a pair of measurements $(i,j)$ at redshifts $(z_i,z_j)$ and angular separation $\theta_{ij}$ as
\begin{align}
r_\parallel &= \left[D_C(z_i) - D_C(z_j)\right] \cos \left( \theta_{ij}/2 \right) ~, \nonumber \\
r_\perp &= \left[D_M(z_i) + D_M(z_j)\right] \sin \left( \theta_{ij}/2 \right) ~,
\label{eqn:comobile-separation}
\end{align}
where $D_C(z)$ is the comoving distance and $D_M(z)$ the angular (or transverse) comoving distance. For our fiducial cosmology with $\Omega_k=0$, they are identical.\\

The estimator for the correlation function of the transmitted flux field is a simple weighted average.
For this purpose, we define the following weight for the  flux decrement $\delta_q(\lambda)$,

\begin{equation} \label{eqn:weights}
w_q(\lambda) = \left(\frac{1+z_\lambda}{1+z_0}\right)^{\gamma_{\alpha}-1} \left[ \eta_{\rm pip}(\lambda) \left( \frac{\sigma_{{\rm pip}, q}(\lambda)}{\overline{F}C_{q}(\lambda)}\right)^2 +  \eta_{\rm LSS} \, \sigma^2_{\rm LSS}(\lambda)  \right]^{-1} ~.
\end{equation}

The right hand term is the inverse of the variance of the flux decrement (\cref{eqn:delta_variance}), but with an additional scaling term, $\eta_{\rm LSS}$, that increases the contribution of the large scale structure variance compared to the pipeline noise term.
This correction was introduced by \cite{2023MNRAS.tmp.3626R} to improve the precision of the correlation function measurement\footnote{$\eta_{\rm LSS}$ was referred to as $\sigma_{\rm mod}^2$ in \cite{2023MNRAS.tmp.3626R}.}.
Our estimator is sub-optimal because it does not take into account the correlations between neighboring flux decrement values caused by the large scale structure.
The adjustment of weights with $\eta_{\rm LSS}$ is a simple way to avoid over-weighting pixels that have a high signal to noise ratio but correlated information content.
We use a value of $\eta_{\rm LSS}=7.5$, which was found to minimise the covariance matrix of the \lya auto-correlation given our wavelength bin size of 0.8\AA\ \citep{2023MNRAS.tmp.3626R}.
The left hand term is a scaling factor to account for the variation of the amplitude of the correlation function with redshift. It is the optimal term for the measurement of the shape of the correlation function.
We set the bias evolution index $\gamma_{\alpha}=2.9$ as in previous studies, following \citep{McDonald2006} (the value of $z_0$ does not affect the results as it cancels in the estimator of the correlation function).

In order to estimate the \lya fluctuations, we have divided the observed flux by a model of the continuum of each quasar (see \cref{eqn:delta_definition}).
As noted in \cref{sec:cont_fit}, the parameters $(a_q, b_q)$ of this model have been fitted using all the \lya pixels in the spectrum and this makes our estimated fluctuation at a given pixel ($\delta_i$) dependent on the flux of the other pixels in the spectrum, and therefore on their true \lya fluctuation ($\delta_j^t$) \cite{Slosar2011,FontRibera2012b}.
Following earlier work \cite{Bautista2017,dMdB2017}, we linearise this relation and approximate the distortion\footnote{Here we ignore uncorrelated sources of noise, like random continuum errors or instrumental noise, since these would not bias the measurement of the 3D correlations.}
as $\delta_i = \eta^{\rm cont}_{ij}~\delta^t_j$, with $\eta^{\rm cont}$ defined as
\begin{equation}
\eta^{\rm cont}_{ij} =  \delta^K_{ij} - \frac{w^{\rm cont}_j}{\sum_k w^{\rm cont}_k} - \frac{w^{\rm cont}_j \Lambda_i \Lambda_j}{
\sum_k w^{\rm cont}_k \Lambda_k^2} ~,
\label{eqn:projection}
\end{equation}
where $\delta^K_{ij}$ is the Kronecker delta function, $w^{\rm cont}_q = \left(\overline{F}C_{q}\right)^2 / \sigma^2_q $ are the weights used in continuum fitting (\cref{eqn:aqbq_likelihood}) and $\Lambda_i \equiv \log(\lambda_i) - \sum_k w^{\rm cont}_k \log(\lambda_k)/\sum_k w^{\rm cont}_k$.

It is clear from \cref{eqn:projection} that during continuum fitting the weighted mean and the linear trend over each \lya forest are set to zero.
In order to make sure that these are also zero when using the weights $w$ defined in \cref{eqn:weights}, we apply a similar linear projection to the measured fluctuations $\delta_j$ and define a projected field $\tilde \delta_i = \eta_{ij} ~ \delta_j$, where $\eta$ is equivalent to $\eta^{\rm cont}$ but using the weights $w$ instead of $w^{\rm cont}$ in \cref{eqn:projection}.
One can show that, to first order in $\delta^t$, $\eta$ is also the relation between the projected field and the true fluctuation\footnote{Picture a scatter plot of (x,y) points from which one subtracts to y the mean and slope as a function of x. The result is obviously independent from any prior subtraction of any mean and slope.}:
\begin{equation}
    \label{eqn:projection2}
    \tilde \delta_i = \eta_{ij} ~ \delta_j = \eta_{ij} ~ \eta_{jk}^{\rm cont} ~ \delta^t_k = \eta_{ij} ~\delta^t_j ~.
\end{equation}
As a result we do not need to know the exact values of $\eta^{\rm cont}_{ij}$, but only those of $\eta_{ij}$.
In \cref{subsec:dmat} we will use this linear projection operator $\eta$ to model the correlations of the projected field $\tilde \delta_q(\lambda)$.

The correlation function is measured in bins of $4~\hMpc$ ranging from 0 to $200~\hMpc$ along both $r_\parallel$ and $r_\perp$ for a total of $2\,500$ bins.
Calling $M$ a two-dimensional bin of comoving separation, the estimator of the correlation function averaged in the separation bin $M$ is the weighted average:
\begin{equation}
\xi_M = \sum_{(i,j) \in M} w_i w_j \tilde\delta_i \tilde\delta_j / \sum_{(i,j) \in M} w_i w_j \label{eqn:auto_corr}
\end{equation}
where $(i,j) \in M$ are all the pairs of projected transmitted flux measurements $\tilde\delta_i$ and $\tilde\delta_j$ from quasar lines of sight separated by an angle $\theta_{ij}$ and at redshift (or wavelength) $z_i$ and $z_j$, such that their comoving separation is in $M$. We only consider pairs from different quasar spectra to avoid the contribution of correlated residuals within a spectrum caused by continuum fitting errors.

\begin{figure}[h]
\centering
\includegraphics[width=0.9\textwidth]{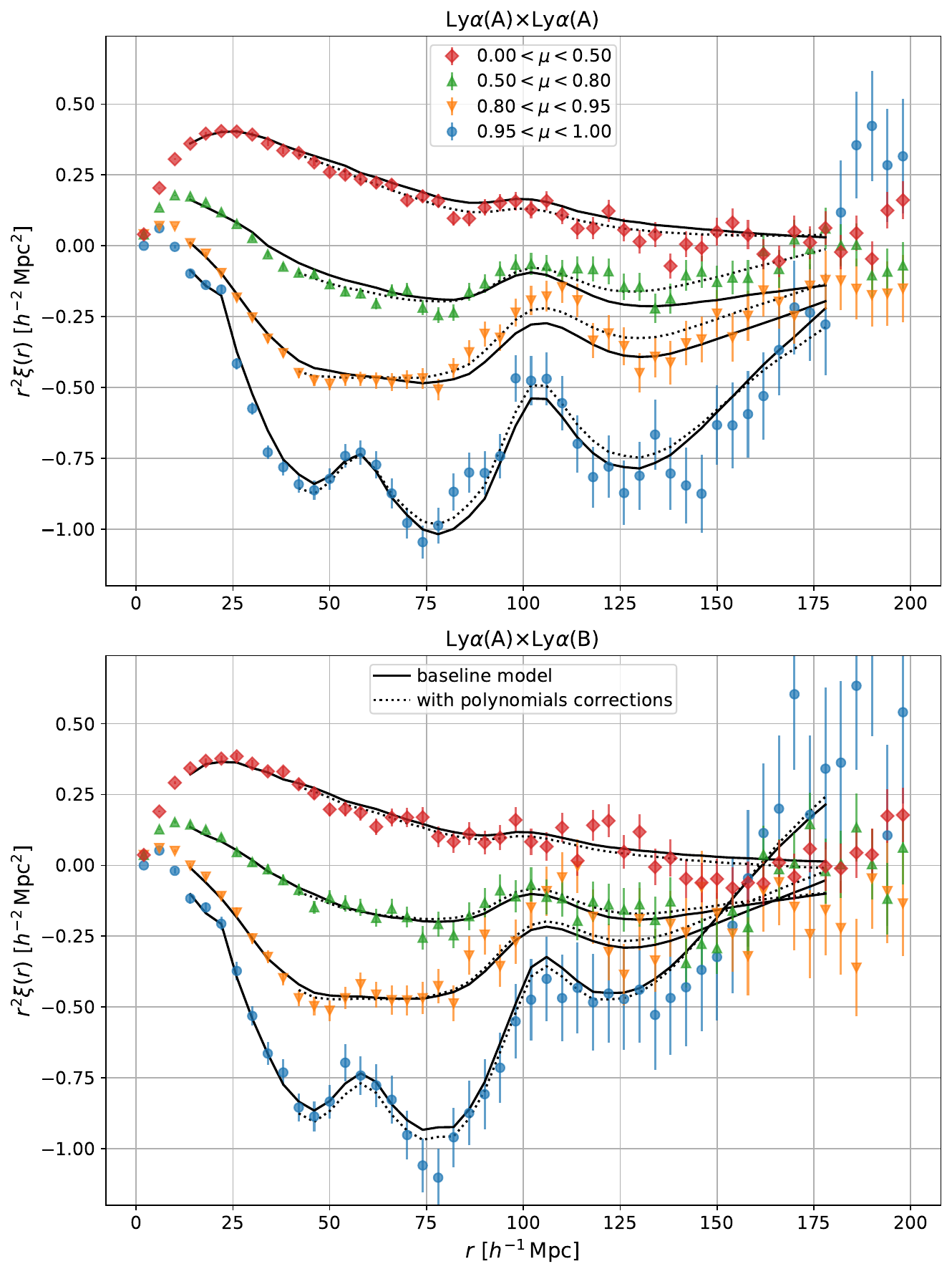}
  \caption{Measured \lya auto-correlation when using pixels from region A (top, colored markers) and when correlating pixels from region A with pixels from region B (bottom), along with the best fit model (solid black curves), described in \cref{sec:model}.
  The different colors and markers correspond to different orientations with respect to the line-of-sight, with blue correlations being close to the line-of-sight $0.95<\mu<1$. The dotted curves show the best fit model with additive polynomial corrections (see \cref{subsec:broadband-validation}).
  \label{fig:baseline-correlation-wedges}}
\end{figure}

In \cref{fig:baseline-correlation-wedges} we show the measurement of the \lya forest auto-correlation when using pixels from region A (\lyaxlyaA, top panel) and when correlating pixels from region A with pixels in region B (\lyaxlyaB, bottom panel).
We show the measurement as a function of the total separation $r = (r_\parallel^2 + r_\perp^2 )^{1/2}$ in wedges defined by a range of the cosine between the orientation of the pair and the line-of-sight, $\mu = r_\parallel / r$. This is obtained by resampling the original data, which has the form of a rectangular 2D grid of $(r_\parallel,r_\perp)$ bins, into $(r,\mu)$ bins. We evaluate the uncertainties in those bins using the covariance matrix, the computation of which is described in \cref{subsec:covariance}. The resampling results in additional correlation between neighboring $(r,\mu)$ bins that share fractions of the original bins. We note that this resampling into wedges is performed for display purpose only and is not used for the fit, which is realized on the original rectangular grid.

\subsection{Measurement of the cross-correlation}

With the same notations as in the previous section, we use the following estimator \cite{FontRibera2012b} for the quasar Lyman-$\alpha$ cross-correlation in a separation bin $M$:

\begin{equation}
\xi_M = \sum_{(i,j) \in M} w_i w^{\rm Q}_j \tilde\delta_i / \sum_{(i,j) \in M} w_i w^{\rm Q}_j ~. \label{eqn:cross_corr}
\end{equation}
Here $(i,j) \in M$ stands for pairs made of a projected transmitted flux measurement $\tilde\delta_i$ at a redshift $z_i$ along a quasar line of sight, and another quasar $j$ at another redshift $z_j$ separated by an angle $\theta_{ij}$ from the first one, such that their comoving separation is in the bin $M$ (once the redshifts and $\theta_{ij}$ are converted to comoving separation using equation~\ref{eqn:comobile-separation}).
We use for quasars the weights
\begin{equation}
 w^Q_j = \left[(1+z_ Q)/(1+z_0)\right]^{\gamma_Q-1}
\end{equation}
where $z_{Q}$ is the quasar redshift, and we choose $\gamma_Q=1.44$ which follows closely the measured bias evolution of quasars \citep{dMdB2019}.

We use the same bin size of $4~\hMpc$ for the cross-correlation but we differentiate positive and negative longitudinal separation, giving us a range from $-200$ to $+200~\hMpc$ for $r_\parallel$ because the cross-correlation is asymmetric. We define the sign of $r_\parallel$ such that $r_\parallel<0$ for pairs where the quasar is behind the \lya transmitted flux measurement.  We have $5\,000$ cross-correlation bins.

\begin{figure}[h]
\centering
\includegraphics[width=0.9\textwidth]{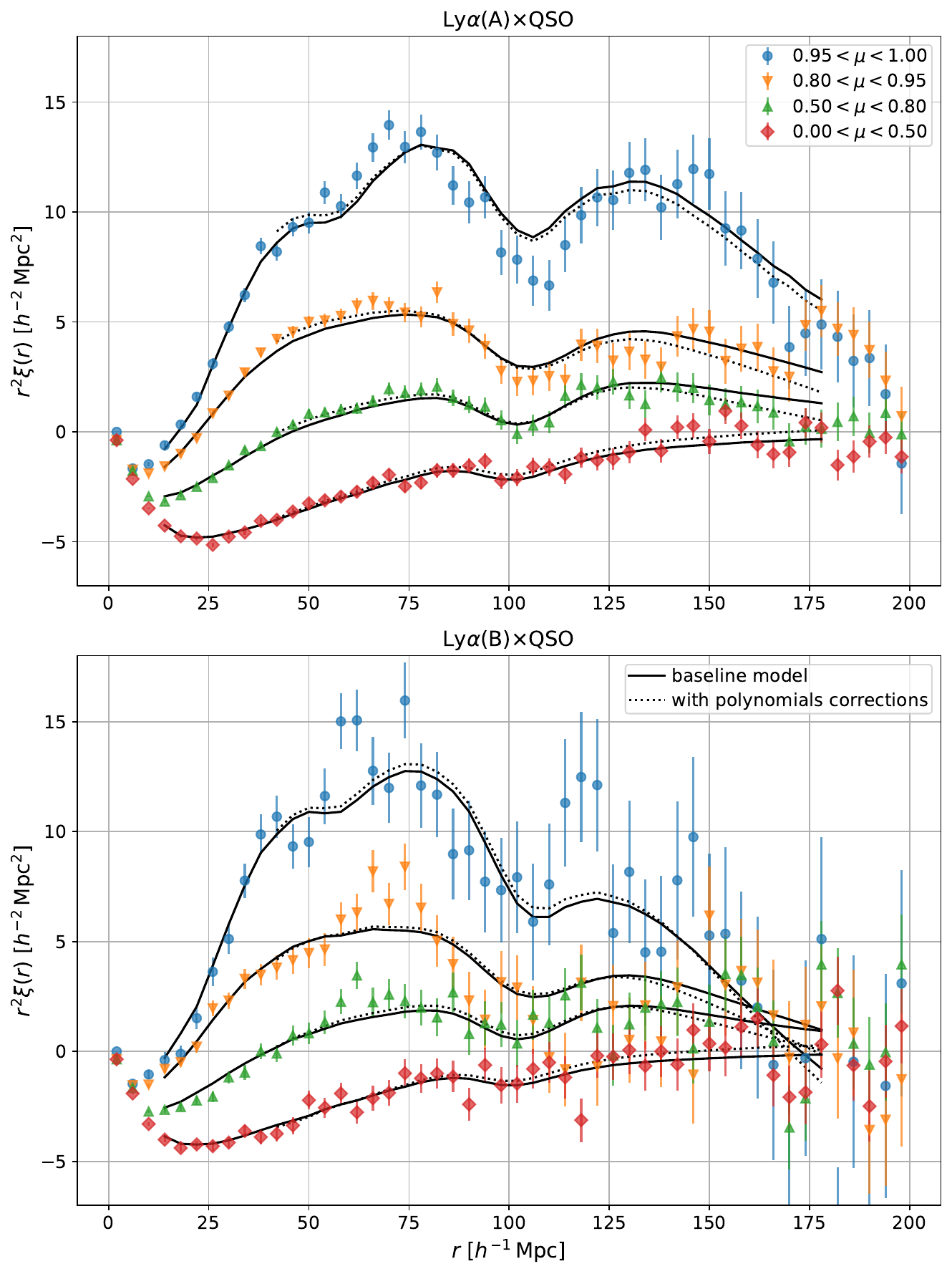}
  \caption{Measured \lyaxqso cross-correlation functions in region A (top, colored markers) and region B (bottom) along with the best fit model (solid black curves), described in \cref{sec:model}.
  The different colors and markers correspond to different orientations with respect to the line-of-sight, with blue correlations being close to the line-of-sight $0.95<\mu<1$. The dotted curves show the best fit model with additive polynomial corrections (see \cref{subsec:broadband-validation}).
  \label{fig:baseline-correlation-lyaqso-wedges}}
\end{figure}

In \cref{fig:baseline-correlation-lyaqso-wedges} we show the measurement of the cross-correlation of quasars with \lya pixels in region A (\lyaxqsoA, top panel) and with \lya pixels in region B (\lyaxqsoB, bottom panel).

\subsection{Covariance matrix}
\label{subsec:covariance}

We use the method described in more detail in \dMdB to compute the covariances.
In brief, the correlation function is first measured independently in sub-samples defined by \texttt{HEALpix} pixels on the sky. Each sub-sample correlation is saved with its weights $W_M$ in each bin $M$, which are denominators in \cref{eqn:auto_corr,eqn:cross_corr}. We do not lose or double-count pairs because each possible pair is assigned a unique sub-sample. The combined correlation function is simply the weighted mean of the sub-sample correlations, and its covariance is determined by replacing the unknown covariance of each sub-sample by the square of its difference with the mean. We ignore the cross-covariance between sub-samples, which is negligible for our scales of interest given the size of a \texttt{HEALpix} pixel of about $(250~\hMpc)^2$ at $z \sim 2.3$ for our choice of \texttt{NSIDE}=16.
The correlation is measured in 1028 pixels, with more than 400 valid forests per pixel on average.
In \cref{fig:var_all_1d} we show that using \texttt{NSIDE}=32 instead has a negligible impact on the BAO parameters.

This noisy estimate of the covariance $C$ is then smoothed. We replace all the non-diagonal elements of the correlation matrix ${\rm Corr}_{MN} \equiv C_{MN}/(C_{MM} C_{NN})^{1/2}$ in which indices correspond to the same differences $|r_\parallel(M)-r_\parallel(N)|$ and $|r_\perp(M)-r_\perp(N)|$ by their average.
This method has proven to be a good approximation of the covariance when compared with other methods, like a Gaussian covariance computed with the Wick expansion (see \cite{Delubac2015} and Appendix C of \dMdB), and it is discussed in detail in the companion paper \cite{KP6s6-Cuceu}.

\begin{figure}
\includegraphics[width=0.99\textwidth]{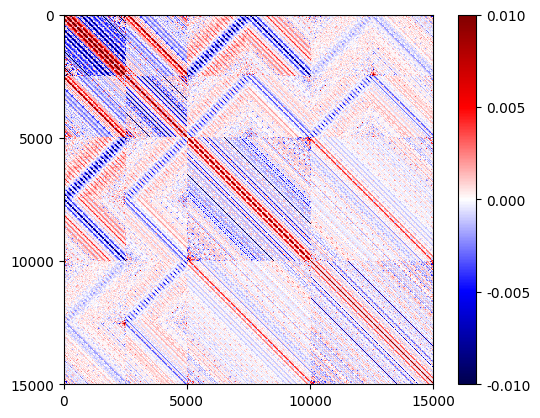}
  \caption{
  Global correlation matrix for the four 2pt functions included in this analysis, as measured from the scatter between correlations measured in more than 1000 HEALPix pixels, after applying the smoothing discussed in the main text.
  The first block of $2\,500 \times 2\,500$ in the top left corresponds to the correlation matrix of the \lyaxlyaA measurement, while the second block of the same size corresponds to the \lyaxlyaB one.
  The third, larger block of $5\,000 \times 5\,000$ corresponds to the \lyaxqsoA\ cross-correlation, with positive and negative values of $r_\parallel$, and the last block (bottom right) is the correlation matrix of the \lyaxqsoB  measurement.
  It is clear in that cosmic variance (in the form of off-diagonal stripes in red and blue) is relevant, even thought it is only detected at the sub-percent level.
  ote that we only have strong correlations ($>10 \%$) for bins with the same value of $r_\perp$ and neighbouring value of $r_\parallel$, and that we limit the range of the color scale to 1\% for visualization purposes.
  }
  \label{fig:global_cov}
\end{figure}

We measure with this sub-sampling technique the covariance of the full data set composed of the four correlation functions, \lyaxlyaA, \lyaxlyaB, \lyaxqsoA, and \lyaxqsoB. Previous works (including \dMdB) measured the  covariances of each correlation function in isolation, ignoring the cross-covariances between them.
Indeed, we show in \cref{subsec:covariance-validation} that ignoring the cross-covariance between the correlation functions results in a 10\% underestimate of uncertainties on the BAO parameters.
The new correlation matrix is shown in \cref{fig:global_cov}. This full covariance estimation represents the most important change in the methodology from previous analyses of the 3D correlations in the \lya\ forest.

\section{Modelling of correlations}
\label{sec:model}

In this section we describe how the correlations are modeled.
The approach is similar to the one used in previous BOSS and eBOSS analyses (see \dMdB for the latest results from eBOSS) except for few methodological changes that are summarised below, and that we describe in more detail in companion papers \cite{2023JCAP...11..045G,KP6s5-Guy,KP6s6-Cuceu}.
We now use the software \texttt{Vega}\footnote{\url{https://github.com/andreicuceu/vega}. We used version \href{https://github.com/andreicuceu/vega/tree/v1.0.0}{v1.0.0}. }, which is an improved version of the \texttt{picca-fitter2} software used in these previous \lya BAO analyses.

From the point of view of this analysis, our model has 2 important (BAO) parameters and 15 nuisance parameters that we marginalize over.
We start in \cref{subsec:decomposition,subsec:zevol,subsec:dmat} with a simplified version of the model that will allow us to introduce some of the key aspects of a \lya BAO analysis, before describing the modelling of contaminants in \cref{subsec:metals,subsec:sky,subsec:hcds,subsec:zerrors,subsec:proximity}.
The priors, best-fit values and uncertainties for the nuisance parameters are presented in \cref{app:nuisance}.

\subsection{Isolating the BAO information}
\label{subsec:decomposition}

We start by building a model for the (anisotropic) large-scale power spectrum of fluctuations in the \lya forest, based on linear perturbation theory.
We use the linear power spectrum of matter fluctuations for our fiducial cosmology (see \cref{tab:fid_cosmo}) evaluated at our effective redshift (see below), and linear bias ($b_F$) and redshift-space distortion ($\beta_F$) parameters.
When modelling the cross-correlation with quasars, we introduce another linear bias parameter ($b_Q$) and we follow \cite{Kaiser1987} to model the linear redshift-space distortions of quasars with $\beta_Q=f/b_Q$, where $f$ is the logarithmic growth rate
\footnote{Note that the same relation does not apply to the \lya forest, and $\beta_F$ is an independent parameter \cite{McDonald2000,McDonald2003}.}.

We then compute its inverse Fourier transform to obtain a model for the correlation function $\xi(r_\perp,r_\parallel)$ that we can compare to our measurement. 
In our fiducial cosmology, there is an excess correlation (the BAO \textit{peak} \footnote{Note that in the \lyaxqso cross-correlation we expect a \textit{BAO trough}, instead of a \textit{BAO peak}.}) at around $100~\hMpc$.
We then introduce two scaling parameters, $\at$ and $\ap$, which multiply $r_\perp$ and $r_\parallel$ respectively in our model, and we vary them in order to better match the BAO peak seen in the measured correlations. 

In order to make sure that we only extract BAO information from the fits, we decompose the model of the correlations into a \textit{peak} and a \textit{smooth} component following \cite{Kirkby2013}, and the scaling parameters ($\at$, $\ap$) are only applied to the peak component
\footnote{We validate this in \cref{subsec:broadband-validation}, where we present an alternative analysis where we add up to 48 extra free parameters describing a flexible \textit{broadband} component.}.
Following \cite{Kirkby2013} again, we apply a Gaussian smoothing to the peak component in order to model the non-linear broadening of the BAO feature caused by non-linear growth of structure.

\subsection{Redshift evolution}
\label{subsec:zevol}

When computing the binned correlations in \cref{eqn:auto_corr,eqn:cross_corr}, we use the same weights to compute the mean separations ($r_\perp$, $r_\parallel$) and to compute the mean redshift ($z$) of each bin.
We evaluate the model at these coordinates.
The redshift evolution of the model is captured by the linear growth of the matter power spectrum ($\sigma_8(z)$), the logarithmic growth rate ($f(z)$), and the redshift evolution of the bias parameters.
We define our parameters at an effective redshift ($\zeff=2.33$, see below), and model the redshift evolution of the bias parameters with a power law, $b(z) = b(\zeff) \left[ (1+z)/(1+\zeff)\right]^\gamma$.
Following \cite{2023JCAP...11..045G,dMdB2020}, we use $\gamma_Q=1.44$ to describe the redshift evolution of the quasar bias, $\gamma_{\alpha}=2.9$ for bias of the \lya forest, and we assume that the RSD parameter of the \lya forest ($\beta_{\alpha}$) does not vary with redshift.

In order to estimate the effective redshift of our BAO measurement, we compute the mean of the redshifts of each correlation bin in the range $80 \, \hMpc < r < 120 \,\hMpc$, weighted with their inverse variance.
We do this separately for the auto-correlation ($\zeff=2.339$) and for the cross-correlation ($\zeff=2.325$), and we then compute a simple mean of those two values for the combined BAO measurement (rounded to two decimal values) to obtain the effective redshift of our BAO measurement ($\zeff=2.33$).

\subsection{Distortion matrix}
\label{subsec:dmat}

As discussed in \cref{sec:correlations}, the distortions introduced with the continuum fitting have led us to the use of the projected field $\tilde \delta$ defined by \cref{eqn:projection}.
We must therefore fit the correlations of the projected field $\tilde \delta$ with a model that has suffered the same projection.
This is performed with the ``distortion matrix'' formalism introduced in \cite{Bautista2017}, $\xi_M = \sum_N D_{MN} ~ \xi_N^\prime$, where $\xi_M$ refers to a ($r_\perp$, $r_\parallel$) bin of the distorted model and $\xi_N^\prime$ to a bin of the undistorted model.
The matrices $D_{MN}$ are constructed using the same linear operators $\eta$ used to compute the projected field (\cref{eqn:projection}).
The elements of the matrices are given by equations (21) and (22) of \dMdB for, respectively, the auto- and cross-correlations.

Two improvements have been made to the distortion matrix treatment over that of \dMdB.
First, the undistorted model bins are now calculated on a grid of $2~\hMpc$ (instead of the $4~\hMpc$ bins used to measure the correlations).
Second, in the $r_\parallel$ directions, we extend the modelling of the undistorted correlations to $300~\hMpc$ rather than to $200~\hMpc$.
This improves the accuracy of the distortion calculation at high $r_\parallel$, but has a negligible impact on the BAO results (see \cref{subsec:variations}).

The computation of the distortion matrix $D_{MN}$ is computationally intensive, and following previous work we only use a small fraction of the dataset to approximate it \cite{Bautista2017,2023JCAP...11..045G}.
By default we use 1\% of the \lya pixels, but in \cref{subsec:variations} we show that doubling that number does not affect the BAO results.

\subsection{Metal contamination}
\label{subsec:metals}

The \lya forest is contaminated by absorption from atomic transitions of elements other than neutral hydrogen. The auto-correlation of those absorbers and their cross-correlation with \lya, QSOs, or other transitions can contribute to the measured \lya\ auto-correlation and its cross-correlation with QSOs, and has to be taken into account. We call them {\it metal} absorbers.
Labeling $\delta_{m}$ their contribution to the total flux decrement $\delta$,
the \lya\ forest auto-correlation will have contributions of the form
\begin{equation}
\langle \delta \delta \rangle = \langle \delta_{\alpha} \delta_{\alpha} \rangle + \sum_m \langle \delta_{m} \delta_m \rangle + \sum_m \langle \delta_{\alpha} \delta_m \rangle + \sum_m \sum_{m^{\prime} \ne m} \langle \delta_m \delta_{m^{\prime}} \rangle ~.
\end{equation}

The second term is the contribution of the metal auto-correlations that are present for all foreground absorbers with $\lambda_{m}> \lambda_{\rm min}$ where $\lambda_{\rm min}$ is the minimum wavelength of the forest in the quasar rest-frame.
 The third term is the contribution of the cross-correlation of metal absorbers with \lya. 
 Only transition wavelengths close to the \lya line will have a significant contribution for the range of longitudinal separation we are studying. 
 The fourth term is the cross-correlation of metals, which can introduce a lot of complexity in the interpretation of the measured correlation function. 
 Fortunately, the metal absorbers are much less abundant than neutral hydrogen and hence have much smaller absorption. We include in the fit the cross-correlation of Si II and Si III absorbers, whose cross-correlation with \lya is also observed, but we neglect the cross-correlation of other foreground absorbers.
 For the \lya quasar cross-correlation, the situation is simpler, as we have only contributions from absorbers with transition wavelengths close to the Lyman-$\alpha$ line.

Previous analyses based on measurements of the \lya\ 1D correlation \citep{Bautista2017}, or cross-correlation with strong absorbers \cite{Pieri2014,Yang2022,Morrison2023}, have shown that only a few transitions had to be considered for the cross-correlation terms cited above. 
Those are \ion{Si}{II} lines at $1190\angs$, $1193\angs$, and $1260\angs$, and one \ion{Si}{III} line at $1207\angs$. 
Other lines are present but can be neglected.
While all of the above transitions will also have an auto-correlation term that we account for in the modeling, it is their cross-correlation with \lya\ that will allow us to differentiate their signals and measure their biases.
Other metal transitions at longer wavelengths and lower redshifts only contribute significantly with their auto-correlation. 
Those cannot be easily separated from the \lya\ auto-correlation signal. 
In a companion paper \cite{KP6s5-Guy}, we show that they are dominated by the \ion{C}{IV} absorption, and that one can measure their contribution from the auto-correlation in the side bands (at wavelengths larger than the \lya\ line in the quasar rest-frame). 
Following a first estimate from their analysis\footnote{A slightly lower bias value was found from a revised analysis. We show in \cref{subsec:variations} that this has a negligible impact on the best fit BAO parameters and their uncertainties.}, we use a prior on the effective \ion{C}{IV} bias of $b_{\rm CIV}^{\rm eff} = -0.0243 \pm 0.0015$ with $\beta_{\rm CIV}=0.5$ which combines the signal from \ion{C}{IV} and other transitions (notably \ion{Mg}{II} and \ion{Si}{IV}).

We use the same set of metal absorbers and priors when modelling the \lyaxlyaB correlations.
Absorbers with wavelength close to the Ly$\beta$ line like the \ion{O}{VI} lines at $1032\angs$ and $1038\angs$ do not contribute in cross-correlation with \lya\ to our measurements because their wavelength is much smaller that the \lya\ line. 
Their auto-correlation, peaking at zero separation, does not contribute either because we do not measure the auto-correlation of pixels in the B region, \lyaBxlyaB.

As explained in \cite{2023JCAP...11..045G}, for each pair of absorbers $(m,n)$ we compute a metal matrix that provides the mapping between the true co-moving separation ($r_{\parallel} , r_{\perp}$) of the two absorbers and their apparent separation when assuming both are caused by the Lyman-$\alpha$ transition. 
In previous works this mapping was computed numerically for a small fraction of the total number of pairs in the sample. This estimation was not precise enough at small separation and expensive in computing time. We now compute uniquely the shift along $r_{\parallel}$ and ignore the few percent change in $r_{\perp}$. This simplification allows us to measure more precisely the effect with one dimensional integrals using the sum of the weights as a function of wavelength.

\subsection{Correlated noise from the data processing}
\label{subsec:sky}

The data processing \citep{Spectro.Pipeline.Guy.2023} introduces correlated noise among the spectra from fibers of the same spectrograph, which correspond to fibers from a unique petal in the focal plane\footnote{The focal plane is segmented into 10 petals or wedges, see \cite{DESI2022.KP1.Instr}.}.

This contamination is studied in a companion paper \cite{KP6s5-Guy}, where we show that the dominant contribution is the sky background model noise (as in BOSS/eBOSS, see \cite{Bautista2017} and \dMdB).
We find that the following expression is sufficient to describe this contamination to the DESI \lya\ auto-correlation function
\begin{equation}
 \xi_{\rm cont}(r_\parallel,r_\perp) =  a_{\rm noise} \,\delta^K(r_\parallel) \, f(r_\perp) \equiv a_{\rm noise} \, \xi_{\rm noise}(r_\parallel,r_\perp)
\end{equation}
with $a_{\rm noise}$ an amplitude and $f(r_\perp)$ a decreasing function of $r_\perp$ proportional to the fraction of pairs at $r_\perp$ that belong to the same petal.
This function is evaluated numerically in \cite{KP6s5-Guy} assuming pairs at $z=2.4$ for the fiducial cosmology to convert angles to co-moving separations. 
We have $f(r_{\perp}>110~\hMpc)=0$ when the separation exceeds the size of a petal. We chose an arbitrary normalization $f(0)=1$ such that $a_{\rm noise}$ is close to the value of this contamination in the first $(r_\parallel,r_\perp)$ bin.

\subsection{HCD contamination}\label{subsec:hcds}

The presence of High Column Density systems (HCDs) in the quasar spectra, including Lyman Limit Systems (LLS, $\log N_{\rm HI} > 17.2$) and Damped Lyman $\alpha$ systems (DLA, $\log N_{\rm HI} > 20.3$),  complicates the modeling of 3D correlations in the \lya forest \cite{McQuinn2011,FontRibera2012a,Rogers2018}.
Like the \lyaf itself, HCDs are tracers of the underlying matter density and on very large scales (larger than the width of their absorption profiles) their contamination is limited to a change in the linear bias parameters of the \lyaf (see section 4.2 in \cite{FontRibera2012a}).
However, the damping wings of DLAs can extend to fairly large line-of-sight separations, and adds an extra scale dependence to the correlation function on scales of tens of megaparsecs.

To diminish the effect of the HCDs on the correlation function, we mask the highly absorbed  wavelength range of identified Damped Lyman alpha systems (DLAs).
As described in \cref{sec:data}, we have good efficiency for identifying DLAs in the spectra with high signal-to-noise.
However, the smoothing effect of unidentified HCDs remains, and this must be modeled.

We use a scale-dependent \lya bias of the form
$b^\prime_{\alpha}=b_{\alpha} + b_{\rm HCD}F_{\rm HCD}(k_\parallel)$
and a similar form for the RSD parameter $\beta_{\rm \alpha}$.
As discussed in \cref{app:voigt_hcd} (see also \cite{Tan2024}), the form of $F_{\rm HCD}(k_\parallel)$ and the magnitude of $b_{\rm HCD}$ can be related to the column density distribution of unmasked HCDs and the bias of their host halos.
The $k_\parallel$ dependence is given by the Fourier transform of the HCD Voigt profiles (an absorption profile with a Gaussian core and Lorentzian tails).
If the column-density distribution of the unmasked HCDs was known,
it would then be possible to calculate $b_{\rm HCD}F_{\rm HCD}(k_\parallel)$ .
Unfortunately, at present we do not know precisely the efficiency of the DLA detection in noisy spectra, and we only know approximately the bias of halos hosting DLAs \cite{FontRibera2012b,PerezRafols2018,PerezRafols2023}. 

Because of this, and motivated by the fact that the Fourier transform of a Lorentzian is an exponential function, we model the contamination with $F(k_\parallel)=\exp(-L_{\rm HCD}k_\parallel)$ and treat $b_{\rm HCD}$ and $L_{\rm HCD}$ as free parameters of the model.
The parameter $\beta_{\rm HCD}$ is poorly determined in the fits
and we choose to use a prior $\beta_{\rm HCD}=0.5\pm0.1$, a value
motivated by the measured bias of $b_{\rm DLA} \sim 2$ of the DLA hosts.
In the \dMdB analysis, $L_{\rm HCD}=10~\hMpc$ was fixed, but in the
present analysis we vary this parameter and find $L_{\rm HCD}=\left( 6.51 \pm 0.9 \right)~\hMpc$.
As discussed in \cref{subsec:variations}, variations in the modelling of the HCD contamination have a negligible impact on the BAO results.

\subsection{Quasar redshift errors}
\label{subsec:zerrors}

Similar to the impact of random peculiar velocities (or ``Fingers of God"), random errors in the estimation of quasar redshifts dilute the clustering of quasars along the line of sight.
This has an impact in the cross-correlation of quasars and the \lya forest, as first discussed in \cite{FontRibera2013}.
Following \dMdB, in our main analysis we model this smoothing with a Lorentzian with free parameter $\sigma_z$, but in \cref{subsec:variations} we show that using a Gaussian instead has a negligible impact on the BAO results.

A small systematic error in the quasar redshifts would be very difficult to detect in the auto-correlation of quasars.
However, such an offset would shift the \lyaxqso cross-correlation such that it would no longer peak at $r_\parallel=0$ \footnote{Remember that the cross-correlation is measured for positive and negative values of $r_\parallel$.}.
We parameterize this shift with a free parameter $\Delta r_\parallel$.

The impact of redshift errors in the \lyaxqso cross-correlation can be seen as a nuisance in \lya BAO studies, but as discussed in \cite{KP6s4-Bault} it also provides a great diagnosis tool to better calibrate the redshifts of quasars.

We find that the quasar redshifts are unbiased, with $\Delta r_\parallel = (0.066 \pm 0.058) \hMpc$.
We also find that the combination of random peculiar velocities and redshift errors are described by a Lorentzian with $\sigma_z=(3.67 \pm 0.14) \hMpc$, significantly smaller than the value reported in the eBOSS analysis of \dMdB, $\sigma_z=(6.86 \pm 0.27) \hMpc$.
The reduced redshift errors can be explained by the updated quasar templates used in \texttt{Redrock} (see table 6 of \cite{RedrockQSO.Brodzeller.2023}).

\subsection{Quasar radiation (proximity effect)}\label{subsec:proximity}

Quasars are some of the brightest objects in the Universe.
Therefore, we expect them to significantly ionize their surroundings, an effect sometimes referred to as the \textit{proximity effect}. 
The \lya forest of neighbouring quasars is therefore affected by two competing effects: the gas density is higher than average (quasars live in high-density regions), but the neutral fraction is lower than average (due to the quasar radiation).

Following \cite{FontRibera2013}, we use a simple model to account for the proximity effect in the \lyaxqso cross-correlation.
In particular, we use the implementation of \dMdB that assumes isotropic radiation from the quasar, a mean-free path of UV photons of $\lambda_{\rm UV}=300~\hMpc$, and has a single free parameter $\xi_0^{\rm TP}$ that sets the amplitude of the contamination.

\subsection{Small-scales corrections}

The BAO parameters only shift the peak component of the model, and as discussed in \cite{Kirkby2013} this is by construction zero on scales smaller than $80~\hMpc$.
Therefore, one could decide to limit the BAO analysis to these very large separations.
However, some of the nuisance parameters described in this section can only be constrained when extending the analysis to smaller separations. 
This is the case for the parameters describing quasar redshift errors, or the parameter describing the \ion{Si}{III} line at $1207\angs$ that causes a sharp feature at ($r_\perp \sim 0$, $r_\parallel \sim 20~\hMpc$).
Moreover, the distortion matrix discussed in \cref{subsec:dmat} spreads the impact of some of these small-scale effects to larger separations (in particular along the line of sight).
For this reason, in our main analysis we include the measurement of correlations down to $r > 10~\hMpc$, and in \cref{subsec:variations} we show that the BAO results do not depend on the minimum separation
\footnote{The nuisance parameters do change with the minimum separation included in the fits.}.

In order to fit the \lya auto-correlation to $10~\hMpc$, we follow previous \lya BAO analyses and use a multiplicative correction to the model of the \lya power spectrum.
In particular we use the correction from \cite{Arinyo2015}, calibrated with hydrodynamical simulations, that models both the effect of non-linearities in the densities and velocities, but also thermal effects in the IGM (thermal broadening, pressure)\footnote{We use the values from the \textit{Planck} simulation in Table 7 of \cite{Arinyo2015}, interpolated to our effective redshift.}.
In \cref{subsec:variations} we show that this correction has a negligible impact on our BAO parameters.

An equivalent model was proposed in \cite{Givans2022} for the cross-correlation of the \lya forest with halos of intermediate masses. 
However, the simulations used were too small to contain enough massive halos to accurately study the clustering of quasars.
We decided to not include this correction in our model.
However, as discussed in \cref{subsec:zerrors} we do take into account the impact of non-linear peculiar velocities on the \lyaxqso cross-correlation.

Finally, following \cite{2023JCAP...11..045G} and previous \lya BAO analyses, we take into account the finite size of our correlation function bins.
\section{Measurement of Baryon Acoustic Oscillations} 
\label{sec:results}

After presenting the measured correlations in \cref{sec:correlations} and discussing the model we used to describe them in \cref{sec:model}, in this section we summarise the statistical method used to fit the measurement and present the main results of this paper: the measurement of the BAO scale along and perpendicular to the line of sight. 

We use the \texttt{Vega} package both for the modelling of the correlations and for the parameter inference.
We use a Gaussian likelihood, and the main results presented in this section were obtained using the Nested Sampler \texttt{Polychord} \cite{Handley:2015a,Handley:2015b}.
The best-fit values are the mean of the posteriors, and the reported uncertainties are the 68\% credible intervals.
However, this analysis is computationally intensive, and in most of the tests in \cref{sec:validation} we use instead a simpler method: 
we use the \texttt{iminuit} software \cite{iminuit,James:1975dr} to find the maximum of the likelihood, and use the derivatives of the likelihood around the best-fit point to estimate a Gaussian posterior.
As discussed in \cref{app:sampling}, both BAO estimates are very similar.

We start with a data vector composed of 4 different 2-point functions, two of them with $2\,500$ data points (\lyaxlyaA and \lyaxlyaB) and two of them with $5\,000$ data points (\lyaxqsoA and \lyaxqsoB), for a total of $15\,000$ data points. 
While these are mostly independent, as discussed in \cref{sec:correlations} 
we include their small cross-covariance, and therefore we use a $15\,000 \times 15\,000$ covariance matrix. 
However, following \dMdB we limit the range of separations used in the fits to $10 < r < 180 ~\hMpc$, reducing the number of data points used in the combined fit to 9540 (see \cref{tab:bao})\footnote{As discussed in \cref{subsec:variations}, the results are not sensitive to the exact range of separations used.}.

\begin{figure}
\includegraphics[width=0.9\textwidth]{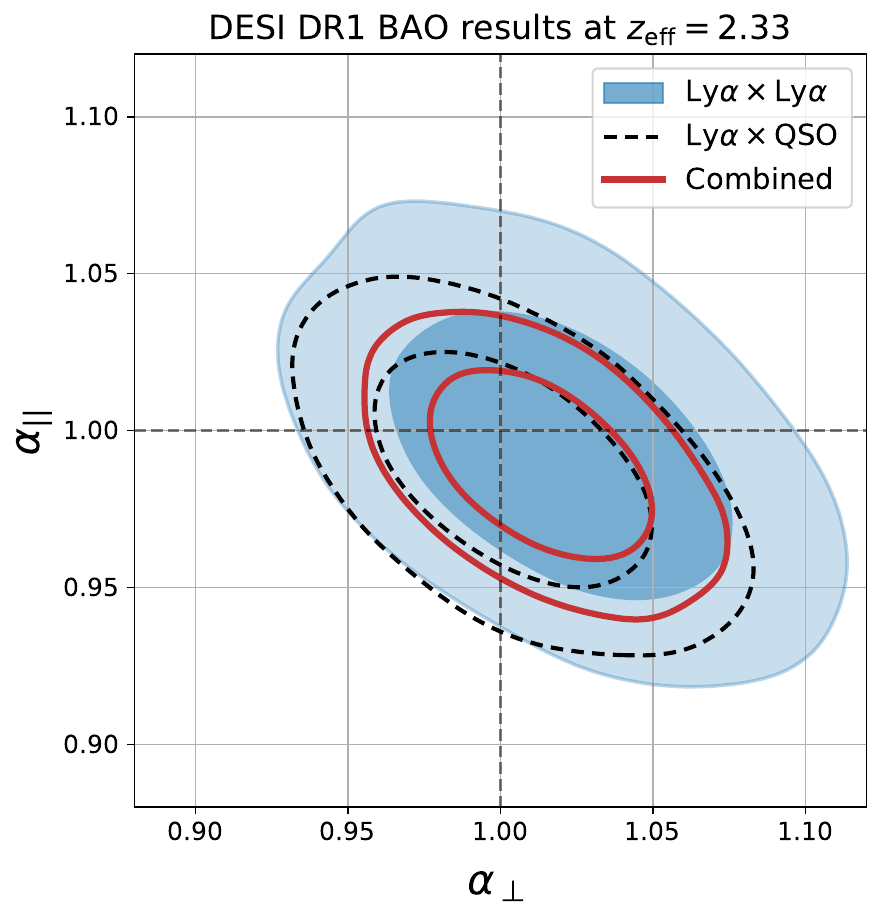}
  \caption{Measurements of the BAO parameters along the line of sight ($\alpha_\parallel$) and across the line of sight ($\alpha_\perp$), with contours corresponding to the 68\% and 95\% confidence regions.
  The auto-correlation results (filled blue contours) are the combined measurement of the \lya forest auto-correlations in the regions A and B. 
  The cross-correlation results (dashed black) are the correlations of the forest in these two regions with quasars.
  The combined results (solid red) simultaneously fit all four correlations taking into account their cross-covariance, and are the main result of this publication. 
  }
  \label{fig:bao}
\end{figure}

The constraints on the BAO parameters are listed in \cref{tab:bao} and shown in \cref{fig:bao}, where we show constraints from the \lya auto-correlation (in filled blue contours, including both \lyaxlyaA and \lyaxlyaB) as well as constraints from the cross-correlation with quasars (dashed black, including both \lyaxqsoA and \lyaxqsoB).
Both measurements are consistent, and their combined constraints (shown in solid red lines) are $\at = 1.013 \pm 0.024$ and $\ap = 0.989 \pm 0.020$, with a correlation coefficient of $\rho = -0.48$.

\begin{table}
\centering
\renewcommand{\arraystretch}{1.2}
\begin{tabular}{c|ccc}
Parameter                                & \multicolumn{3}{c}{Best fit}\\
                                         & Combined                    & \lyaxlya            & \lyaxqso           \\
\hline

$\alpha_{\parallel}$                     & $0.989 \pm 0.020$           & $0.993^{+0.029}_{-0.032}$   & $0.988^{+0.024}_{-0.025}$   \\
$\alpha_{\perp}$                         & $1.013 \pm 0.024$           & $1.020^{+0.036}_{-0.037}$   & $1.005 \pm 0.030$           \\
$\rho_{\alpha_\parallel, \alpha_\perp}$  & -0.48                       & -0.46                       & -0.50                       \\
\hline
$N_{\rm bin}$                            & 9540                        & 3180                        & 6360                        \\
$N_{\rm param}$                          & 17                          & 12                          & 14                          \\
$\chi^2_{\rm min}$                       & 9624.36                     & 3183.79                     & 6427.41                     \\
p-value                                  & 0.23                        & 0.42                        & 0.23                        \\

\end{tabular}
\caption{Best fit BAO parameters (mean of the posterior), uncertainties (68\% credible intervals) and correlation coefficient $\rho$ from the three main analyses: auto-correlations (\lyaxlyaA and \lyaxlyaB), cross-correlations (\lyaxqsoA and \lyaxqsoB) and their combination. 
All parameters  are given at $\zeff = 2.33$.
The \textit{p-value} is only accurate for the combined analysis, because in the other analyses we fix the value of one of the nuisance parameters to the best-fit value in the combined analysis (see discussion in \cref{app:nuisance}).
}
\label{tab:bao}
\end{table}

As discussed in \cref{sec:model}, in addition to the two BAO parameters our model has 15 nuisance parameters that we marginalise over.
The number of degrees of freedom in the combined fit is 9523 (9540-17), and the $\chi^2$ of the best-fit model is 9624.36, with a probability of having a value larger than this of 23\%.
The best-fit values of the nuisance parameters are discussed in \cref{app:nuisance}.
Some of these nuisance parameters only affect the auto-correlation or the cross-correlation, and are therefore ignored when fitting these correlations individually (see \cref{tab:bao} and \cref{tab:nuisances}).
Moreover, when analysing these correlations separately we are not able to break some of the degeneracies between nuisance parameters, and we use extra priors as described in \cref{tab:nuisances}.

In the latest \lya BAO analyses from eBOSS, the auto-correlation provided $\sim 20\%$ better constraints on $\ap$ than the cross-correlation, while providing $\sim 10 \%$ weaker constraints on $\at$ (see figure 12 in \dMdB). 
Redshift space distortions in the quasar dataset ($\beta_{\rm Q} \sim 0.3$) are milder than in the \lya forest ($\beta_{\alpha} \sim 1.7$), reducing the constraining power along the line-of-sight direction. 
On the other hand, the \lya BAO measurement from DESI DR1 seems to be dominated by the cross-correlation for both $\at$ and $\ap$ (see \cref{fig:bao}).
Most \lya quasars in DESI DR1 have only one observation (see right panel of \cref{fig:eboss_desi_footprint}).
In future data releases their signal-to-noise will increase as we collect more observations, and the constraining power of the auto-correlation (doubly affected by noise in the spectra) will increase more than that of the cross-correlation.


\section{Analysis validation}
\label{sec:validation}

Baryon Acoustic Oscillations (BAO) imprint a characteristic three-dimensional feature in the measured correlations.
On the other hand, most spurious correlations from instrumental systematics and astrophysical contaminants are smooth and featureless
\footnote{An important exception is the contamination from metal lines (mostly Silicon) that cause characteristic \textit{bumps} in line-of-sight correlations, but with a very different angular ($\mu$) dependence that allows us to distinguish between them and the BAO parameters.}.
This makes BAO measurements particularly robust.

However, some of the analysis choices presented in \cref{sec:data} cause small changes in the dataset that introduce statistical fluctuations in the BAO measurement.
Examples of these are the observed wavelength range, the rest-frame limits of the \lya regions A and B, or the masking of pixels (due to sky lines, DLAs or BAL features).
Moreover, differences in the quasar redshift estimators cause pixels to fall in and out of the A and B regions, adding an extra source of statistical fluctuations that is also seen in mocks (see appendix B of \dMdB).

\subsection{Blinding strategy}

In order to avoid unconscious or confirmation biases, the development and testing of the analysis framework defined in \cref{sec:data,sec:correlations,sec:model,sec:results} was done using synthetic datasets and blinded measurements.
We considered several blinding strategies for the DESI \lya BAO analysis, including the possibility of blinding the data at the catalog level as done in the galaxy BAO analysis of DESI \citep{DESI2024.III.KP4}.
However, the presence of known sky lines in the spectra, as well as the presence of absorption lines with small restframe wavelength separations from \lya, made it challenging to apply a robust blinding to the data at the catalog level.

For this reason, we opted instead for blinding the measured correlation functions following a simple method described in \cref{app:blinding}.
In short, we applied an additive correction to the measured correlation functions to mimic a blind shift in the BAO parameters.

We defined a list of tests that we needed to pass in order to validate the analysis before \textit{unblinding} the measurement.
First, we validated the analysis using synthetic datasets (or mocks), as explained in detail in a companion paper \cite{KP6s6-Cuceu}, and as summarised in \cref{subsec:test_mocks}.
Second, we studied the consistency of the results under various data splits, as discussed in \cref{subsec:data_splits}.
Finally, we tested the robustness of the results under variations in the analysis setup, as described in \cref{subsec:variations}.
We report here test results applied to the unblinded data, but the same tests were first performed on the blinded data set.

\subsection{Validation using synthetic data}
\label{subsec:test_mocks}

A detailed description of the procedure to generate synthetic DESI spectra (or mocks) for \lya studies can be found in \cite{2024arXiv240100303H}.
The analysis validation of the \lya\ BAO measurement using mocks is presented in a companion paper \cite{KP6s6-Cuceu}.
Here we give a brief summary of the mocks, and we show some of the main tests validating the analysis.

\subsubsection{DESI \lya mocks}

We generated DESI mocks from two different sets of fast simulations: 100 realisations of \texttt{LyaCoLoRe} mocks \cite{Farr2020_LyaCoLoRe,Ramirez2022} and 50 realisations of \texttt{Saclay} mocks \cite{Etourneau2023}.
Both sets of mocks use a log-normal description of the density field, and use simplified recipes to distribute quasars and simulate the optical depth of \lya absorption in redshift space.
These recipes were calibrated in order to approximately reproduce the mean flux, the 1D power spectrum, and the large-scale biases of the \lya forest as measured by the eBOSS Collaboration.
The \texttt{LyaCoLoRe} simulations were also used in the final \lya BAO analysis of eBOSS presented in \dMdB.

As described in \cite{2024arXiv240100303H}, these simulations are post-processed with the script \texttt{quickquasars} of the \texttt{desisim} package\footnote{\url{https://github.com/desihub/desisim}. We used version \href{https://github.com/desihub/desisim/tree/0.38.0}{v0.38.0}.}, where the DESI specificities are introduced, namely the footprint, signal to noise, spectrograph resolution, quasar redshift errors, etc.
At the same time, astrophysical contaminants are introduced such as Damped Lyman alpha systems (DLAs), Broad Absorption Line features (BALs), and absorption from metal lines.

We have made two small changes in the procedure with respect to the description in \cite{2024arXiv240100303H}.
First, we have improved the way in which we imprint the footprint inhomogeneities caused by the survey strategy of DESI.
Second, we have slightly modified the recipes to add metal absorption to the mocks in order to better match the amount of metal contamination seen in the data.
Both of these changes are discussed in more detail in \cite{KP6s6-Cuceu}.

We analyse these mocks using the same analysis applied to the data, with the following minor exceptions in the modelling: (i) we ignore the contamination from CIV, the transverse proximity effect and correlated sky residuals, since these are not included in the mocks;
(ii) we include an extra smoothing to the model to account for pixelisation effects coming from the finite cell size of the log-normal simulations; (iii) we do not smooth the peak component of our model since we use lognormal simulations that do not capture the non-linear broadening of the BAO peak.
These differences are also discussed in more detail in \cite{KP6s6-Cuceu}.

The contaminants that we cannot study with synthetic data are studied with variations in their modelling in \cref{subsec:variations} or discussed in a companion paper. For instance, the CIV contamination and the correlated signal from the sky subtraction are studied in \cite{KP6s5-Guy} as discussed in \cref{subsec:sky}. 
Given that the fitted amplitudes of these contaminants are uncorrelated with $\ap$ and $\at$ (see \cref{fig:bao_corr} of \cref{app:nuisance}), we are confident that they do not bias our estimate of the BAO parameters.
The lack of non-linear broadening of the BAO peak in the mocks results in an underestimation of the BAO uncertainties measured from mocks (see \cref{fig:sigma_bao}). 
As discussed in \cref{sec:model}, when fitting the DESI DR1 measurement we model the broadening using calculations based on Lagrangian Perturbation Theory \cite{Kirkby2013}.
This model reproduces the shape of the BAO peak in the correlations measured from \lya mocks constructed from N-body simulations \cite{Hadzhiyska2023}.

Finally, the algorithm used to identify DLAs in the data requires a significant amount of computing time, and therefore we decided to not run it on the many realisations of mocks.
Instead, we assume that we find all HCDs with $\log N_{\rm HI} > 20.3$, and none below this column density, and we mask them in the analysis.
Similarly, we assume that we can find all BAL features in the data, and mask the corresponding region of spectra accordingly. The impact of BAL completeness is studied in detail in \cite{KP6s9-Martini}.

\subsubsection{Validation of the covariance matrix}
\label{subsec:covariance-validation}

In order to validate the covariance matrices of the correlation functions for the purpose of measuring the BAO scale, we measured the quantity
\begin{equation}
\delta_\alpha \equiv \left( \partial_{\alpha} M ^ T C^{-1} \partial_{\alpha} M \right)^{-1} \partial_{\alpha} M ^ T C^{-1} R
\end{equation}
for each mock, where $\partial_{\alpha} M$ is the derivative of the best fit model from a stack of mocks with respect to a BAO scale parameter $\alpha$, $C$ the covariance matrix of the correlation functions and $R$ the data minus the best fit model from the stack. The covariance $C$ is determined for each mock independently with the sub-sampling and smoothing method described in \cref{subsec:covariance}, so the noise of the covariance itself is directly comparable to that of the true data.

This is not a true fit (which involves a non-linear model plus many nuisance parameters, see \cref{sec:model}) but a straightforward compression of the data and its covariance that is best suited to describe the fluctuations that matter for measuring $\alpha$.
We also measure the uncertainty on this parameter from the covariance,
\begin{equation}
\sigma_\alpha \equiv \left( \partial_{\alpha} M ^ T C^{-1} \partial_{\alpha} M \right)^{-1/2} ~.
\end{equation}

We then measure the rms of the distribution of $ \left( \delta_\alpha / \sigma_\alpha \right)$. If the covariance matrix is correct we expect this distribution to be Gaussian with a rms of 1. We looked at $\alpha_{\parallel}$ and $\alpha_{\perp}$ and linear combinations of both (major and minor axes of 2D uncertainty contour).
When using the full covariance matrix (including the cross-covariance between the correlation functions), we find a scatter of $0.96 \pm 0.05$ ($0.99 \pm 0.06$) for $\alpha_\parallel$ ($\alpha_\perp$) for the \texttt{LyaCoLoRe} mocks, and $1.07 \pm 0.07$ ($1.03 \pm 0.07$) for the \texttt{Saclay} mocks.
This shows that the statistical uncertainties derived from the covariance matrix are a good estimate of the dispersion among random mock realizations. This gives us confidence in the covariance matrix at the scales of interest for the measurement of the BAO.
As discussed in more detail in \cite{KP6s6-Cuceu}, the scatter was 10\% larger for both sets of mocks when ignoring the cross-covariance between the correlation functions.

\subsubsection{Validation of BAO estimates}

\begin{figure}
\centering
\includegraphics[width=0.49\textwidth]{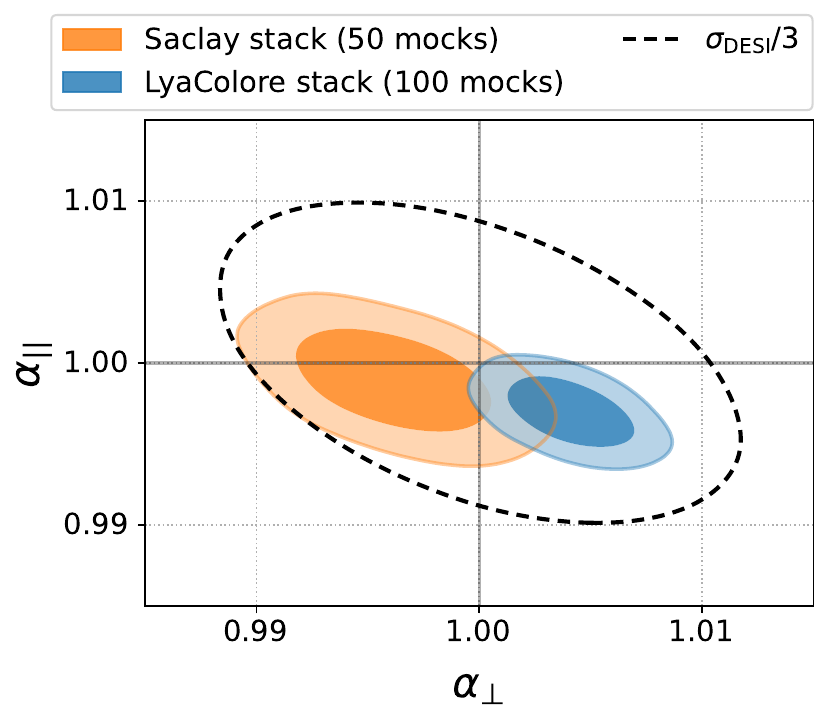}
\includegraphics[width=0.49\textwidth]{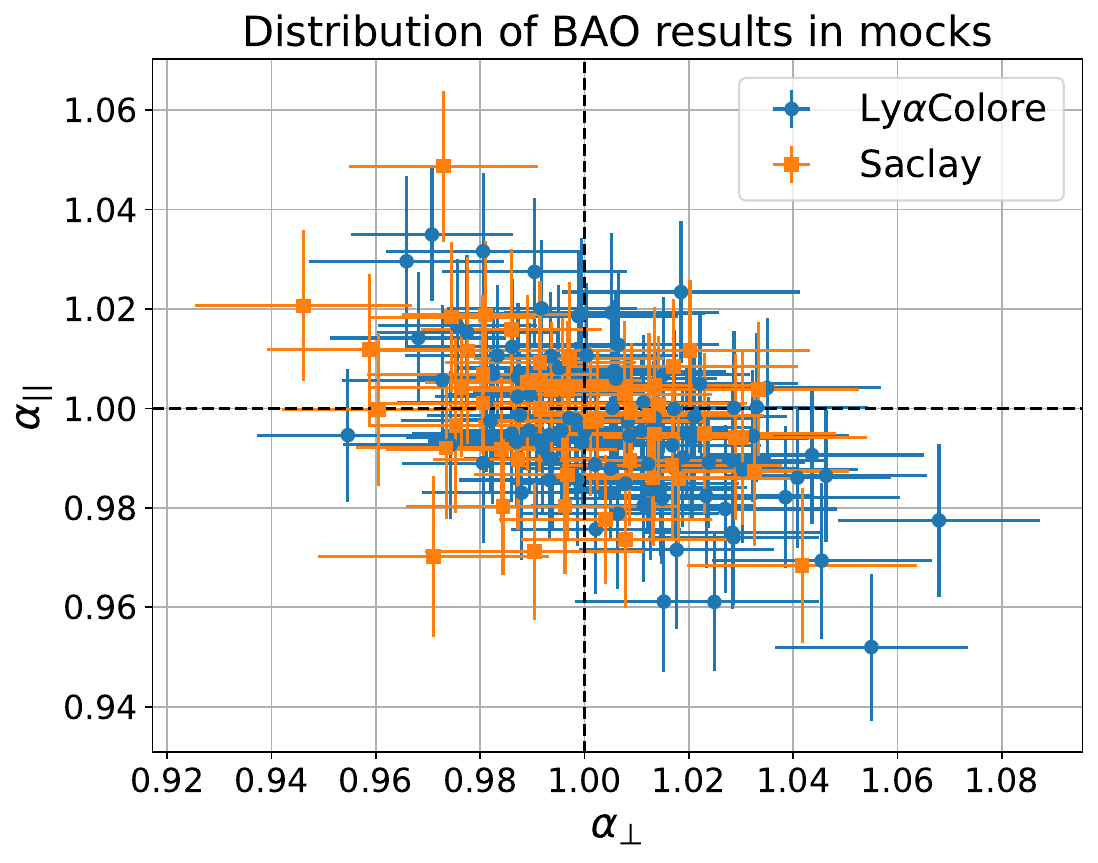}
\caption{Left: BAO constraints from the ``stack" of 100 \texttt{LyaColore} (blue) and 50 \texttt{Saclay} (orange) DESI DR1 synthetic datasets, compared to the constraints from data (scaled down by a factor of 3, dashed black ellipse).
Right: scatter plot of the best fit $\alpha_{\perp}$ and $\alpha_{\parallel}$ from each of the \texttt{LyaColore} (blue) and \texttt{Saclay} (orange) mocks.
Note the difference of scale between the two plots.
}
\label{fig:stack_bao}
\end{figure}

We measured the correlations in 100 \texttt{LyaCoLoRe} mocks and 50 \texttt{Saclay} mocks mimicking the DESI DR1 dataset, and combine their measurements of the correlations into ``stacks of correlations", with a statistical power much larger than that of the final DESI dataset.
The BAO constraints from these stacks are shown in the left panel of \cref{fig:stack_bao}.
We used the same cosmology to generate the mocks and to analyse them, so we should expect to recover values of ($\at$, $\ap$) consistent with unity.

In order to proceed with the unblinding, we had set a requirement that the measurement on the stack of many mocks could not present a bias larger than a third of the statistical uncertainty obtained when fitting the blinded data\footnote{Note that the dashed black contour in \cref{fig:stack_bao} shows the uncertainties on data \textit{after unblinding}, and these are slightly larger than the blinded ones.}.
This corresponded to a tolerance of $\sim 0.005$ in $\alpha_{\parallel}$ and $\sim 0.007$ in $\alpha_{\perp}$.

There is a small bias in the measurement of BAO from the stack of 100 \texttt{LyaCoLoRe} mocks (blue contours), but it is smaller than the requirement accuracy (black dashed contours).
The difference between the results from the two sets of mocks also suggests that any offsets seen due to analysis problems are at the same level as systematics in the creation of the mocks.
Moreover, combining these with the results from the stack of 50 \texttt{Saclay} mocks (orange contours) would further reduce the bias.

\subsubsection{Validation of BAO uncertainties}

We discuss here the distribution of BAO results when fitting each of the DESI DR1 mocks individually, and use them to validate several aspects of our analysis.
In the right panel of \cref{fig:stack_bao} we show the best-fit values for the BAO parameters ($\alpha_\parallel$ and $\alpha_\perp$) for the 50 \texttt{Saclay} and the 100 \texttt{LyaCoLoRe} mocks, with errorbars representing the 1-$\sigma$ uncertainties.
In the following we combine the results from these 150 mocks.

\begin{figure}
\centering
\includegraphics[width=0.98\textwidth]{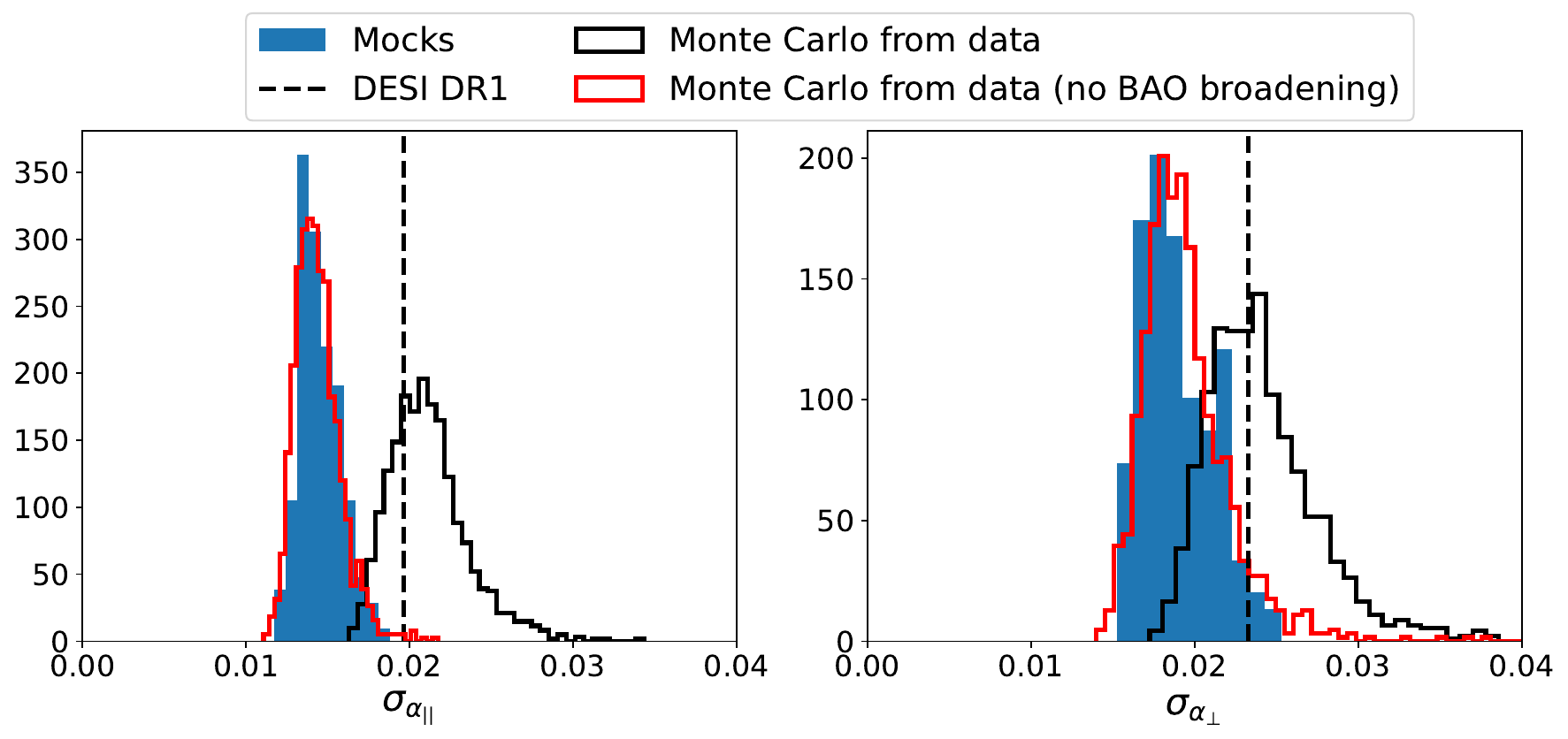}
\caption{Distributions of BAO uncertainties along ($\sigma_{\alpha_{||}}$, left) and across ($\sigma_{\alpha_\bot}$, right) the line of sight.
The blue histogram shows the distribution from the analysis of 150 DESI DR1 mocks, while the vertical dashed lines are the uncertainties measured in the data.
The solid black line shows the distribution of BAO uncertainties from Monte Carlo realisations of the data covariance matrix, when using the best-fit model.
The solid red line, on the other hand, shows an equivalent distribution from Monte Carlo realisations generated around a linear model that does not include the non-linear broadening of the BAO peak.
These Monte Carlo realisations are discussed in detail in \cite{KP6s6-Cuceu}.
}
\label{fig:sigma_bao}
\end{figure}

\Cref{fig:sigma_bao} shows the distribution of uncertainties on $\alpha_\parallel$ and $\alpha_\perp$ in the mocks (blue histogram) and compares it to the uncertainty measured in DESI DR1 (vertical dashed line).
One can see that the BAO uncertainties vary significantly from mock to mock, as expected, but that the uncertainties from DESI DR1 are larger than those from mocks.
The BAO uncertainties from analyses of the DESI DR1 mocks are expected to be a bit smaller than in the data, since the mocks used in this analysis do not have non-linear broadening of the BAO peak.
In order to study this, we generated $1\,000$ Monte Carlo (MC) realisations of the correlation functions using the covariance matrix from the data, adding the random fluctuations to the best-fit model from our main analysis
\footnote{These Monte Carlo realisations are discussed in more detail in \cite{KP6s6-Cuceu}.}.
The distribution of BAO uncertainties from these MC realisations is shown in black in \cref{fig:sigma_bao}.
It clearly shows that the DESI DR1 \lya BAO result is not an outlier, and that the constraining power of the mocks is larger than that of the data.
In the same figure we show (in red) the distribution of uncertainties from another set of $1\,000$ MC realisations, where the fluctuations have now been added to a model that ignores the non-linear broadening of the BAO peak.
This distribution is in very good agreement with the distribution of uncertainties from the fits on individual mocks, and confirms the hypothesis that the non-linear broadening of the BAO peak degrades the \lya BAO result significantly.

As discussed in \cite{KP6s6-Cuceu}, the $\chi^2$ value from the data is consistent with the distribution of $\chi^2$ in the mocks.
In the same publication, we also look at the distribution of BAO residuals ($\Delta \alpha_{\parallel}/\sigma_{\alpha_{\parallel}}$ and  $\Delta \alpha_{\perp}/\sigma_{\alpha_{\perp}}$).
Their rms is found to be of $1.01 \pm 0.07$ for $\alpha_\parallel$ and $1.11 \pm 0.06$ for $\alpha_\perp$, with uncertainties obtained through bootstrap. Those values which are close to one validate our error propagation. We note that this is a more comprehensive test than the one presented in \cref{subsec:covariance-validation} because it is based on the results from the full non-linear fit that includes numerous nuisance parameters.

From the tests discussed above we conclude that our BAO estimates on mocks are unbiased at the level of precision required by the current dataset, and that the scatter of best-fit values is consistent with the reported uncertainties.

\subsection{Data splits}
\label{subsec:data_splits}

\begin{figure}
\centering
\includegraphics[width=0.99\textwidth]{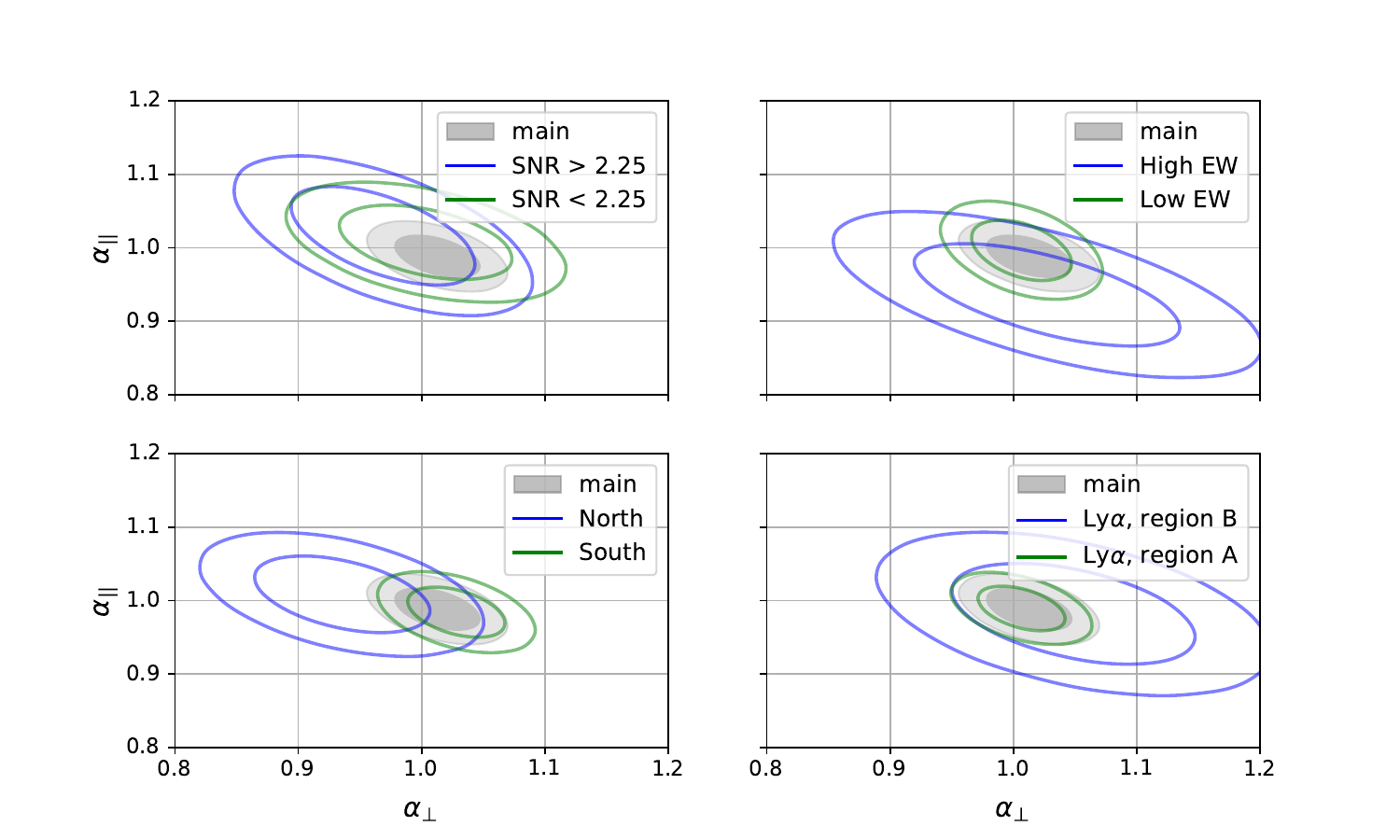}
\caption{BAO constraints from the main analysis (grey) and from data splits.
Top left: low (green) vs high (blue) SNR in the quasar spectrum.
Top right: low (green) vs high (blue) CIV equivalent width (EW) in the quasar spectrum.
Bottom left: South (green) vs North (blue) imaging used in the quasar target selection.
Bottom right: correlations from region A (green) and region B (blue); the A region shows the combined measurement from the auto-correlation of the forest measured in the A region (\lyaxlya) and the cross-correlation of this region with quasars (\lyaxqsoA).
The contours labeled region B show the combined measurement of the forest auto-correlation measured in the B region (\lyaxlyaB) and the cross-correlation of this region with quasars (\lyaxqsoB).
}
\label{fig:data_splits}
\end{figure}

The second set of tests that we use to validate the analysis are data splits, where we measure BAO using different subsets of the data.
A first data split was already shown in \cref{fig:bao}, where we presented the consistency of BAO measurements from the auto-correlations (including both the \lyaxlyaA and the \lyaxlyaB correlations) and from the cross-correlations (including \lyaxqsoA and \lyaxqsoB).
In the bottom right panel of \cref{fig:data_splits} we group them instead in correlations that only use pixels in the  A region (\lyaxlyaA and \lyaxqsoA, in green) and correlations that use pixels in the B region (\lyaxlyaB and \lyaxqsoB, in blue).
The B region does not contain as much information as the A region for several reasons: only quasars at higher redshift have this region in the DESI spectrograph, so the number of \lya\ lines-of-sight in the B region is smaller; the B region, as defined in \cref{subsec:forests}, is significantly shorter than the A region, so there are fewer pixels per line-of-sight; finally, the B region is affected by other Lyman lines that add extra variance to the fluctuations.

We will now discuss the consistency of the BAO constraints when splitting the quasar catalog in three ways: by imaging survey used in the target selection; by \ion{C}{IV} emission line equivalent width (EW); and by signal-to-noise in the spectrum.
In these cases we run alternative end-to-end analyses starting from new sub-catalogs, i.e., fitting new quasar continua, measuring correlations and fitting them separately for each subset of the data\footnote{Note that when we split the data set in two by \ion{C}{IV} EW or SNR, the density of quasars in each subset is a factor of two lower, so the number of pixel pairs or pixel quasar pairs is about four times smaller.}.

We start in the bottom left panel of \cref{fig:data_splits} by splitting the catalog based on the imaging survey that was used for target selection. Most of the DESI footprint was observed with the DECam camera on the Blanco telescope in Chile, including the entire South Galactic Cap and the southern fraction of the North Galactic Cap \citep{LS.Overview.Dey.2019,BASS.Zou.2017,LS.dr9.Schegel.2024}. We refer to this subset of the data as ``South", while we designate the area that was imaged in the BASS and MzLS surveys at $\delta \textgreater 32.375^{\circ}$ as ``North". The South sample is significantly larger, as it contains 82\% of the quasars (579\,166 quasars with $z>1.77$ in the South and 130,399 in the North).

In the top right panel of \cref{fig:data_splits} we look at a second catalog split based on the \ion{C}{IV} EW. We do this split because we expect the shape of the quasar spectral energy distribution to depend on EW, due to the well known anti-correlation between the EW of quasar emission lines and the continuum luminosity known as the Baldwin Effect \citep{Baldwin1977}.
We measure the \ion{C}{IV} EW of the quasars with \texttt{fastspecfit} \footnote{\url{https://github.com/desihub/fastspecfit}. We use version \href{https://github.com/desihub/fastspecfit/tree/2.4.2}{v2.4.2}.}, finding a median of 37.3\,\AA\ for all quasars, and a median of 41.6\,\AA\ for $3\sigma$ measurements of the \ion{C}{IV} emission line.
We split the sample at 39\,\AA, which produce a low (high) EW sample of 371\,751 (337,814) quasars at $z>1.77$.
As expected from the Baldwin Effect, the low EW sample is somewhat higher luminosity and has a somewhat higher effective redshift of 2.36 compared to 2.29 for the high EW sample.

Finally, in the top left panel of \cref{fig:data_splits} we present the third quasar catalog split, based on mean signal-to-noise ratio (SNR) in the spectra.
Instead of splitting the quasar catalog into two subsets of equal size, we chose a SNR threshold of 2.25 such that both subsets have the same weight in the measurement of the auto-correlation function.
There are different ways of estimating the SNR of quasar spectra, and we decided to use the mean value of SNR per pixel averaged over the \lya\ region, as reported by the \texttt{picca} code at the end of the continuum fitting process of the main analysis.
This results in a lower SNR catalogue with 321\,767 quasars and a higher SNR catalogue with 106\,636 quasars.
The sum of these catalogues does not match the total size of the catalogue used in the main analysis (709,565 quasars at $z>1.77$), since we do not detect the forest continuum at $z < 2$ and therefore do not have a SNR estimate.

With the exception of the North vs South data split, the subsets discussed here share the same footprint and redshift range.
However, cosmic variance is a very small component of the covariance matrix and to a first approximation the data splits can be considered as independent.
Taking this into account, the BAO constraints on the various data splits are consistent with statistical fluctuations.

\subsection{Alternative analyses}
\label{subsec:variations}

In this section we show a final set of validation tests: robustness of the BAO measurement under variations in the analysis configuration.
We set a requirement for unblinding that none of the variations would cause a shift on the BAO parameters
larger than a third of the statistical uncertainty from the main (blind) analysis.
This corresponded to a tolerance of $\sim 0.005$ in $\alpha_{\parallel}$ and $\sim 0.007$ in $\alpha_{\perp}$.
However, some of the analysis variations result in a small change in the size of the dataset.
In these cases we relax the requirement and take into account the probability of the measured shift being caused by statistical fluctuations.
These are discussed in \cref{app:bao_shifts}.

In order to reduce the amount of computing time needed in these alternative analyses, we do not run the nested sampler algorithm and only report the maximum likelihood values and Gaussian errors computed by \texttt{iminuit}.
As discussed in \cref{app:sampling}, the BAO results do not vary significantly with the sampling method used.
In some of the alternative analyses \texttt{iminuit} has difficulty breaking internal degeneracies between nuisance parameters, particularly between $L_{\rm HCD}$ and some of the bias parameters.
In order to avoid this, we fix $L_{\rm HCD}$ in all the alternative analyses to the best-fit value of the main analysis ($6.51~\hMpc$), and we show in the variation ``vary $L_{\rm HCD}$" that this has a negligible impact on the BAO results.

\begin{figure}
\centering
\includegraphics[width=0.79\textwidth]{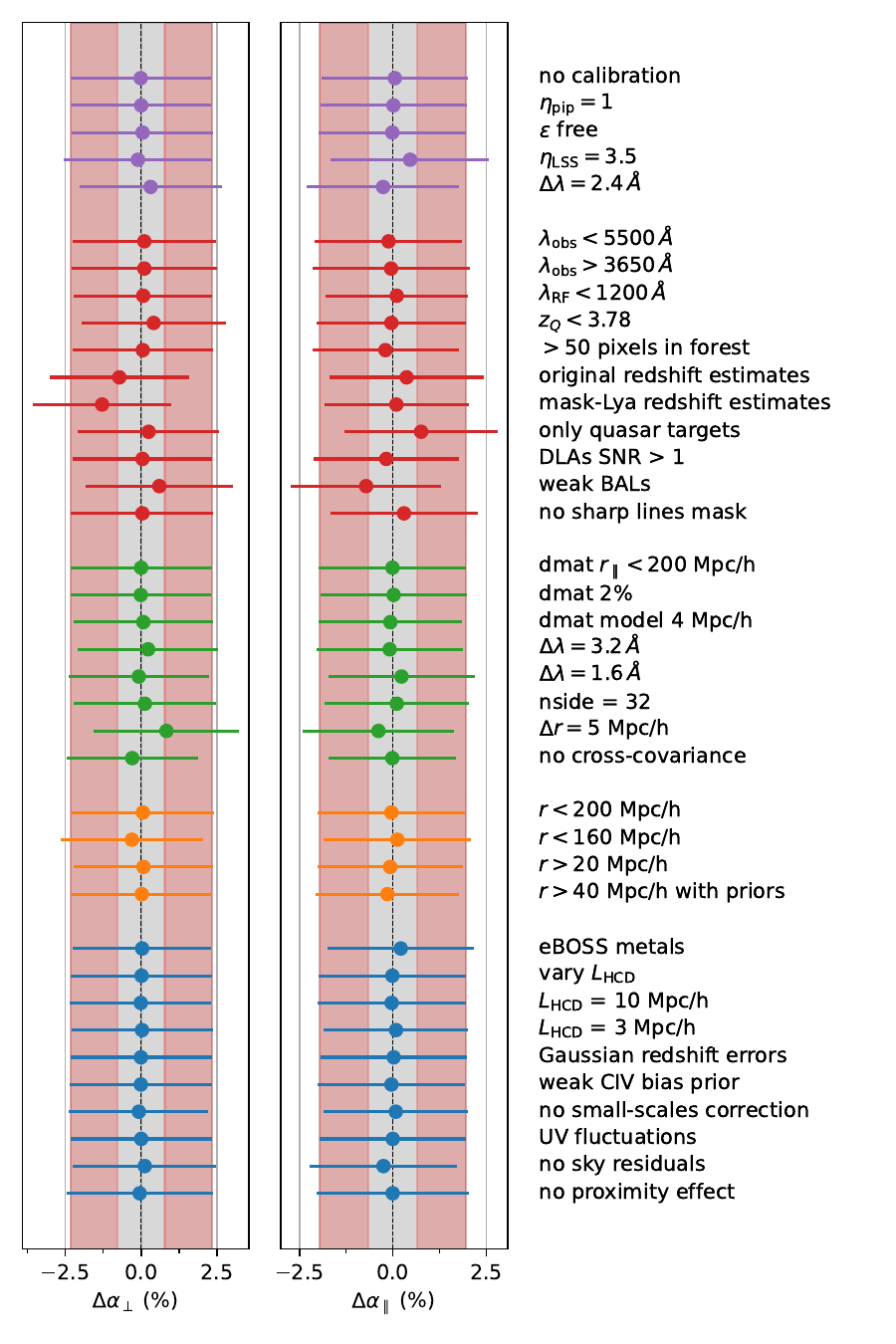}
\caption{Shifts in the BAO parameters from alternative analyses, including variations in the method to estimate the fluctuations (purple);
variations in the dataset (red);
variations in the method to compute correlations and covariances (green);
variations in the range of separations used (orange);
and variations in the modelling (blue).
The red shaded contours show the one $\sigma$ uncertainty from the main analysis, while the smaller gray area shows the threshold set to these tests ($\sigma / 3$).
Note that the two parameters are anti-correlated ($\rho=-0.48$).
Variations of the dataset (in red) are subject to statistical fluctuations as described in \cref{app:bao_shifts}.}
\label{fig:var_all_1d}
\end{figure}

\subsubsection{Variations in the estimation of the fluctuations}

The method to estimate the \lya fluctuations starting from the observed quasar spectra is described in \cref{sec:cont_fit}.
In the first set of variations shown (in purple) in \cref{fig:var_all_1d} we quantify the impact on the BAO results when we vary different aspects of the method.
In particular we look at the following variations:

\begin{itemize}

     \item no calibration: We do not re-calibrate the spectra with the \ion{C}{III} region mentioned in \cref{subsec:forests} and described in \cite{2023MNRAS.tmp.3626R}.

     \item $\eta_{\rm pip} = 1$: We do not apply the re-calibration of the instrumental noise $\eta(\lambda)$ mentioned in \cref{subsec:forests} and described in \cite{2023MNRAS.tmp.3626R}.

     \item $\epsilon$ free: We include an extra term ($\epsilon$ in \dMdB) in the computation of the \lya\ weights to try to capture quasar diversity.

     \item $\eta_{\rm LSS} = 3.5$: We reduce the contribution from the intrinsic \lya\ forest variance to the weights by a factor of two.

     \item $\Delta \lambda = 2.4\angs$: We coadd three pixels into one before the continuum fitting step (as done in \dMdB).
     In this variation we also use $\sigma^2_{\rm mod} = 3.1$ as suggested by \cite{2023MNRAS.tmp.3626R} when coadding DESI data by three pixels.

\end{itemize}

We continue with a second set of variations (in red) in \cref{fig:var_all_1d} where we look at variations that cause changes in the dataset by removing (or adding) pixels or entire quasars.
As discussed in \cref{app:bao_shifts}, these can cause statistical fluctuations in the BAO results.
In particular we look at the following variations:

\begin{itemize}

     \item $\lambda_{\rm obs} < 5500\angs$: We use only \lya\ pixels below this observed wavelength ($\lambda_{\rm obs} < 5577\angs$ in the main analysis).

     \item $\lambda_{\rm obs} > 3650\angs$: We use only \lya\ pixels above this observed wavelength ($\lambda_{\rm obs} > 3600\angs$ in the main analysis).

     \item $\lambda_{\rm RF} < 1200\angs$: We use only \lya\ pixels below this rest-frame wavelength ($\lambda_{\rm RF} < 1205\angs$ in the main analysis).

     \item $z_Q < 3.78$: We use only quasars with $z_Q < 3.78$ (highest redshift included in the mocks discussed in \cref{subsec:test_mocks}).

     \item $> 50$ pixels in forest: We include lines-of-sight with more than 50 valid \lya\ pixels (150 in the main analysis).

     \item original redshift estimates: We use the quasar redshifts in the original run of \texttt{Redrock} (slightly biased as discussed in \cite{RedrockQSO.Brodzeller.2023}).

     \item mask-Lya redshift estimates: We use a different estimator for the quasar redshifts, re-running \texttt{Redrock} with only wavelengths longer than the \lya\ emission line (as done in \dMdB).

     \item only quasar targets: We use only quasars that were considered quasar targets.

     \item DLAs ${\rm SNR}>1$: We mask DLAs in spectra with ${\rm SNR}>1$ (${\rm SNR}>3$ in the main analysis).

     \item weak BALs: We remove spectra with strong BAL features ($A_{\rm I} > 840$, 50\% percentile of strongest BALs in eBOSS \cite{Ennesser2022}).

     \item no sharp lines mask: We do not mask the 4 sharp lines discussed in \cref{subsec:forests}.

\end{itemize}

Some of the variations (like ``mask-Lya redshift estimates") are slightly outside the threshold of $1/3\, \sigma$.
As we discuss in \cref{app:bao_shifts}, these shifts can be explained by statistical fluctuations caused by the addition or subtraction of \lya pixels from the dataset.

\subsubsection{Variations in the measurement of correlations}

We now move to the third set of variations (in green) shown in \cref{fig:var_all_1d}, where we look at the impact of varying the setup in the measurement of correlations, their covariances and the distortion matrices described in \cref{sec:correlations}.
In particular we look at the following variations:
\begin{itemize}

    \item dmat $r_\parallel < 200$ Mpc/h: We model the distortion matrix up to $r_\parallel = 200~\hMpc$ as done in \dMdB ($r_\parallel = 300~\hMpc$ in the main analysis).

    \item dmat 2\%: We use 2\% of pixels to compute the distortion matrix (1\% in the main analysis).

    \item dmat model 4 Mpc/h: We model the distortion matrix using the same $4~\hMpc$ binning as the correlation function ($2~\hMpc$ in the main analysis).

    \item $\Delta \lambda = 3.2$ \AA: We rebin the \lya fluctuations in groups of 4 pixels (3 in the main analysis).

    \item $\Delta \lambda = 1.6$ \AA: We rebin the \lya fluctuations in groups of 2 pixels (3 in the main analysis).

    \item nside $= 32$: We measure the correlations in HEALPix pixels defined by nside=32 (16 in the main analysis).

    \item $\Delta r = 5$ Mpc/h: We use $5~\hMpc$ bins in the correlation functions ($4~\hMpc$ in the main analysis).

    \item no cross-covariance: We ignore the cross-covariance of the different correlation functions (as done in eBOSS, \dMdB).

\end{itemize}

We do not see any problematic variation related to the measurement of correlations and their covariances.

\subsubsection{Variations in the parameter estimation}

We move now to the impact of variations in the parameter estimation.
We start with the four set of variations (orange) shown in \cref{fig:var_all_1d} by looking at the impact of the range of separations used:

\begin{itemize}

    \item $r < 200$ Mpc/h: We fit separations with $r < 200~\hMpc$ ($180~\hMpc$ in the main analysis).

    \item $r < 160$ Mpc/h: We fit separations with $r < 160~\hMpc$ ($180~\hMpc$ in the main analysis).

    \item $r > 20$ Mpc/h: We fit separations with $r > 20~\hMpc$ ($10~\hMpc$ in the main analysis).

    \item $r > 40$ Mpc/h with priors: We fit separations with $r > 40~\hMpc$ ($10~\hMpc$ in the main analysis).
    Without the smaller scales we are not able to constraint several nuisance parameters, and therefore we add the informative priors described in \cref{tab:tight_priors}.

\end{itemize}

\begin{table}
\centering
\begin{tabular}{c|c}
Parameter                                          & Prior                                             \\
\hline
$b_{HCD}$                                          & $\mathcal{N}(-0.0556, 0.0034)$                    \\
$10^3 b_{\rm SiIII(1207)}$                         & $\mathcal{N}(-9.78, 0.56)$                        \\
$\sigma_v (h^{-1} {\rm Mpc})$                      & $\mathcal{N}(3.66, 0.14)$                         \\
$\Delta r_{\parallel} (h^{-1} {\rm Mpc})$     & $\mathcal{N}(0.067, 0.058)$                       \\
$\xi_0^{\rm TP}$                                   & $\mathcal{N}(0.399, 0.051)$                       \\
\end{tabular}
\caption{Extra Gaussian priors added to some of the variations discussed in \cref{subsec:variations}.
They correspond to the best-fit values and uncertainties from the main analysis as reported in \cref{tab:nuisances}.
}
\label{tab:tight_priors}
\end{table}

Finally, in the last set of variations (in blue) shown in \cref{fig:var_all_1d} we look at the impact of different modelling choices.
In particular we look at the following variations
\footnote{The last two variations in this list were added during the refereeing process and were not part of the original analysis validation.}:
\begin{itemize}

    \item eBOSS metals: We model the contamination by Silicon lines (metals) following the method used in eBOSS \cite{dMdB2020} instead of the new method described in \cref{subsec:metals}.

    \item vary $L_{\rm HCD}$: We vary the parameter $L_{\rm HCD}$ in the model of the contamination by HCDs as done in the main analysis.
    This parameters was fixed to $L_{\rm HCD}=6.51~\hMpc$ in the other variations to minimise the degeneracies between this and other nuisance parameters.
    See the related discussion in \cref{app:voigt_hcd}.

    \item $L_{\rm HCD} = 10$ Mpc/h: We use a fixed value of $L_{\rm HCD}=10~\hMpc$ to model HCD contamination (free parameter in the main analysis), as was done in \dMdB.

    \item $L_{\rm HCD} = 3$ Mpc/h: We use a fixed value of $L_{\rm HCD}=3~\hMpc$ to model HCD contamination (free parameter in the main analysis).

    \item Gaussian redshift errors: We use a Gaussian distribution to model quasar redshift errors and quasar peculiar velocities (a Lorentzian distribution is used in the main analysis).

    \item weak CIV bias prior: We use a significantly weaker flat prior on the CIV bias parameter of $-0.03 < b_{\rm CIV} < 0$ (instead of $-0.0243 \pm 0.0015$ in the main analysis).

    \item no small-scales correction: We ignore the small-scales multiplicative correction from \cite{Arinyo2015} in the modelling of the \lya auto-correlation.

    \item UV fluctuations: Following the prescription of \cite{Bautista2017}, we model the impact of fluctuations in the UV background \cite{Pontzen2014,Gontcho2014} on the \lya forest auto-correlation.
    We do not detect these fluctuations in our analysis.

    \item no sky residuals: We ignore the contamination from correlated sky residuals in the \lya auto-correlation, discussed in \cref{subsec:sky}.

    \item no proximity effect: We ignore the \textit{proximity effect}, the impact of quasar radiation in the cross-correlation discussed in \cref{subsec:proximity}.

\end{itemize}

None of the variations in the parameter estimation cause a significant shift in the inference of the BAO parameters.
Some of the nuisance parameters, on the other hand, are more sensitive to changes in the analysis setup.

\subsubsection{Broadband polynomial corrections}
\label{subsec:broadband-validation}

As a final test to demonstrate the robustness of the BAO measurement, we run an alternative analysis where we introduce a flexible but smooth additive component to each of the four modelled correlations.
In particular, we follow the procedure of \cite{Bautista2017,dMdB2020} and use (for each correlation) Legendre polynomials $L_j(\mu)$ of order $j=0, 2, 4$ and $6$ to describe the angular dependence of the additive terms, divided by powers of $r^i$ with $i=0, 1, 2$ (corresponding to a parabola in $r^2 \xi(r)$).
The total number of broad-band parameters is therefore 48 (12 for each of the four correlations).

We then fit the baseline model with those additional parameters in the limited separation range 40 Mpc/h $< r <$ 180 Mpc/h while adding extra priors on nuisance parameters as described in \cref{tab:tight_priors} to break degeneracies between the polynomial coefficients and other parameters.
The best fit model with and without those broadband corrections is compared to the data in \cref{fig:baseline-correlation-wedges,fig:baseline-correlation-lyaqso-wedges}. The shifts in the best fit BAO parameters when adding broadband terms are $\Delta \alpha_\parallel = + 0.001$ and $\Delta \alpha_\perp = - 0.001$, and the change in the uncertainties is negligible ($|\Delta \sigma_\alpha| <0.001$, both for $\alpha_\parallel$ and $\alpha_\perp$). We also note that the value of $\chi^2$ is reduced by only 47.5 points when adding 48 new parameters.

\subsection{Conclusion on the validation tests}
\label{subsec:validation-conclusion}

We have presented numerous validation tests of the analysis using mocks, data splits, variations in choices made for the analysis, and adding ad-hoc broadband terms to improve the fit of the correlations around the BAO peak. All tests where the data set is left unchanged result in variations much smaller than 1/3 of the final statistical error.
All data splits and other tests where the data set was altered are consistent with statistical fluctuations.
%
%

Adding broadband terms moved the best fit by less than a tenth of the statistical error. The largest offsets were found with mocks with average shifts of $\Delta \alpha_{\parallel} = -0.003 \pm 0.0014$ and  $\Delta \alpha_{\perp} = +0.004 \pm 0.0018$  for the LyaColore mocks (see \cref{fig:stack_bao}). 
Because we find similar discrepancies among the two sets of mocks (LyaColore and Saclay), we can not use those offsets to correct the measurements and we hence have to treat them as systematic uncertainties. Adopting a conservative systematic uncertainty of 0.5\%, this results in a 3\% (2\%) increase of the total uncertainties on the longitudinal (transverse) BAO measurement when combined quadratically with the statistical errors. We consider that this increase in uncertainties is small enough to be ignored. As a result we only report statistical uncertainties in the following sections.

\section{Discussion} 
\label{sec:discussion}

In \cref{sec:results} we presented a measurement of the BAO parameters ($\at$, $\ap$) at $\zeff=2.33$ from DESI DR1. 
Thanks to the peak-smooth decomposition introduced in \cref{subsec:decomposition}, we interpret these parameters as:

\begin{align}
\ap &= \frac{D_H(\zeff) / r_d}{\left[ D_H(\zeff) / r_d \right]_{\rm fid}} ~, \nonumber \\ 
\at &= \frac{D_M(\zeff) / r_d}{\left[( D_M(\zeff) / r_d \right]_{\rm fid}} ~,
\end{align}
where $D_M(z)$ is the transverse comoving distance, $D_H(z)=c/H(z)$ and $r_d$ is the sound horizon at the drag epoch. Quantities with $\left[ \right]_{\rm fid}$ are computed with the fiducial cosmology (see \cref{tab:fid_cosmo}).
In combination with the $z<2$ BAO measurements from \cite{DESI2024.III.KP4}, our BAO measurement enables the state-of-the-art cosmological constraints presented in \cite{DESI2024.VI.KP7A}.

\subsection{Cosmological distances}

Substituting values from our fiducial cosmology, we can rewrite the ($\ap$, $\at$) constraints as the following constraints on ratios of distances:
\begin{align}
 D_H(\zeff) / r_d & = 8.52 \pm 0.17 ~, \nonumber \\
 D_M(\zeff) / r_d & = 39.71 \pm 0.95 ~,
\end{align}
with a correlation coefficient of $\rho = -0.48$.
For a given value of the sound horizon $r_d$, these translate into a measurement of the expansion rate at $\zeff=2.33$:
\begin{equation}
 H(\zeff) = \left( 239.2 \pm 4.8 \right) \frac{147.09~\Mpc}{r_d} \, \rm{km/s/Mpc}
 \end{equation}
and a measurement of the comoving transverse distance to $\zeff$:
\begin{equation}
 D_M(\zeff) = \left( 5.84 \pm 0.14 \right) \frac{r_d}{147.09~\Mpc} \, \rm{Gpc} ~.
\end{equation}

It is also convenient to report the BAO information as an isotropic dilation parameter
\begin{equation}
D_V(\zeff)/ r_d = \left(\zeff D_M^2 D_H \right)^{1/3}/ r_d = 31.51 \pm 0.44 ~,
\end{equation}
and an anisotropic (or Alcock-Paczy\'nski \cite{AP1979}) parameter $f_{\rm AP}$:
\begin{equation}
f_{\rm AP}(\zeff) = \frac{D_M(\zeff)}{D_H(\zeff)} = 4.66 \pm 0.18 ~.
\end{equation}
However, the ratio $D_V / r_d$ is only the optimal definition of the isotropic BAO parameter in the absence of redshift space distortions. 
Every BAO measurement will have a different combination of ($D_H$,$D_M$) that will minimise the correlation with $f_{\rm AP}$ and will therefore have a smaller relative uncertainty.
From the posterior of the \lya BAO measurement of DESI DR1, we find that this combination is approximately:
\begin{align}    
D_M(\zeff)^{9/20} D_H(\zeff)^{11/20} / r_d &= 17.03\pm 0.19 ~,
\end{align}
which corresponds to a 1.1\% measurement of the (isotropic) BAO scale at $\zeff=2.33$.

\subsection{Comparison with previous measurements from SDSS}

\begin{figure}
\centering
\includegraphics[width=0.8\textwidth]{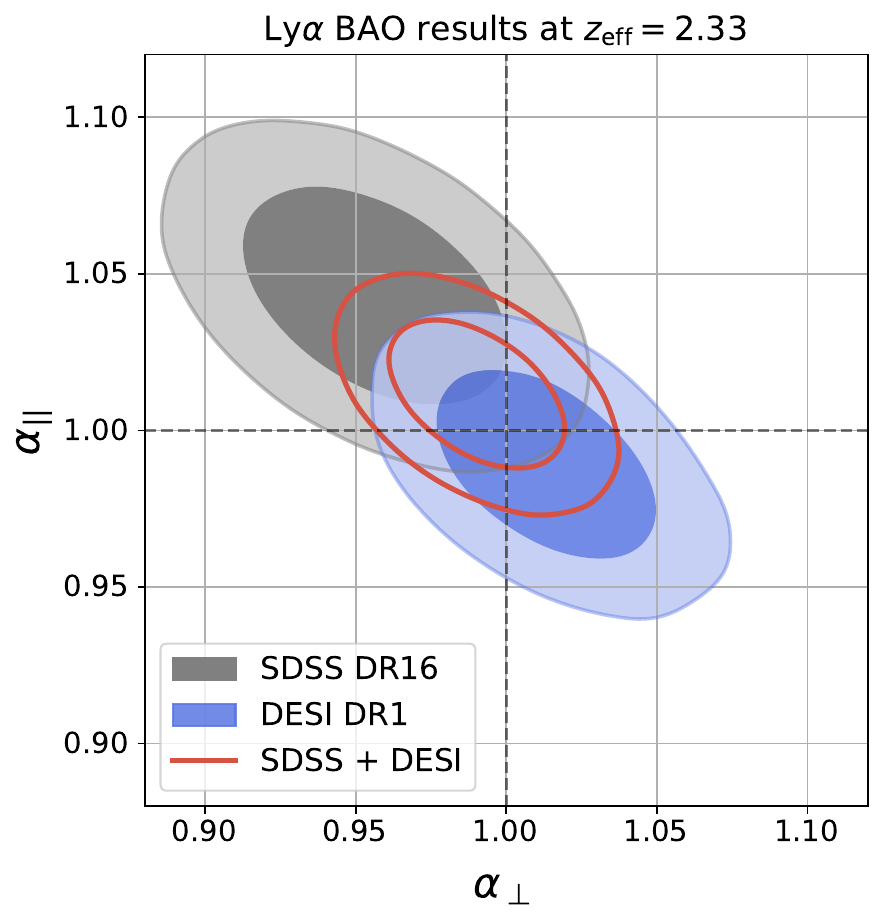}
\caption{\lya BAO measurement from DESI DR1 (blue) compared to the equivalent measurement from eBOSS using data from SDSS DR16 (\dMdB, gray).
The red contours show the combined measurement taking into account the cross-survey covariance as discussed in \cref{app:eboss_covar}.}
\label{fig:eboss_desi_bao}
\end{figure}

Prior to this work, the BAO scale at $z \sim 2.3$ had been measured by the BOSS and eBOSS collaborations with successively larger \lya forest datasets \cite{Busca2013,Slosar2013,Kirkby2013,FontRibera2014,Delubac2015,Bautista2017,dMdB2017,dSA2019,Blomqvist2019,dMdB2020}.
The BAO measurements from SDSS DR11 \cite{FontRibera2014,Delubac2015} showed a mild $\sim 2.3 \sigma$ tension with the best-fit $\Lambda$CDM model of Planck. 
As discussed in \dMdB, this tension gradually disappeared with the addition of more data, and in the last measurement from eBOSS with SDSS DR16 the disagreement was only at the $1.5 \sigma$ level. 
In \cref{fig:eboss_desi_bao} we compare the BAO measurements from this work (solid blue) with those from \dMdB using the quasar catalog of SDSS DR16 (solid gray). 

The differences between the analyses have been discussed in the previous sections, but we summarise them here. 
The main difference is the input quasar catalog.
There are $\sim 150\,000$ quasars with $z>1.77$ in common in both datasets, representing $\sim 45\%$ and $\sim 25\%$ of the quasars in SDSS DR16 and DESI DR1 respectively. 
DESI targeted quasars down to $r<23$, compared to $g<22$ in BOSS and $g<22.5$ in eBOSS.
We expect most \lya quasars in DESI to receive four observations by the end of the survey, at which point the distribution of signal-to-noise per Angstrom in DESI and SDSS quasars will be similar, but in DESI DR1 the majority of quasars have only been observed once and are therefore on average noisier than in SDSS DR16. 

The spectrographs are also different: in the blue arm (relevant for \lya science) DESI has a resolving power R in the range 2000-3000, compared to 1500-2000 in BOSS/eBOSS.
While DESI pixels have a constant width of $0.8\angs$, SDSS pixels were constant in $\log \lambda$ with widths ranging from $0.83$ to $1.27\angs$ in the relevant range of wavelengths. The spectro-photometry is also greatly improved thanks to an atmospheric dispersion corrector, and the spectrograph optics are much more stable, allowing a finer 2D spectral extraction and improved sky subtraction \citep{Spectro.Pipeline.Guy.2023}. 
The higher quality of the DESI spectra allows us to better determine quasar redshifts, as captured by the $\sigma_v$ parameter in our fits that is almost a factor of two smaller than the one reported in \dMdB (see \cref{subsec:zerrors,app:nuisance}).

The \textit{continuum fitting} of quasars for both this work and \dMdB was performed using the \texttt{picca} software, but the code has been re-written and re-structured in a more modular way since the analysis of \dMdB.
There have been several minor changes in the methodology as well: we now use the CIII region to re-calibrate the spectra (instead of a region redder than the MgII emission line); we extended the \lya region to $1205\angs$ (instead of $1200\angs$); we include quasars with BAL features and mask them following the prescription by \cite{Ennesser2022} (instead of rejecting the entire line of sight); we use modified \lya weights, presented in \cite{2023MNRAS.tmp.3626R}. 
These changes were motivated by the studies of \cite{2023MNRAS.tmp.3626R} using early DESI data, and the (minor) impact of these changes are discussed in the variations of \cref{subsec:variations}.

The correlations in both analyses were measured with \texttt{picca} as well, and the only methodological change is the inclusion of the cross-covariance between correlations discussed in \cite{KP6s6-Cuceu}.
There are also some minor changes in the modelling, discussed in \cite{2023JCAP...11..045G,KP6s6-Cuceu}: 
the distortion matrix is now modelled at higher resolution and up to $300~\hMpc$ (instead of $200~\hMpc$); 
the contamination of metals is modelled in a slightly different way; 
the model describing the contamination of HCDs now has a free parameter $L_{\rm HCD}$ (fixed at $10~\hMpc$ in \dMdB);
the correlated sky residuals are now modelled with an improved method described in \cite{KP6s5-Guy} that only needs a single free parameter (versus four in \dMdB). 
While these changes are a clear improvement in the modelling of the correlations, in \cref{fig:var_all_1d} of \cref{subsec:variations} we show that their impact on the BAO results is negligible. 

Finally, an important difference between the analyses is the effort that we did to validate the analysis pipeline using only blinded measurements and synthetic datasets. 
As described in \cref{sec:validation}, we only \textit{unblinded} the measurements once we had passed a long list of robustness and consistency tests, including variations of the analysis and data splits.

\subsection{DESI - SDSS cross-survey covariance}

Even though there is a large overlap in redshift range and footprint between the \lya samples of SDSS DR16 and DESI DR1 (see \cref{fig:eboss_desi_data}), the contribution from cosmic variance to these measurements is small, and their cross-survey covariance is smaller than one might naively think.
We quantify the cross-covariance in \cref{app:eboss_covar}, where we also use it to combine both results into a ``DESI + SDSS" \lya BAO measurement:
\begin{align}
    \ap^{\rm DESI + SDSS} &= 1.012 \pm 0.016 ~, \nonumber \\  
    \at^{\rm DESI + SDSS} &= 0.990 \pm 0.019 ~,  
\end{align}
with a correlation coefficient of $\rho=-0.47$.
These contours are also plotted as red contours in \cref{fig:eboss_desi_bao}, and can be used in cosmological inference assuming a Gaussian posterior. Given our fiducial cosmology, this gives
the following constraints on ratios of distances:
\begin{align}
 D_H(\zeff) / r_d & = 8.72 \pm 0.14 ~, \nonumber \\
 D_M(\zeff) / r_d & = 38.80 \pm 0.76 ~.
\end{align}
It is important to note that we have not redone the SDSS analysis with our analysis pipeline.
We have taken the ($2 \times 2$) \lya BAO posterior from \dMdB, and combined it with our own posterior after taking into account an approximate cross-survey covariance that was computed as described in \cref{app:eboss_covar}.
Ignoring the cross-survey covariance would lead to an underestimate of the covariance of the combined result by 10\%.

One could do a joint analyses of both surveys, starting by co-adding the roughly $100\,000$ spectra of quasars at $z>2.1$ observed with both telescopes.
However, the modelling would be complicated, since there are several differences in the contaminants of both surveys. 
These differences include: correlated sky residuals extend to different transverse separations; DESI observes fainter quasars, and the effective quasar bias could be different; thanks to the higher spectral resolution of DESI, quasar redshift errors are smaller than in SDSS; the efficiency of the DLA finder might also be quite different, impacting the level of HCD contamination.

The combined ``DESI + SDSS" BAO measurement has an uncertainty 20\% smaller than the one from DESI DR1.
As the DESI survey observes more and more data, the SDSS dataset will gradually add less and less to the joint analysis.

\subsection{Future work}

There are several known issues that we have not addressed in this publication, and that we leave for future work. 

\begin{itemize}
  \item Non-linear evolution causes a small systematic shift on the BAO peak \cite{Crocce2008,Smith2008,Seo2008}. At low redshift, this is expected to be a sub-percent bias, and at high redshift it is expected to be even lower, below our current systematic uncertainties of 0.5\% coming from the analysis of mocks (it is a conservative estimate, see \cref{subsec:validation-conclusion}).
  
  \item Relative velocities between dark matter and baryons can bias the position of the BAO peak \cite{Tseliakhovich2010}.
  Studies using hydrodynamical simulations have shown that the bias should be small in the \lya forest auto-correlation \cite{Hirata2018,Givans2020}, but we do not have equivalent studies for the cross-correlation with quasars.
  
  \item The combination of quasar redshift errors and quasar continuum errors can lead to spurious correlations in 3D analyses of the \lya forest \cite{Youles2022}.
  In a companion paper \cite{KP6s6-Cuceu}, we have analysed synthetic data to show that the impact of this systematic on BAO measurements is limited to $0.1 \sigma$ of the current statistical uncertainties. 
  However, this could become a problem for the final analysis of DESI, or for other \lya studies that obtain cosmological information from the full shape of the correlations \cite{FontRibera2018,Cuceu2021,Cuceu2023a,Cuceu2023b,Gerardi2023}.
  
  \item In this work we used mocks (synthetic datasets) derived from log-normal density fields.
  We are working on the development of more complex mocks, including mocks based on N-body simulations \cite{Hadzhiyska2023} and mocks based on Lagrangian Perturbation Theory; 
  these mocks will include the non-linear broadening of the BAO peak discussed in \cref{fig:sigma_bao}.

  \item BAO analyses using galaxy samples at low redshift often use a \textit{reconstruction} algorithm to undo some of the non-linear effects and improve the accuracy and precision of the BAO measurements \cite{2007ApJ...664..675E,KP4s4-Paillas}.
  However, the reconstruction method is not recommended for surveys limited by shot noise (as opposed to cosmic variance), since the estimates of the displacement field would be noisy and could degrade the BAO precision. 
  While this will be an interesting possibility for future \lya surveys with higher quasar densities, we do not expect this method to improve the precision of the BAO measurement with the current DESI data set.
  
\end{itemize}

\section{Conclusions} 
\label{sec:conclusions}

We have measured the three-dimensional correlations in the \lya forest dataset from the first year of DESI data, as well as its cross-correlation with the position of DESI quasars.
Using these correlations we have measured the BAO scale parallel ($\ap$) and perpendicular ($\at$) to the line of sight with a precision of 2.0\% and 2.4\% respectively. 
The statistical uncertainties on the BAO parameters from just one year of DESI observations are already smaller than the ones from \dMdB, who used 10 years of BOSS and eBOSS observations. 

This analysis is the first \lya BAO measurement that was fully blinded\footnote{The BOSS DR9 analysis of \cite{Busca2013, Slosar2013, Kirkby2013} was  partially blinded, in the sense that during several months the BAO peak was masked when plotting the correlation function.}.
It is also the first time that the (small) cross-covariance between the auto-correlation of \lya fluctuations and its cross-correlation with quasars is taken into account (see \cref{subsec:covariance}).
In a companion paper \citep{KP6s6-Cuceu}, we validated the analysis with 150 mocks mimicking the DESI DR1 \lya sample.
We also characterized the correlated noise introduced by the data processing pipeline and the imprint of foreground absorbers on the \lya auto-correlation in \cite{KP6s5-Guy}.  
In parallel, we set a long list of robustness tests that we needed to pass before we could \textit{unblind} the measurements. 
These are discussed in \cref{subsec:data_splits,subsec:variations}.

The BAO measurement presented here can be translated into constraints on the following ratio of distances: 
$D_H(\zeff=2.33) / r_d = 8.52 \pm 0.17$ and $D_M(\zeff=2.33) / r_d = 39.71 \pm 0.95$, where $D_H=c/H(z)$, $D_M(z)$ is the transverse comoving distance and $r_d$ is the sound horizon at the drag epoch.

This publication is part of a series of publications presenting clustering measurements of the different tracers in DESI DR1 \citep{DESI2024.I.DR1}. Besides this \lya BAO measurement at $z=2.33$, the clustering of galaxies and quasars at $z<2$ is presented in \cite{DESI2024.II.KP3}, the BAO measurements derived from this data set are described in \cite{DESI2024.III.KP4}, and the cosmological constraints from the combined galaxy, quasar and \lya forests BAO can be found in \cite{DESI2024.VI.KP7A}. A complementary analysis of the two points clustering statistics of galaxies and quasars using a larger range of scales beyond the BAO scale will be presented in \cite{DESI2024.V.KP5} with their corresponding cosmological constraints in \cite{DESI2024.VII.KP7B}.

In the near future, we will also present other cosmological analyses using the \lya forest dataset from DESI DR1.
We will extract non-BAO information from the full shape of 3D correlations on large, linear scales (see \cite{Cuceu2021,Cuceu2023a,Cuceu2023b,Gerardi2023}), which should further improve the precision of these results. 
We will also constrain the linear matter power spectrum on small scales using the 1D power spectrum of fluctuations in the \lya forest (see \cite{Ravoux2023, Karacayli2024}). DESI has completed approximately three years of observations and has collected more than half of its planned dataset. Upon completion of the five year survey, we expect the precision of the BAO results to improve by a factor of two. 

\section*{Data Availability}

The data used in this analysis will be made public along the Data Release 1 (\url{https://data.desi.lbl.gov/doc/releases/}). The data points corresponding to the most relevant figures in this paper are available at \texttt{Zenodo}\footnote{\url{ https://doi.org/10.5281/zenodo.10799350}}
\acknowledgments
This material is based upon work supported by the U.S. Department of Energy (DOE), Office of Science, Office of High-Energy Physics, under Contract No. DE–AC02–05CH11231, and by the National Energy Research Scientific Computing Center, a DOE Office of Science User Facility under the same contract. Additional support for DESI was provided by the U.S. National Science Foundation (NSF), Division of Astronomical Sciences under Contract No. AST-0950945 to the NSF’s National Optical-Infrared Astronomy Research Laboratory; the Science and Technology Facilities Council of the United Kingdom; the Gordon and Betty Moore Foundation; the Heising-Simons Foundation; the French Alternative Energies and Atomic Energy Commission (CEA); the National Council of Humanities, Science and Technology of Mexico (CONAHCYT); the Ministry of Science and Innovation of Spain (MICINN), and by the DESI Member Institutions: \url{https://www.desi.lbl.gov/collaborating-institutions}. Any opinions, findings, and conclusions or recommendations expressed in this material are those of the author(s) and do not necessarily reflect the views of the U. S. National Science Foundation, the U. S. Department of Energy, or any of the listed funding agencies.

The authors are honored to be permitted to conduct scientific research on Iolkam Du’ag (Kitt Peak), a mountain with particular significance to the Tohono O’odham Nation.

\bibliographystyle{JHEP.bst}
\bibliography{main,DESI2024}

\appendix

\section{Alternative modelling of HCD contaminations} 
\label{app:voigt_hcd}

As discussed in \cref{subsec:hcds}, the presence of high column density
systems (HCDs) changes the flux correlation function by smearing
the $\delta$ field in the radial direction \cite{McQuinn2011,FontRibera2012a,Rogers2018}.
The contamination can be described by introducing a scale-dependent component to the bias that depends on $k_\parallel$ that is added to the normal IGM-induced \lya bias.
Following \cite{Bautista2017} we write this term as $b_{\rm HCD}F_{\rm HCD}(k_\parallel)$ where
\begin{equation} \label{eq:voigt_hcd}
    F_{\rm HCD}(k_\parallel,z)= A(z) ~ \int dN_{\rm HI} ~ f(N_{\rm HI},z) ~ V(N_{\rm HI},k_\parallel,z)
\end{equation}
where $V(N_{\rm HI},k_\parallel)$ is the Fourier transform of 
the Voigt profile for an HCD of column density $N_{\rm HI}$, 
and $f(N_{\rm HI})$ is the column-density distribution of HCDs.
The normalization factor $A(z)$ can be chosen so that $F_{\rm HCD}(k_\parallel=0)=1$, in which case $b_{\rm HCD}$ is proportional to the product of the HCD halo bias and the mean absorption caused by HCDs (see appendix B in \cite{McQuinn2011} and eq. 4.19 in \cite{FontRibera2012a}).

Given our lack of precise knowledge of the HCD distribution $f(N_{\rm HI})$, following \dMdB we model $F_{\rm HCD}(k_\parallel)=\exp(-L_{\rm HCD} k_\parallel)$ as an exponential with unknown scale parameter $L_{\rm HCD}$ that characterizes the typical size of unmasked HCDs. 

\begin{figure}
\centering
\includegraphics[width=0.75\textwidth]{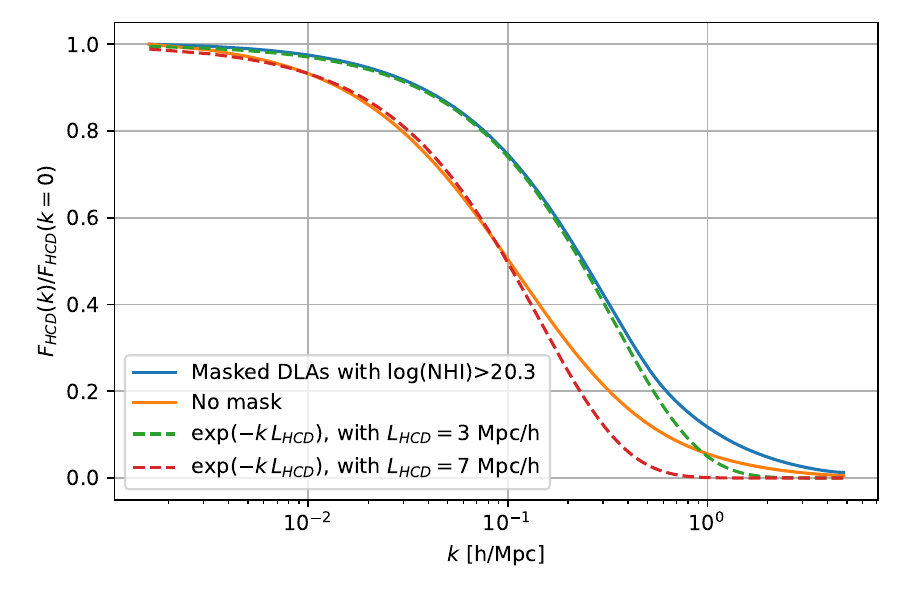}
  \caption{Comparison of the Voigt model for HCD contamination (solid curves) and the exponential model used in this work (dashed curves),
  for the HCD column-density distribution predicted by \cite{Prochaska2014}.
  The orange solid line includes HCDs of all column densities
  and the blue solid line includes only those with
  $\log(N_{\rm HI})<20.3$, which is appropriate for perfect masking of DLAs.
  The dashed lines show that exponential models can reproduce the Voigt model, except at large $k_\parallel$
  where the Voigt model has a longer tail.
  }
  \label{fig:voigt_hcd}
\end{figure}

We compare the functional forms of $F_{\rm HCD}(k_\parallel)$ in \cref{fig:voigt_hcd}.
The solid orange line shows the computation from \cref{eq:voigt_hcd} when we use the column density distribution $f(N_{\rm HI})$ from \cite{Prochaska2014}.
The solid blue line uses the same model, but it only integrates up to $\log(N_{\rm HI})=20.3$ to mimic the effect of perfectly masking all DLAs.
The dashed red and green lines show the exponential model for $L_{\rm HCD}=7$ and $L_{\rm HCD}=3~\hMpc$, respectively, and they capture fairly well the suppression of power from \cref{eq:voigt_hcd}. 

\section{Nuisance parameters}
\label{app:nuisance}

As discussed in \cref{sec:model}, our model to describe the measured correlations has 17 free parameters, including the two BAO parameters ($\at$, $\ap$) and 15 nuisance parameters that we marginalize over. 
The best-fit values and uncertainties for all 17 parameters are shown in \cref{tab:nuisances}, together with the priors used in the analysis. 

While most of the parameters are well constrained by the data and the priors are uninformative, we use informative Gaussian priors for two parameters.
These are the linear bias of CIV absorption, $b_{\rm CIV(eff)}$, and the RSD parameter of absorption caused by HCDs, $\beta_{\rm HCD}$.
When fitting the auto- and cross-correlation independently, we are not able to break the degeneracy between $L_{\rm HCD}$ and other nuisance parameters, and we fix this value to the best-fit value of the combined analysis ($L_{\rm HCD}=6.51~\hMpc$).
Moreover, when fitting the cross-correlation alone we also need to add an informative Gaussian prior on the quasar bias \footnote{Motivated by preliminary studies of quasar clustering in DESI DR1 \cite{ChaussidonY1fnl}.}
of $b_{\rm Q}=3.5 \pm 0.1$  to break the degeneracy between this parameter and the bias of the \lya forest ($b_{\alpha}$). 

\begin{table}
\centering
\renewcommand{\arraystretch}{1.2}
\begin{tabular}{c|c|ccc}
Parameter                                      & Priors                                         & \multicolumn{3}{c}{Best fit}\\
                                               &                                                & Combined                         & \lyaxlya                 & \lyaxqso                \\
\hline

$\alpha_{\parallel}$                           & $\mathcal{U}[0.01, 2.00]$                      & $0.989 \pm 0.020$                & $0.993^{+0.029}_{-0.032}$        & $0.988^{+0.024}_{-0.025}$        \\
$\alpha_{\perp}$                               & $\mathcal{U}[0.01, 2.00]$                      & $1.013 \pm 0.024$                & $1.020^{+0.036}_{-0.037}$        & $1.005 \pm 0.030$                \\
$b_{\alpha}$                                 & $\mathcal{U}[-2.00, 0.00]$                     & $-0.1078^{+0.0045}_{-0.0054}$    & $-0.1078 \pm 0.0036$             & $-0.099^{+0.015}_{-0.013}$       \\
$\beta_{\alpha}$                             & $\mathcal{U}[0.00, 5.00]$                      & $1.743^{+0.074}_{-0.100}$        & $1.745^{+0.076}_{-0.088}$        & $2.07^{+0.23}_{-0.35}$           \\
$10^3 b_{\rm SiII(1190)}$                      & $\mathcal{U}[-500.00, 0.00]$                   & $-4.50 \pm 0.64$                 & $-5.53^{+0.85}_{-0.82}$          & $-3.5 \pm 1.1$                   \\
$10^3 b_{\rm SiII(1193)}$                      & $\mathcal{U}[-500.00, 0.00]$                   & $-3.05^{+0.63}_{-0.62}$          & $-4.16^{+0.81}_{-0.79}$          & $-1.76^{+1.13}_{-0.81}$          \\
$10^3 b_{\rm SiII(1260)}$                      & $\mathcal{U}[-500.00, 0.00]$                   & $-4.02 \pm 0.62$                 & $-4.63 \pm 0.91$                 & $-4.00 \pm 0.82$                 \\
$10^3 b_{\rm SiIII(1207)}$                     & $\mathcal{U}[-500.00, 0.00]$                   & $-9.79 \pm 0.68$                 & $-10.80^{+0.74}_{-0.75}$         & $-8.7 \pm 1.1$                   \\
$10^3 b_{\rm CIV(eff)}$                        & $\mathcal{N}(-24.3, 1.5)$*                     & $-24.3 \pm 1.5$                  & $-24.5 \pm 1.6$                  &                                  \\
$b_{HCD}$                                      & $\mathcal{U}[-0.20, 0.00]$                     & $-0.0563^{+0.0045}_{-0.0036}$    & $-0.0582 \pm 0.0037$             & $-0.053^{+0.013}_{-0.014}$       \\
$\beta_{HCD}$                                  & $\mathcal{N}(0.500, 0.090)$                    & $0.625 \pm 0.080$                & $0.588^{+0.082}_{-0.081}$        & $0.528^{+0.088}_{-0.094}$        \\
$L_{\rm HCD} (h^{-1} {\rm Mpc})$          & $\mathcal{U}[0.00, 40.00]$                     & $6.51^{+0.82}_{-0.96}$           &                                  &                                  \\
$b_{\rm Q}$                                      & $\mathcal{U}[0.00, 10.00]$*                    & $3.408 \pm 0.048$                &                                  & $3.49 \pm 0.10$                  \\
$\Delta r_{\parallel} (h^{-1} {\rm Mpc})$ & $\mathcal{U}[-3.00, 3.00]$                     & $0.066 \pm 0.058$                &                                  & $0.077^{+0.061}_{-0.062}$        \\
$\sigma_z (h^{-1} {\rm Mpc})$                  & $\mathcal{U}[0.00, 15.00]$                     & $3.67 \pm 0.14$                  &                                  & $4.12^{+0.42}_{-0.43}$           \\
$\xi_0^{\rm TP}$                               & $\mathcal{U}[0.00, 2.00]$                      & $0.395 \pm 0.051$                &                                  & $0.320^{+0.082}_{-0.083}$        \\
$10^4 a_{\rm noise}$                           & $\mathcal{U}[0.00, 100.00]$                    & $3.54 \pm 0.16$                  & $3.57 \pm 0.17$                  &                                  \\
\hline
$N_{\rm bin}$                                  & --                                             & 9540                             & 3180                             & 6360                             \\
$N_{\rm param}$                                & --                                             & 17                               & 12                               & 14                               \\
$\chi^2_{\rm min}$                             & --                                             & 9624.36                          & 3183.79                          & 6427.41                          \\
p-value                                        & --                                             & 0.23                             & 0.42                             & 0.23                             \\
\end{tabular}
\caption{Priors, best-fit values (mean of the posterior) and uncertainties (68\% credible intervals) for the 17 free parameters in the fits. 
When analysing the auto-correlation or the cross-correlation alone, we fix $L_{\rm HCD}$ to the best-fit value of the combination ($L_{\rm HCD}=6.51~\hMpc$).
This is necessary to break internal degeneracies, but it makes the p-value of these analyses difficult to interpret.
When analysing the cross-correlation alone we also use an extra prior on the quasar bias parameter ($b_{\rm Q}=3.5 \pm 0.1$) to break the degeneracy with the \lya biases.
Some parameters are not needed when fitting the auto-correlation or the cross-correlation alone.}
\label{tab:nuisances}
\end{table}

\begin{figure}
\includegraphics[width=0.9\textwidth]{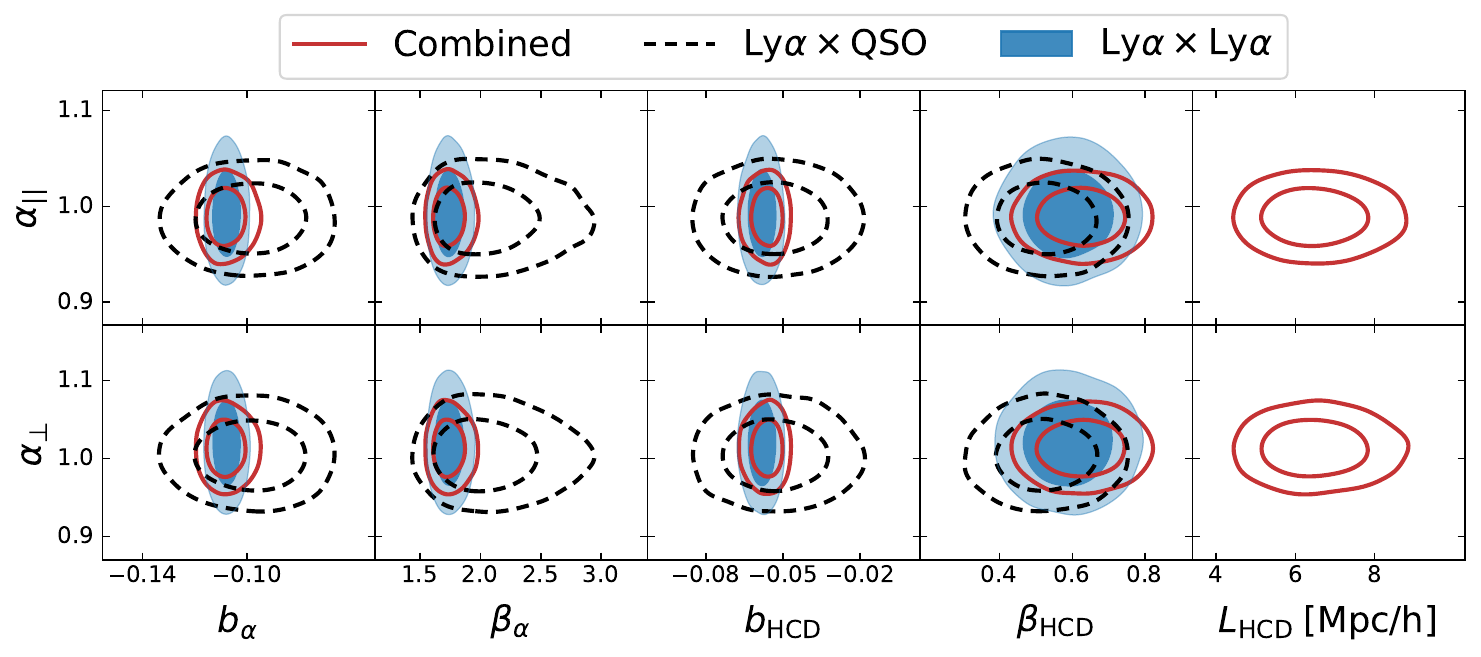}
\includegraphics[width=0.9\textwidth]{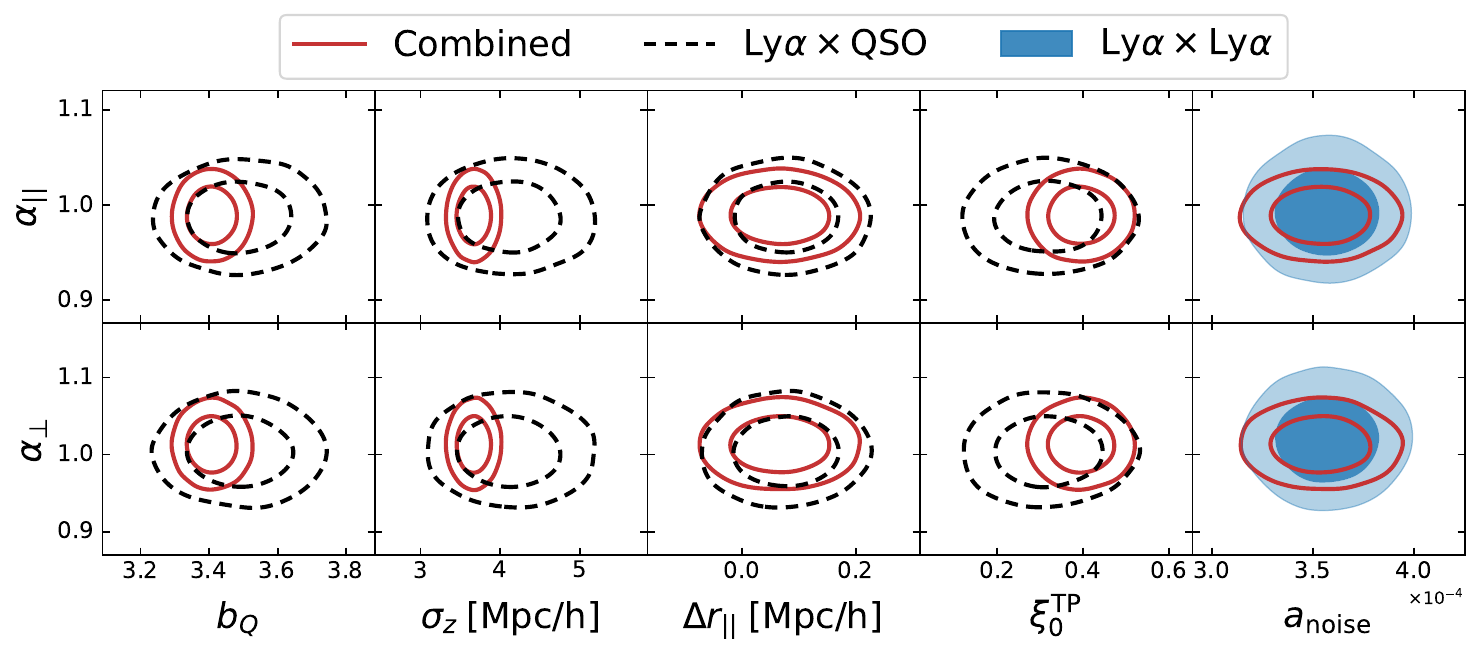}
\includegraphics[width=0.9\textwidth]{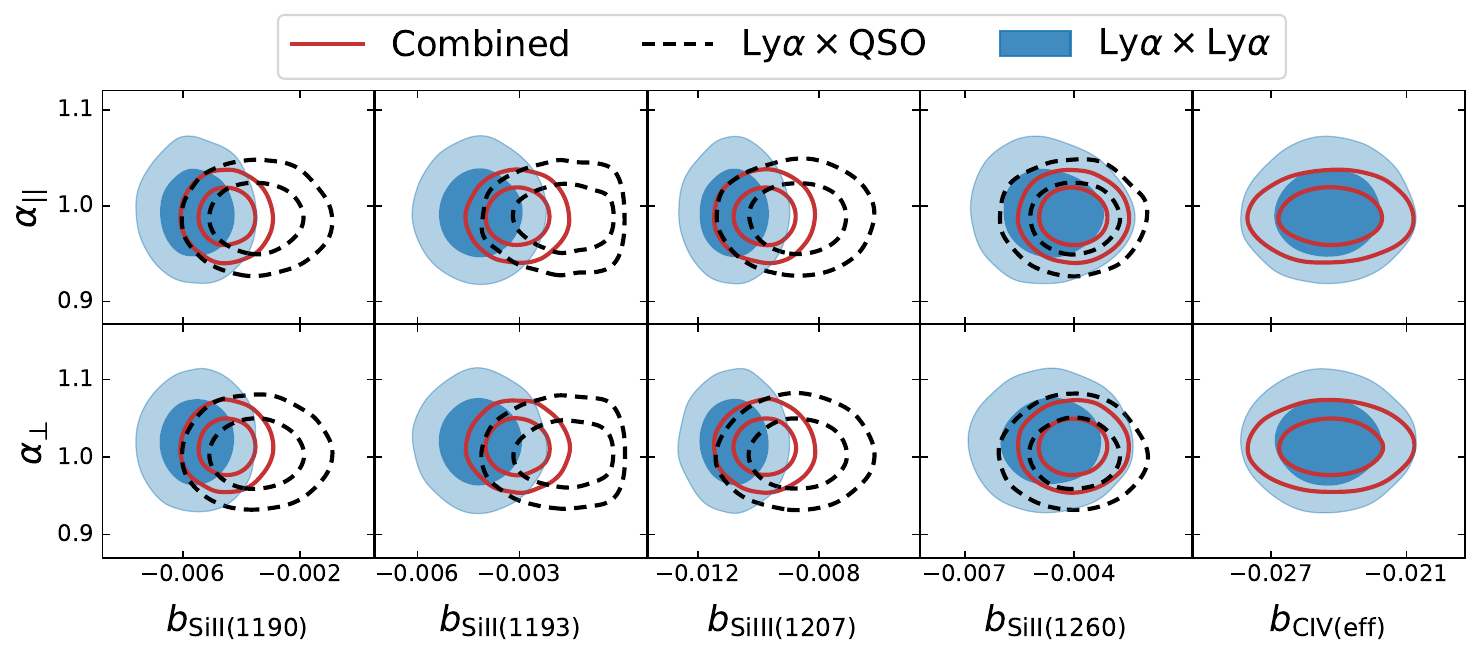}
  \caption{Correlation between the BAO parameters ($\at$, $\ap$) and the 15 nuisance parameters, for the combined analysis (solid red), the cross-correlation alone (dashed black) and the auto-correlation alone (filled blue).
  Not all nuisance parameters are varied when fitting the auto- and cross-correlations separately.
  The BAO parameters are not strongly correlated with any of the nuisance parameters. 
  }
  \label{fig:bao_corr}
\end{figure}

Comparison of the different columns in \cref{tab:nuisances} shows that the best-fit values from the auto-correlations alone (including regions A and B) are in agreement with those from the fit of the cross-correlations with quasars (also including both regions).
However, we would like to add a word of caution when interpreting these nuisance parameters. 
While we have extensively tested the robustness of the BAO results under different data splits and analysis settings, some of the nuisance parameters do vary significantly with reasonable changes in the analysis choices. 
For instance, depending on how aggressively we mask DLAs, we obtain different values for the parameters that model the contamination by HCDs ($b_{\rm HCD}$, $\beta_{\rm HCD}$ and $L_{\rm HCD}$), but the differences also propagate to other parameters that are degenerate with these, including $b_{\alpha}$ and $\beta_{\alpha}$.

Finally, in \cref{fig:bao_corr} we show that none of the 15 nuisance parameters is correlated with any of the BAO parameters ($\at$, $\ap$).

\section{Blinding}
\label{app:blinding}

Our analysis validation was performed entirely on blinded data, with clearly defined requirements that needed to be achieved before unblinding (\Cref{sec:validation}). The main goal of our blinding strategy was to blind the BAO measurement in a way that does not impede the analysis process. We considered multiple blinding methods, including blinding at the catalog level  (as done for the DESI galaxy BAO analyses, see \cite{DESI2024.III.KP4}). 
However, the disadvantage of catalog-level blinding is that metal contamination results in very well-measured peaks along the line-of-sight due to Ly$\alpha\times$Metal correlations (see \cref{subsec:metals}). 
Any catalogue-level blinding would have also shifted the positions of these peaks, giving away the direction and magnitude of the blinding. We therefore instead developed a blinding method that only shifts the position of the BAO peak at the level of the measured correlation functions.

Our blinding strategy starts with a correlation function model $\xi_t$ with all nuisance parameters set to their best-fit values from \cite{dMdB2020}. 
We then used \texttt{Vega} to compute the blinding template:
\begin{equation}
    \xi_b = \xi_t(\alpha_{||}=1+\Delta\alpha_{||},\;\alpha_\bot=1+\Delta\alpha_\bot) - \xi_t(\alpha_{||}=1,\;\alpha_\bot=1),
\end{equation}
where $\Delta\alpha_{||}$ and $\Delta\alpha_\bot$ were randomly drawn from Gaussian distributions with mean zero and variance equal to two times the uncertainties on $\alpha_{||}$ and $\alpha_\bot$ from \cite{dMdB2020}. 
The random values used in the DR1 BAO analysis were 
$\Delta \alpha_{||}=-0.011$ and $\Delta\alpha_\bot=-0.084$, i.e., a fairly small shift along the line of sight and a very large blinding shift in the transverse BAO measurement, close to the maximum allowed by the blinding strategy.
These values of $\Delta\alpha_{||}$ and $\Delta\alpha_\bot$ were unknown to us and never stored anywhere. 
We did save the template\footnote{The format used was carefully chosen to eliminate the possibility of the template being viewed accidentally.} and our pipeline (\texttt{picca}) automatically added this $\xi_b$ template to the correlation function at the moment of writing it to file, effectively blinding the BAO scale.

\begin{figure}
\includegraphics[width=0.8\textwidth]{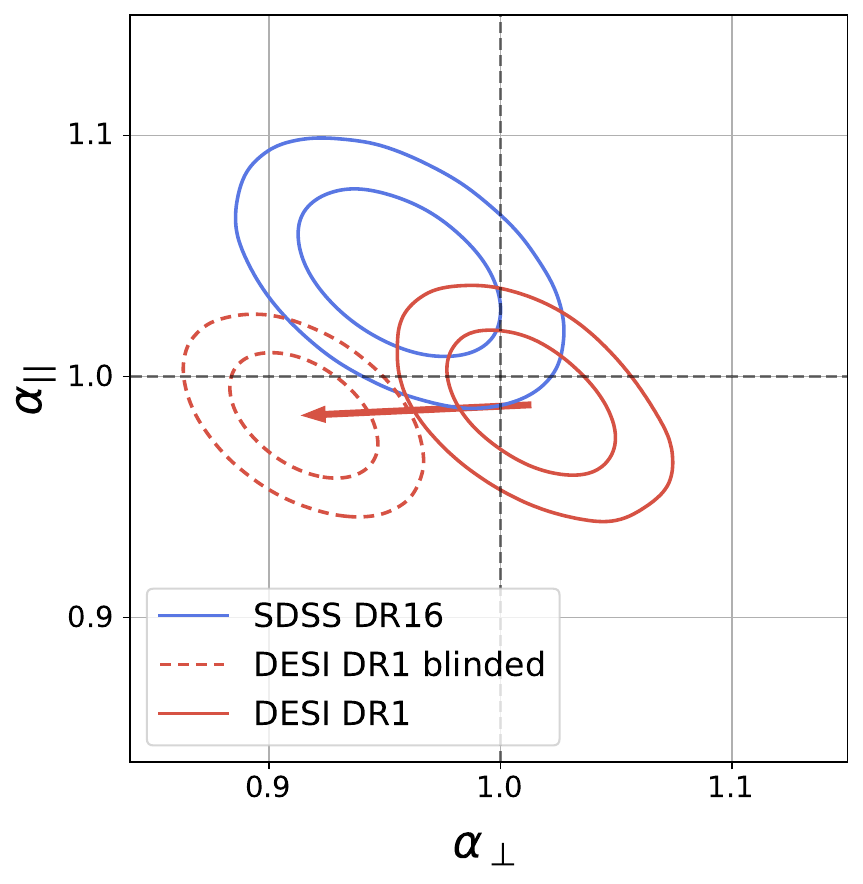}
  \caption{BAO parameters along ($\ap$) and across the line of sight ($\at$) from DESI DR1 (solid red) and from its blinded measurement (dashed red).
  The blinded measurement was in tension with Planck (dashed gray) and with the results from \dMdB using SDSS DR16 (solid blue).}
  \label{fig:blinding}
\end{figure}

The random shift applied to the blinded measurements can be seen in \cref{fig:blinding}, where we show the DESI \lya BAO results before (dotted red) and after (solid red) unblinding.
The difference in the best fit BAO parameters before and after unblinding ($\Delta \alpha_{||}=-0.005$, $\Delta\alpha_\bot=-0.098$, red arrow in the figure) is consistent with the random shifts applied in the blinding ($\Delta \alpha_{||}=-0.011$, $\Delta\alpha_\bot=-0.084$).

We unblinded our analysis on December 8th, 2023, after passing the extensive validation tests described in \Cref{sec:validation}. We only made two minor changes in the methodology after unblinding.
First, we fixed a small bug in \texttt{picca} related to the masking of BAL features in the \lya region B, with an impact on the best-fit BAO parameters smaller than 0.1\%.
Second, we obtained a more accurate measurement of the \ion{C}{IV} bias parameter from \cite{KP6s5-Guy}, with no impact on the BAO results.

\section{Comparison of sampling methods} 
\label{app:sampling}

\begin{figure}
\includegraphics[width=0.8\textwidth]{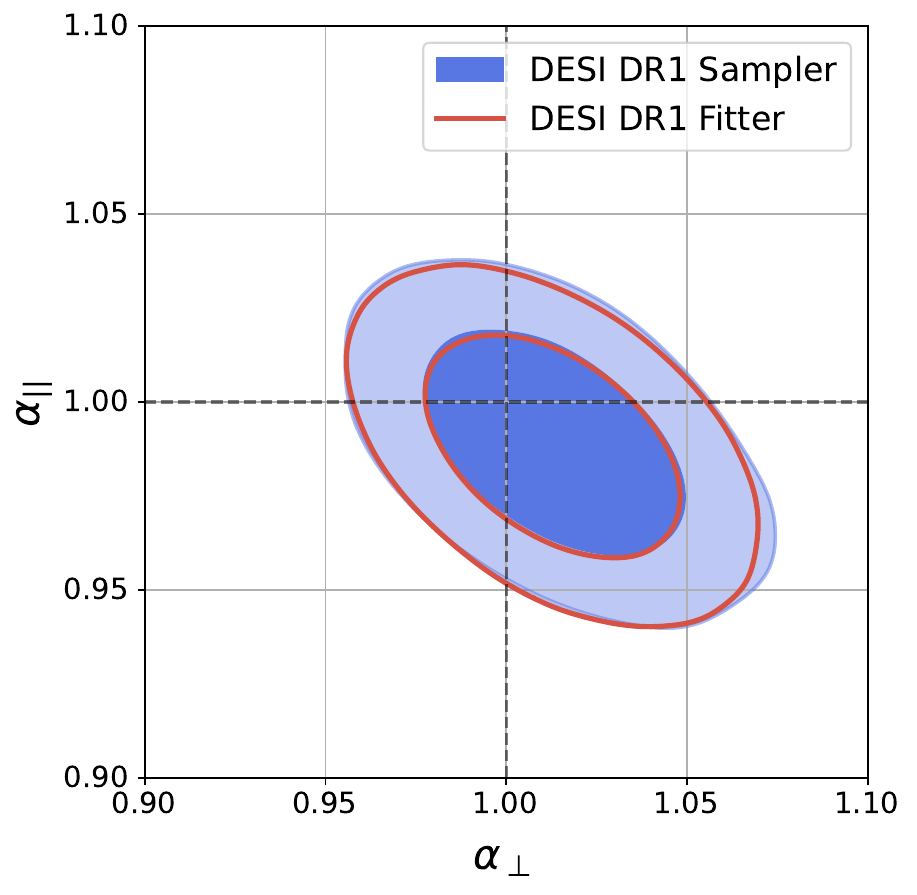}
  \caption{$68\%$ and $95\%$ credible regions of BAO parameters along the line-of-sight ($\alpha_\parallel$) and across the line-of-sight ($\alpha_\perp$). The filled blue contours are based on the full-posterior distribution computed by \texttt{Polychord} and the solid red contours are based on the approximate Gaussian fit computed by \texttt{iminuit}.
  }
  \label{fig:sampling}
\end{figure}

The main results in this publication (including \cref{fig:bao}) were obtained using the Nested Sampler \texttt{Polychord} \cite{Handley:2015a,Handley:2015b}. The quoted parameter values are given by the mean of the posterior distribution, and the reported uncertainties are the 68\% credible regions. However, computing the full posterior distribution is computationally intensive, and in most of the tests in \Cref{sec:validation} we instead use a faster approximate method. This involves the use of the \texttt{iminuit} package \cite{iminuit} to find the maximum likelihood (minimum $\chi^2$) point in parameter space. Approximate Gaussian uncertainties are then computed by taking the second derivative with respect to parameters around the best-fit point \cite{James:1975dr}. In \cref{fig:sampling} we show that both methods lead to very similar BAO contours. Therefore, the faster approximate method is good enough to check for shifts in the BAO position as done in \Cref{sec:validation}.

In contrast, previous Ly$\alpha$ forest BAO analyses used a frequentist approach to obtain their main results \cite[see e.g.][]{Bautista2017,dMdB2017,dMdB2020}. This involved using the Profile Likelihood method to create a two-dimensional $\chi^2$ grid of $\alpha_{||}$ and $\alpha_\bot$, and then calibrating the size of contours based on $\Delta\chi^2$ values obtained from large sets of Monte Carlo simulations of the correlation functions. \cite{Cuceu2020} found that BAO measurements obtained with this method agree well with Bayesian results based on the full posterior distribution. Therefore, in this work, we rely on the simpler and often faster method of sampling the full posterior for our main results.

\section{Significance of BAO shifts} 
\label{app:bao_shifts}

In \cref{subsec:variations} we present an exhaustive list of robustness tests, where we look at the impact on the BAO parameters when changing different settings in the analysis.
In most of these variations, the dataset is exactly the same, and therefore we expect shifts in the BAO parameters to be caused exclusively by the analysis settings.
Before \textit{unblinding} the measurements we required that none of these variations caused a shift in the BAO parameters larger than a third of the statistical uncertainty from the results on (blinded) data.

In the second set of alternative analyses (in red) in \cref{fig:var_all_1d}, however, we also showed variations that caused minor changes in the dataset, by adding / removing quasars from the sample or by adding / removing \lya pixels from the data vector.
In these variations we relaxed the requirement, since changes in the dataset will cause statistical fluctuations in the measurement of the BAO parameters.
This explains why the shifts shown in red in \cref{fig:var_all_1d} are somewhat larger than those in the other variations.

Two of the larger shifts correspond to the variations ``only quasar targets" and ``weak BALs" (see \cref{subsec:variations} for details on the variations).
These variations cause the datasets to be 11\% and 8\% less constraining than the main analysis (based on the increase in the errorbars on the measured correlations).
Following \cite{Gratton2020}, we estimate the statistical fluctuations for these variations to be on average $0.28$ and $0.33~\sigma$ respectively.
We conclude that the shifts on BAO parameters in these variations are therefore consistent with statistical fluctuations.

There is another variation that has caused a shift larger than the requirement, and this is the ``mask-Lya redshift estimates" ($\Delta \ap=0.001$, $\Delta \at=-0.013$).
In this variation, we have used an alternative redshift estimator that masks wavelengths bluer than the \lya emission line of the quasar.
The rms of the differences in redshifts above $z>2$ is 443 km/s
\footnote{The distribution is not quite Gaussian, and 5.5\% of the redshifts have changed by more than 1 000 km/s.}
, causing differences in pixel-quasar pairs of order $4.2~\hMpc$ (using the fiducial cosmology to compute $H(\zeff=2.33)$) and causing statistical fluctuations in the measurement of the cross-correlation (see appendix B of \dMdB).
Besides the expected effect in the cross-correlation, differences in the redshift estimates have a more subtle effect in the auto-correlation as well. 
The rest-frame wavelength of a given pixel changes with the quasar redshift, and we find that on average 2\% of the pixels of a given \lya forest are moved in or out of the restframe wavelength range used in the analysis, causing further statistical fluctuations.

Finally, we want to assess whether the shift caused by the use of a different redshift estimates is a systematic shift or if it can be explained by statistical fluctuations.
For this purpose, we perform a bootstrap analysis based on a random sampling with replacement of the \texttt{HEALpix} pixels used to compute the average correlation functions and we generate 1000 bootstrap samples for each of the two analyses.
We then fit the BAO parameters for each of these samples and look at the distribution of shifts between the two analyses.
We estimate with this technique that the statistical uncertainties on the shifts of $\alpha_{\parallel}$ and $\alpha_{\perp}$ are of 0.007 and 0.009, respectively. We conclude from this study that the observed shifts of $\Delta \ap=0.001$ and $\Delta \at=-0.013$ are consistent with statistical fluctuations.


\section{Estimation of the DESI - SDSS covariance}
\label{app:eboss_covar}

\begin{figure}
\includegraphics[width=0.99\textwidth]{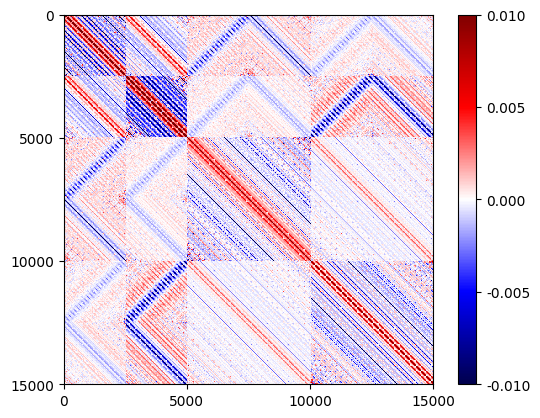}
  \caption{Correlation matrix corresponding to the cross-covariance of SDSS DR16 and DESI DR1 correlations.
  The first block of $2\,500 \times 2\,500$ in the top left corresponds to the correlation matrix of the \lyaxlyaA measurement of SDSS, while the second block of the same size corresponds to the same measurement in DESI.
  The third, larger block of $5\,000 \times 5\,000$ corresponds to the \lyaxqsoA cross-correlation in SDSS, and the last block (bottom right) has the same measurement in DESI.
  }
  \label{fig:cov_eboss_desi}
\end{figure}

The cross-covariance of the four correlation functions measured with the DESI DR1 \lya forest dataset is shown in \cref{fig:global_cov}.
As described in \cref{subsec:covariance}, we compute a noisy estimate of this $15\,000 \times 15\,000$ covariance from the scatter of the measurements obtained in different \texttt{HEALPix} pixels.
Following \cite{dMdB2020,2023JCAP...11..045G}, we smooth the correlation matrix so that it is invertible and we can use it to define a likelihood function.

Here we use the same method to compute the cross-covariance between the DESI DR1 correlations functions and those measured in \dMdB using the SDSS DR16 \lya dataset.
A matrix describing the covariance of all eight correlations would be $30\,000 \times 30\,000$.
However, given that the measurements using the B region carry a small amount of information (see the bottom right panel in \cref{fig:data_splits}) we ignore them in this appendix.
In \cref{fig:cov_eboss_desi} we show the correlation matrix of SDSS DR16 and DESI DR1 measurements of \lyaxlyaA and \lyaxqsoA.

Using this cross-covariance and the best-fit model from the combined analysis from \cref{tab:nuisances}, we generated $1\,000$ Monte Carlo realisations of the two main correlation functions (\lyaxlyaA and \lyaxqsoA) of SDSS and of DESI, along with the correct cross-covariance between the surveys.
We then minimised the likelihood for the SDSS and DESI correlation functions separately, and looked at the correlation between the best-fit BAO parameters ($\at$, $\ap$) to obtain the correlation matrix of the four BAO parameters:

\begin{gather}
 \begin{pmatrix}
  1 & \rho(\at^{\rm S}, \ap^{\rm S})
    & \rho(\at^{\rm S}, \at^{\rm D})
    & \rho(\at^{\rm S}, \ap^{\rm D}) \\
  \rho(\ap^{\rm S}, \at^{\rm S}) & 1
    & \rho(\ap^{\rm S}, \at^{\rm D})
    & \rho(\ap^{\rm S}, \ap^{\rm D}) \\
  \rho(\at^{\rm D}, \at^{\rm S})
    & \rho(\at^{\rm D}, \ap^{\rm S}) & 1
    & \rho(\at^{\rm D}, \ap^{\rm D}) \\
  \rho(\ap^{\rm D}, \at^{\rm S})
    & \rho(\ap^{\rm D}, \ap^{\rm S})
    & \rho(\ap^{\rm D}, \at^{\rm D})
    &  1
 \end{pmatrix}
 =
 \begin{pmatrix}
  1 & -0.53 & 0.09 & -0.05 \\
  -0.53 & 1 & -0.05 & 0.10 \\
  0.09 & -0.05 & 1 & -0.50 \\
  -0.05 & 0.10 & -0.50 & 1
 \end{pmatrix} ~,
\end{gather}
where the superscript $^{\rm S}$ ($^{\rm D}$) refers to SDSS (DESI) measurements of BAO.

The correlation of the BAO parameters from eBOSS and DESI can be computed separately from their posteriors, as discussed in the main text.
For this reason, we modify the matrix above and instead use $\rho(\at^{\rm S}, \ap^{\rm S})=-0.45$ and $\rho(\at^{\rm D}, \ap^{\rm D})=-0.48$. 
We use this modified correlation matrix, and the variances of each measurement  reported by each survey, to build a $4\times4$ multi-survey covariance $C$ for the multi-survey data vector $d = ( \at^{\rm S}, \ap^{\rm S}, \at^{\rm D}, \ap^{\rm D})$.

We use these results to compute a combined BAO measurement $d^{\rm DS} = (\at^{\rm DS}, \ap^{\rm DS})$ with covariance $C_{\rm DS}$ using linear algebra:
\begin{equation}
    C_{\rm DS}^{-1} = S^T C^{-1} S \qquad \rm{and} \qquad 
    C_{\rm DS}^{-1} d^{\rm DS} = S^T C^{-1} d ~,
\end{equation}
where we have defined the matrix $S$ as:

\begin{gather}
S =
 \begin{pmatrix}
  1 & 0 \\
  0 & 1 \\
  1 & 0 \\
  0 & 1
 \end{pmatrix} ~.
\end{gather}

Using these equations we obtain the following combined BAO measurements:
\begin{align}
    \at^{\rm DESI + SDSS} &= 0.990 \pm 0.019 ~, \nonumber \\
    \ap^{\rm DESI + SDSS} &= 1.012 \pm 0.016 ~,
\end{align}
with correlation coefficient of $\rho=-0.47$.


\section{Author Affiliations}
\label{sec:affiliations}

\noindent \hangindent=.5cm $^{1}${Instituto de F\'{\i}sica Te\'{o}rica (IFT) UAM/CSIC, Universidad Aut\'{o}noma de Madrid, Cantoblanco, E-28049, Madrid, Spain}

\noindent \hangindent=.5cm $^{2}${Lawrence Berkeley National Laboratory, 1 Cyclotron Road, Berkeley, CA 94720, USA}

\noindent \hangindent=.5cm $^{3}${Physics Dept., Boston University, 590 Commonwealth Avenue, Boston, MA 02215, USA}

\noindent \hangindent=.5cm $^{4}${Tata Institute of Fundamental Research, Homi Bhabha Road, Mumbai 400005, India}

\noindent \hangindent=.5cm $^{5}${Centre for Extragalactic Astronomy, Department of Physics, Durham University, South Road, Durham, DH1 3LE, UK}

\noindent \hangindent=.5cm $^{6}${Institute for Computational Cosmology, Department of Physics, Durham University, South Road, Durham DH1 3LE, UK}

\noindent \hangindent=.5cm $^{7}${Department of Physics, University of Michigan, Ann Arbor, MI 48109, USA}

\noindent \hangindent=.5cm $^{8}${Leinweber Center for Theoretical Physics, University of Michigan, 450 Church Street, Ann Arbor, Michigan 48109-1040, USA}

\noindent \hangindent=.5cm $^{9}${IRFU, CEA, Universit\'{e} Paris-Saclay, F-91191 Gif-sur-Yvette, France}

\noindent \hangindent=.5cm $^{10}${Institut de F\'{i}sica d’Altes Energies (IFAE), The Barcelona Institute of Science and Technology, Campus UAB, 08193 Bellaterra Barcelona, Spain}

\noindent \hangindent=.5cm $^{11}${Instituto Avanzado de Cosmolog\'{\i}a A.~C., San Marcos 11 - Atenas 202. Magdalena Contreras, 10720. Ciudad de M\'{e}xico, M\'{e}xico}

\noindent \hangindent=.5cm $^{12}${Instituto de Ciencias F\'{\i}sicas, Universidad Aut\'onoma de M\'exico, Cuernavaca, Morelos, 62210, (M\'exico)}

\noindent \hangindent=.5cm $^{13}${Physics Department, Yale University, P.O. Box 208120, New Haven, CT 06511, USA}

\noindent \hangindent=.5cm $^{14}${Department of Physics and Astronomy, University of California, Irvine, 92697, USA}

\noindent \hangindent=.5cm $^{15}${Aix Marseille Univ, CNRS/IN2P3, CPPM, Marseille, France}

\noindent \hangindent=.5cm $^{16}${Department of Physics, Kansas State University, 116 Cardwell Hall, Manhattan, KS 66506, USA}

\noindent \hangindent=.5cm $^{17}${Department of Physics \& Astronomy, University of Rochester, 206 Bausch and Lomb Hall, P.O. Box 270171, Rochester, NY 14627-0171, USA}

\noindent \hangindent=.5cm $^{18}${Institute for Astronomy, University of Edinburgh, Royal Observatory, Blackford Hill, Edinburgh EH9 3HJ, UK}

\noindent \hangindent=.5cm $^{19}${Dipartimento di Fisica ``Aldo Pontremoli'', Universit\`a degli Studi di Milano, Via Celoria 16, I-20133 Milano, Italy}

\noindent \hangindent=.5cm $^{20}${Centre for Astrophysics \& Supercomputing, Swinburne University of Technology, P.O. Box 218, Hawthorn, VIC 3122, Australia}

\noindent \hangindent=.5cm $^{21}${NSF NOIRLab, 950 N. Cherry Ave., Tucson, AZ 85719, USA}

\noindent \hangindent=.5cm $^{22}${Department of Physics and Astronomy, The University of Utah, 115 South 1400 East, Salt Lake City, UT 84112, USA}

\noindent \hangindent=.5cm $^{23}${Department of Physics \& Astronomy, University College London, Gower Street, London, WC1E 6BT, UK}

\noindent \hangindent=.5cm $^{24}${Department of Astronomy and Astrophysics, University of Chicago, 5640 South Ellis Avenue, Chicago, IL 60637, USA}

\noindent \hangindent=.5cm $^{25}${Fermi National Accelerator Laboratory, PO Box 500, Batavia, IL 60510, USA}

\noindent \hangindent=.5cm $^{26}${Korea Astronomy and Space Science Institute, 776, Daedeokdae-ro, Yuseong-gu, Daejeon 34055, Republic of Korea}

\noindent \hangindent=.5cm $^{27}${Institute of Cosmology and Gravitation, University of Portsmouth, Dennis Sciama Building, Portsmouth, PO1 3FX, UK}

\noindent \hangindent=.5cm $^{28}${Departamento de Astrof\'{\i}sica, Universidad de La Laguna (ULL), E-38206, La Laguna, Tenerife, Spain}

\noindent \hangindent=.5cm $^{29}${Instituto de Astrof\'{\i}sica de Canarias, C/ V\'{\i}a L\'{a}ctea, s/n, E-38205 La Laguna, Tenerife, Spain}

\noindent \hangindent=.5cm $^{30}${Department of Physics and Astronomy, University of Sussex, Brighton BN1 9QH, U.K}

\noindent \hangindent=.5cm $^{31}${Departamento de F\'{i}sica, Instituto Nacional de Investigaciones Nucleares, Carreterra M\'{e}xico-Toluca S/N, La Marquesa,  Ocoyoacac, Edo. de M\'{e}xico C.P. 52750,  M\'{e}xico}

\noindent \hangindent=.5cm $^{32}${Institute for Advanced Study, 1 Einstein Drive, Princeton, NJ 08540, USA}

\noindent \hangindent=.5cm $^{33}${Center for Cosmology and AstroParticle Physics, The Ohio State University, 191 West Woodruff Avenue, Columbus, OH 43210, USA}

\noindent \hangindent=.5cm $^{34}${NASA Einstein Fellow}

\noindent \hangindent=.5cm $^{35}${School of Mathematics and Physics, University of Queensland, 4072, Australia}

\noindent \hangindent=.5cm $^{36}${Departamento de F\'{i}sica, Universidad de Guanajuato - DCI, C.P. 37150, Leon, Guanajuato, M\'{e}xico}

\noindent \hangindent=.5cm $^{37}${Instituto de F\'{\i}sica, Universidad Nacional Aut\'{o}noma de M\'{e}xico,  Cd. de M\'{e}xico  C.P. 04510,  M\'{e}xico}

\noindent \hangindent=.5cm $^{38}${CIEMAT, Avenida Complutense 40, E-28040 Madrid, Spain}

\noindent \hangindent=.5cm $^{39}${Department of Physics \& Astronomy and Pittsburgh Particle Physics, Astrophysics, and Cosmology Center (PITT PACC), University of Pittsburgh, 3941 O'Hara Street, Pittsburgh, PA 15260, USA}

\noindent \hangindent=.5cm $^{40}${Department of Astronomy and Astrophysics, UCO/Lick Observatory, University of California, 1156 High Street, Santa Cruz, CA 95064, USA}

\noindent \hangindent=.5cm $^{41}${Department of Astronomy, School of Physics and Astronomy, Shanghai Jiao Tong University, Shanghai 200240, China}

\noindent \hangindent=.5cm $^{42}${Space Sciences Laboratory, University of California, Berkeley, 7 Gauss Way, Berkeley, CA  94720, USA}

\noindent \hangindent=.5cm $^{43}${University of California, Berkeley, 110 Sproul Hall \#5800 Berkeley, CA 94720, USA}

\noindent \hangindent=.5cm $^{44}${Universities Space Research Association, NASA Ames Research Centre, Moffett Blvd, Mountain View, CA 94035, U.S.A.}

\noindent \hangindent=.5cm $^{45}${Center for Astrophysics $|$ Harvard \& Smithsonian, 60 Garden Street, Cambridge, MA 02138, USA}

\noindent \hangindent=.5cm $^{46}${Department of Physics, The Ohio State University, 191 West Woodruff Avenue, Columbus, OH 43210, USA}

\noindent \hangindent=.5cm $^{47}${The Ohio State University, Columbus, 43210 OH, USA}

\noindent \hangindent=.5cm $^{48}${Kavli Institute for Particle Astrophysics and Cosmology, Stanford University, Menlo Park, CA 94305, USA}

\noindent \hangindent=.5cm $^{49}${SLAC National Accelerator Laboratory, Menlo Park, CA 94305, USA}

\noindent \hangindent=.5cm $^{50}${Instituto de Astrof\'{i}sica de Andaluc\'{i}a (CSIC), Glorieta de la Astronom\'{i}a, s/n, E-18008 Granada, Spain}

\noindent \hangindent=.5cm $^{51}${Ecole Polytechnique F\'{e}d\'{e}rale de Lausanne, CH-1015 Lausanne, Switzerland}

\noindent \hangindent=.5cm $^{52}${Departamento de F\'isica, Universidad de los Andes, Cra. 1 No. 18A-10, Edificio Ip, CP 111711, Bogot\'a, Colombia}

\noindent \hangindent=.5cm $^{53}${Observatorio Astron\'omico, Universidad de los Andes, Cra. 1 No. 18A-10, Edificio H, CP 111711 Bogot\'a, Colombia}

\noindent \hangindent=.5cm $^{54}${Department of Physics, The University of Texas at Dallas, Richardson, TX 75080, USA}

\noindent \hangindent=.5cm $^{55}${Institut d'Estudis Espacials de Catalunya (IEEC), 08034 Barcelona, Spain}

\noindent \hangindent=.5cm $^{56}${Institute of Space Sciences, ICE-CSIC, Campus UAB, Carrer de Can Magrans s/n, 08913 Bellaterra, Barcelona, Spain}

\noindent \hangindent=.5cm $^{57}${Departament de F\'{\i}sica Qu\`{a}ntica i Astrof\'{\i}sica, Universitat de Barcelona, Mart\'{\i} i Franqu\`{e}s 1, E08028 Barcelona, Spain}

\noindent \hangindent=.5cm $^{58}${Institut de Ci\`encies del Cosmos (ICCUB), Universitat de Barcelona (UB), c. Mart\'i i Franqu\`es, 1, 08028 Barcelona, Spain.}

\noindent \hangindent=.5cm $^{59}${Consejo Nacional de Ciencia y Tecnolog\'{\i}a, Av. Insurgentes Sur 1582. Colonia Cr\'{e}dito Constructor, Del. Benito Ju\'{a}rez C.P. 03940, M\'{e}xico D.F. M\'{e}xico}

\noindent \hangindent=.5cm $^{60}${Centro de Investigaci\'{o}n Avanzada en F\'{\i}sica Fundamental (CIAFF), Facultad de Ciencias, Universidad Aut\'{o}noma de Madrid, ES-28049 Madrid, Spain}

\noindent \hangindent=.5cm $^{61}${Excellence Cluster ORIGINS, Boltzmannstrasse 2, D-85748 Garching, Germany}

\noindent \hangindent=.5cm $^{62}${University Observatory, Faculty of Physics, Ludwig-Maximilians-Universit\"{a}t, Scheinerstr. 1, 81677 M\"{u}nchen, Germany}

\noindent \hangindent=.5cm $^{63}${Department of Astrophysical Sciences, Princeton University, Princeton NJ 08544, USA}

\noindent \hangindent=.5cm $^{64}${Kavli Institute for Cosmology, University of Cambridge, Madingley Road, Cambridge CB3 0HA, UK}

\noindent \hangindent=.5cm $^{65}${Department of Astronomy, The Ohio State University, 4055 McPherson Laboratory, 140 W 18th Avenue, Columbus, OH 43210, USA}

\noindent \hangindent=.5cm $^{66}${Department of Physics, Southern Methodist University, 3215 Daniel Avenue, Dallas, TX 75275, USA}

\noindent \hangindent=.5cm $^{67}${Department of Physics and Astronomy, University of Waterloo, 200 University Ave W, Waterloo, ON N2L 3G1, Canada}

\noindent \hangindent=.5cm $^{68}${Perimeter Institute for Theoretical Physics, 31 Caroline St. North, Waterloo, ON N2L 2Y5, Canada}

\noindent \hangindent=.5cm $^{69}${Waterloo Centre for Astrophysics, University of Waterloo, 200 University Ave W, Waterloo, ON N2L 3G1, Canada}

\noindent \hangindent=.5cm $^{70}${Graduate Institute of Astrophysics and Department of Physics, National Taiwan University, No. 1, Sec. 4, Roosevelt Rd., Taipei 10617, Taiwan}

\noindent \hangindent=.5cm $^{71}${Sorbonne Universit\'{e}, CNRS/IN2P3, Laboratoire de Physique Nucl\'{e}aire et de Hautes Energies (LPNHE), FR-75005 Paris, France}

\noindent \hangindent=.5cm $^{72}${Department of Astronomy and Astrophysics, University of California, Santa Cruz, 1156 High Street, Santa Cruz, CA 95065, USA}

\noindent \hangindent=.5cm $^{73}${Department of Astronomy \& Astrophysics, University of Toronto, Toronto, ON M5S 3H4, Canada}

\noindent \hangindent=.5cm $^{74}${University of Science and Technology, 217 Gajeong-ro, Yuseong-gu, Daejeon 34113, Republic of Korea}

\noindent \hangindent=.5cm $^{75}${Departament de F\'{i}sica, Serra H\'{u}nter, Universitat Aut\`{o}noma de Barcelona, 08193 Bellaterra (Barcelona), Spain}

\noindent \hangindent=.5cm $^{76}${Laboratoire de Physique Subatomique et de Cosmologie, 53 Avenue des Martyrs, 38000 Grenoble, France}

\noindent \hangindent=.5cm $^{77}${Instituci\'{o} Catalana de Recerca i Estudis Avan\c{c}ats, Passeig de Llu\'{\i}s Companys, 23, 08010 Barcelona, Spain}

\noindent \hangindent=.5cm $^{78}${Max Planck Institute for Extraterrestrial Physics, Gie\ss enbachstra\ss e 1, 85748 Garching, Germany}

\noindent \hangindent=.5cm $^{79}${Department of Physics and Astronomy, Siena College, 515 Loudon Road, Loudonville, NY 12211, USA}

\noindent \hangindent=.5cm $^{80}${Department of Physics \& Astronomy, University  of Wyoming, 1000 E. University, Dept.~3905, Laramie, WY 82071, USA}

\noindent \hangindent=.5cm $^{81}${National Astronomical Observatories, Chinese Academy of Sciences, A20 Datun Rd., Chaoyang District, Beijing, 100012, P.R. China}

\noindent \hangindent=.5cm $^{82}${Aix Marseille Univ, CNRS, CNES, LAM, Marseille, France}

\noindent \hangindent=.5cm $^{83}${Ruhr University Bochum, Faculty of Physics and Astronomy, Astronomical Institute (AIRUB), German Centre for Cosmological Lensing, 44780 Bochum, Germany}

\noindent \hangindent=.5cm $^{84}${Departament de F\'isica, EEBE, Universitat Polit\`ecnica de Catalunya, c/Eduard Maristany 10, 08930 Barcelona, Spain}

\noindent \hangindent=.5cm $^{85}${Universit\'{e} Clermont-Auvergne, CNRS, LPCA, 63000 Clermont-Ferrand, France}

\noindent \hangindent=.5cm $^{86}${University of California Observatories, 1156 High Street, Sana Cruz, CA 95065, USA}

\noindent \hangindent=.5cm $^{87}${Department of Physics \& Astronomy, Ohio University, Athens, OH 45701, USA}

\noindent \hangindent=.5cm $^{88}${Department of Physics and Astronomy, Sejong University, Seoul, 143-747, Korea}

\noindent \hangindent=.5cm $^{89}${Abastumani Astrophysical Observatory, Tbilisi, GE-0179, Georgia}

\noindent \hangindent=.5cm $^{90}${Faculty of Natural Sciences and Medicine, Ilia State University, 0194 Tbilisi, Georgia}

\noindent \hangindent=.5cm $^{91}${Space Telescope Science Institute, 3700 San Martin Drive, Baltimore, MD 21218, USA}

\noindent \hangindent=.5cm $^{92}${Centre for Advanced Instrumentation, Department of Physics, Durham University, South Road, Durham DH1 3LE, UK}

\noindent \hangindent=.5cm $^{93}${Physics Department, Brookhaven National Laboratory, Upton, NY 11973, USA}

\noindent \hangindent=.5cm $^{94}${Beihang University, Beijing 100191, China}

\noindent \hangindent=.5cm $^{95}${Department of Astronomy, Tsinghua University, 30 Shuangqing Road, Haidian District, Beijing, China, 100190}

\noindent \hangindent=.5cm $^{96}${Physics Department, Stanford University, Stanford, CA 93405, USA}

\noindent \hangindent=.5cm $^{97}${Department of Physics, University of California, Berkeley, 366 LeConte Hall MC 7300, Berkeley, CA 94720-7300, USA}

\end{document}